\pdfoutput=1

\documentclass[prb,aps,amsmath,amssymb,onecolumn]{revtex4}

\usepackage{graphicx,bm,amsmath}
\usepackage{color}
\usepackage{comment}
\usepackage[colorlinks,
citecolor=blue,linkcolor=red,urlcolor=blue]{hyperref}

\newcommand{\be}{\begin{equation}}
\newcommand{\ee}{\end{equation}}
\newcommand{\bea}{\begin{eqnarray}}
\newcommand{\eea}{\end{eqnarray}}

\newcommand{\up}{\uparrow}
\newcommand{\down}{\downarrow}
\newcommand{\avg}[1]{\left< #1 \right>}
\newcommand{\abs}[1]{\left| #1 \right|}
\def\tS{\widetilde{S}}
\def\texo#1{\texorpdfstring{#1}{Lg}}
\def\nn{\nonumber\\}
\def\fr#1{(\ref{#1})}

\begin{document}

\title{Full counting statistics in the spin-1/2 Heisenberg XXZ chain }
\author{Mario Collura}
\affiliation{The Rudolf Peierls Centre for Theoretical Physics,
Oxford University, Oxford, OX1 3NP, UK}
\author{Fabian H.L. Essler} 
\affiliation{The Rudolf Peierls Centre for Theoretical Physics, Oxford
University, Oxford, OX1 3NP, UK}
\author{Stefan Groha} 
\affiliation{The Rudolf Peierls Centre for Theoretical Physics, Oxford
University, Oxford, OX1 3NP, UK}

\begin{abstract}
The spin-1/2 Heisenberg chain exhibits a quantum critical regime
characterized by quasi long-range magnetic order at zero temperature.
We quantify the strength of quantum fluctuations in the ground state
by determining the probability distributions of the components of the
(staggered) subsystem magnetization. Some of these exhibit scaling and
the corresponding universal scaling functions can be determined by
free fermion methods and by exploiting a relation with the boundary
sine-Gordon model. 
\end{abstract}


\maketitle

\section{Introduction}
Universality is a key organizing principle for continuous phase
transitions\cite{cardy,sachdev}. It posits that certain quantities
are independent of microscopic details and coincide in different
physical systems that belong to the same ``universality class''. The
latter are determined by properties such as symmetries and
dimensionality and are amenable to field theory
descriptions. In 1+1 dimensions this permits the exact description of
universal properties such as critical exponents and correlation
functions at conformally invariant quantum critical points. As
emphasized in Ref.~\onlinecite{LF}, less familiar quantities like
the order parameter probability distribution function display
universal scaling as well. In quantum theory these probability
distributions describe the statistics of measurements on identical
systems, which generally give rise to different outcomes. Their
analysis provides very detailed information about the  physical
properties of many-particle systems and has been explored in a variety
of areas including condensed matter\cite{CM,CM2} and cold atom
physics\cite{CA1,CA2,CA3,CA4}. Theoretical results on full counting
statistics in quantum critical systems are relatively scarce. The list
of available results includes phase fluctuations in
Luttinger liquids\cite{gritsev,IGD06,IGD08,lovas}, the order parameter
statistics in the Ising field theory\cite{LF}, the transverse
magnetization in the Ising chain\cite{cd-07} and 
the magnetization in the Haldane-Shastry model\cite{stephan}. Here we
consider the (staggered) subsystem magnetization in the anisotropic
one-dimensional spin-$1/2$ Heisenberg XXZ chain
\be
H = J \sum_{j=1}^{L} 
S^{x}_{j}S^{x}_{j+1} + S^{y}_{j}S^{y}_{j+1} + \Delta\,
S^{z}_{j}S^{z}_{j+1} \ .
\label{H_XXZ}
\ee
The XXZ chain is a paradigmatic model for quantum critical behaviour
in 1+1 dimensions. It features a critical line parametrized by the
exchange anisotropy $-1\leq\Delta\leq 1$. The special values
$\Delta=\pm 1$ correspond to the isotropic antiferromagnet and
ferromagnet respectively. In the regime $-1<\Delta\leq 1$ the
low-energy behaviour of the model \fr{H_XXZ} is described by Luttinger
liquid theory or equivalently a free, compact
boson\cite{LutherPeschel,affleck89,GNT,gia}. The long-distance
asymptotics of spin-spin correlation functions is of the form
\bea
\langle{\rm GS}|S^x_{j+n}S^x_j|{\rm GS}\rangle&=&(-1)^n\frac{A}{4n^\eta}\left(1-\frac{B}{n^{4/\eta-4}}\right)
-\frac{\tilde{A}}{4n^{\eta+1/\eta}}\left(1+\frac{\tilde{B}}{n^{2/\eta-2}}\right)+\ldots 
\ ,\nn
\langle{\rm GS}
|S^z_{j+n}S^z_j|{\rm GS}\rangle&=&-\frac{1}{4\pi^2\eta n^2}\left(1+\frac{\tilde{B}_z}{n^{4/\eta-4}}\frac{4-3\eta}{2-2\eta}\right)+
(-1)^n\frac{A_z}{4n^{1/\eta}}\left(1-\frac{B_z}{n^{2/\eta-2}}\right)+\ldots
\label{corrs2}
\eea
where explicit expressions for the amplitudes in \fr{corrs2} are
known \cite{lukyanov,Affleck98,lukyanov2,lukyanovterras} and
$\eta$ is related to the anisotropy parameter $\Delta$ by
\be
\Delta=-\cos(\pi\eta).
\ee
It follows from \fr{corrs2} that throughout the critical regime the
dominant correlations are those of the staggered magnetizations in the
xy-plane. The XXZ chain thus exhibits antiferomagnetic quasi-long
range order in the XY plane in spin space. Two-point functions such
s \fr{corrs2} are a standard means for characterizing physical
properties and identifying ground state ``phases'' in quantum critical
systems\cite{gia}. A key objective of our work is to provide a
complementary characterization of ground state properties in the
critical XXZ chain by determining the quantum mechanical fluctuations
of the subsystem magnetization in the ground state.
More precisely we consider the probability distributions of the
following observables
\be
S^\alpha(\ell)=\sum_{j=1}^\ell S^\alpha_j\ ,\qquad
N^\alpha(\ell)=\sum_{j=1}^\ell (-1)^jS^\alpha_j\ .
\label{SN}
\ee
The quantities $S^\alpha(\ell)$ and $N^\alpha(\ell)$  describe the
smooth and staggered components of the $\alpha$-component of the
magnetization of the subsystem consisting of sites $1$ to $\ell$,
where $\ell\ll L$. We note that whereas $S^z(L)$ is a conserved
quantity, $S^z(\ell)$ is not. The probabilities of the
observables \fr{SN} taking some value $m$ when the system is prepared
in the ground state and a measurement is then performed are 
\bea
P^\alpha_{S}(m,\ell)&=&\langle{\rm GS}|\delta(S^\alpha(\ell)-m)|{\rm GS}\rangle
=\int_{-\infty}^\infty \frac{d\theta}{2\pi}\ e^{-i\theta m}\ 
\langle{\rm GS}|e^{i\theta S^\alpha(\ell)}|{\rm GS}\rangle\ ,\nn
P_{N}^\alpha(m,\ell)&=&\langle{\rm GS}|\delta(N^\alpha(\ell)-m)|{\rm GS}\rangle
=\int_{-\infty}^\infty \frac{d\theta}{2\pi}\ e^{-i\theta m}\ 
\langle{\rm GS}|e^{i\theta N^\alpha(\ell)}|{\rm GS}\rangle\ .
\label{probdist}
\eea
As we have already mentioned, probability distributions
like \fr{probdist} are experimentally measurable in cold atom
experiments. The central objects of our analysis are the
generating functions of the moments of the probability
distributions \fr{probdist} are  
\bea
G^\alpha_\ell(\theta)\equiv 
\langle{\rm GS}|e^{i\theta S^\alpha(\ell)}|{\rm GS}\rangle\ ,\quad
F^\alpha_\ell(\theta)\equiv 
\langle{\rm GS}|e^{i\theta N^\alpha(\ell)}|{\rm GS}\rangle\ .
\label{defG}
\eea
It is easy to see that they have the following properties
\bea
X^\alpha_\ell(0)&=&1\ ,\quad
X^\alpha_\ell(-\theta)=\Big(X^\alpha_\ell(\theta)\Big)^*\ ,\quad
X^\alpha_\ell(\theta+2\pi)=(-1)^\ell X^\alpha_\ell(\theta)\ ,\quad
X=F,G.
\label{relations}
\eea
The last relation allows us to restrict our attention to the interval 
$0\leq\theta<2\pi$ and can be obtained e.g. from the representation
\be
e^{i\theta S^\alpha(\ell)}=\prod_{j=1}^\ell\left[\cos(\theta/2)
+i\sin(\theta/2)\sigma^\alpha_j\right].
\ee
Defining
\be
\widetilde{X}^\alpha_\ell(r)=\int_{-\pi}^\pi\frac{d\theta}{2\pi}\
e^{-ir\theta}\
X^\alpha_\ell(\theta)\ ,\quad X=F,G,
\ee
the probability distributions of interest can be expressed as
\be
P_{N}^\alpha(m,\ell)=
\begin{cases}
\sum_{r\in\mathbb{Z}}\widetilde{F}^\alpha_\ell(r)\ \delta(m-r)& \text{if\
}
\ell \text{ is even,}\\
\sum_{r\in\mathbb{Z}}\widetilde{F}^\alpha_\ell\big(r+\frac{1}{2}\big)\ 
\delta\big(m-r-\frac{1}{2}\big)& \text{if\
}
\ell \text{ is odd.}
\end{cases}
\label{PxN}
\ee
An analogous equation holds for $P_S^\alpha(m,\ell)$.

\subsection{Moments of the probability distributions}
As we are not imposing a magnetic field and spontaneous symmetry
breaking of the U(1) symmetry of the Heisenberg Hamiltonian is
forbidden in one spatial dimension, translational invariance implies
that the averages of $S^\alpha(\ell)$ and $N^\alpha(\ell)$ vanish
\be
\langle{\rm GS}|S^\alpha(\ell)|{\rm GS}\rangle=0=
\langle{\rm GS}|N^\alpha(\ell)|{\rm GS}\rangle.
\ee
The variances have the following asymptotic expansions for large
sub-system sizes $\ell$
\be
\langle {\rm GS}|\big(S^\alpha(\ell)\big)^2|{\rm GS}\rangle=\ell s_\alpha+o(\ell)\
,\quad
\langle{\rm GS}|\big(N^\alpha(\ell)\big)^2|{\rm GS}\rangle=\ell n_\alpha+o(\ell)\ .
\ee
For sufficiently large values of $\ell$ we expect the coefficients
$s_\alpha$ and $n_\alpha$ to be equal to the corresponding quantities
for the entire system, i.e.
\be
s_x=s_y=\lim_{L\to\infty}\frac{1}{L}\langle{\rm
GS}|\big(S^x_L\big)^2|{\rm GS}\rangle\ ,\quad
n_\alpha=\lim_{L\to\infty}\frac{1}{L}\langle{\rm
GS}|\big(N^\alpha_L\big)^2|{\rm GS}\rangle.
\label{sx}
\ee
As $S^z(L)$ is a conserved quantity and our system is translationally
invariant we have $s_z=0$. It is instructive to consider the
calculation of the variance of the subsystem magnetization by field
theory methods. As the variances are non-universal quantities they are
expected to be susceptible to short-distance physics, and this is
indeed borne out by the explicit calculation summarized in
Appendix \ref{app:variances}. 

While the moments themselves depend on microscopic details, certain
ratios can be universal\cite{privman,goldenfeld,LF}. In particular
one may expect the following ratios to exhibit universal behaviour
\be
\frac{\big\langle\left(S^\alpha(\ell)\right)^{2n}\big\rangle}
{\big\langle\left(S^\alpha(\ell)\right)^{2}\big\rangle^n}\ ,\qquad
\frac{\big\langle\left(N^\alpha(\ell)\right)^{2n}\big\rangle}
{\big\langle\left(N^\alpha(\ell)\right)^{2}\big\rangle^n}\ .
\label{moments}
\ee
If these ratios are universal, the modified generating functions
\be
\langle{\rm GS}|e^{i\theta_Y Y^\alpha(\ell)}|{\rm GS}\rangle\ ,\quad
\theta_Y=\frac{\theta}{\sqrt{\langle{\rm GS}|
\big(Y^\alpha(\ell)\big)^2|{\rm GS}\rangle}}\ ,\quad
Y=S,N,
\label{rescaling}
\ee
will be universal functions of the parameter $\theta$. This means in
particular that they can be calculated by field theory methods. In
practice \fr{rescaling} tells us that the moment generating functions
calculated from field theory and computed directly in the lattice
model should agree up to an overall rescaling of the parameter $\theta$.

\section{Field theory description of the XXZ chain} 
It is well established that the long distance behaviour of local
equal time correlation functions in the critical XXZ chain is well
described by (perturbed) Luttinger liquid
theory\cite{AGSZ89,lukyanov,Affleck98,lukyanovterras,Lyon}. 
In absence of a magnetic field the Hamiltonian can be cast in the form 
\be
{\cal H}(\Delta) =\frac{v}{2} \int dx \left[K (\partial_x\theta)^2
+\frac{1}K (\partial_x\phi)^2\right]+\dots\ ,
\ee
where $\phi$ and $\theta$ are Bose fields with commutation relations 
$[\phi(t,x),\theta(t,y)]=(i/2){\rm sgn}(x-y)$, the dots indicate perturbations that are
irrelevant in the renormalization group sense and
\be
v=  \frac{\pi}{2}\frac{\sqrt{1-\Delta^2}}{\arccos \Delta},\qquad
K  = \frac{\pi}{2}\frac{1}{\pi-\arccos \Delta },
\label{KvBA}
\ee
The bosonization formulas for the spin operators are
\bea
S^z_j&\simeq&-\frac{a_0}{\sqrt{\pi }}\partial_x\phi(x)
+(-1)^j c_1 \,\sin (\sqrt{4 \pi } {\phi(x)})+\ldots,  \label{Szbos} \\
S^{x}_j&\simeq& b_0(-1)^j\cos\big(\sqrt{\pi} \,\theta(x)\big)
+ib_1\sin\big(\sqrt{\pi} \,\theta(x)\big)
 \sin\big({\sqrt{4 \pi }\phi(x)}\big)+\ldots \, \label{S-bos},
\eea
where $a_0$ is the lattice spacing and the amplitudes $b_0$, $c_1$,
$b_1$ are known exactly \cite{lukyanovterras}. For large subsystem sizes we
thus have
\bea
S^z(\ell)&\approx&-\frac{1}{\sqrt{\pi }}\left[\phi(\ell
a_0)-\phi(0)\right]+\dots\ ,\nn
N^z(\ell)&\approx&\frac{c_1}{a_0}\int_0^\ell dx\ \sin\big(\sqrt{4 \pi }
\phi(x)\big)+\dots\
,\nn
N^x(\ell)&\approx&\frac{b_0}{a_0}\int_0^\ell dx\ \cos\big(\sqrt{\pi }
\theta(x)\big)+\dots\ .
\label{SNbos}
\eea
Applying the bosonization prescription to our generating functions and
ignoring subleading terms we obtain
\bea
G^z_\ell(\theta)&\approx&\big\langle 0|e^{-i\frac{\theta}{\sqrt{\pi}}\phi(\ell
a_0)}
e^{i\frac{\theta}{\sqrt{\pi}}\phi(0)}|0\big\rangle\ ,\nn
F^z_\ell(\theta)&\approx&\big\langle  0|
e^{-i\theta\frac{c_1}{a_0}\int_0^\ell dx\ \sin\big(\sqrt{4 \pi }
\phi(x)\big)}|0\big\rangle\ ,\nn
F^x_\ell(\theta)&\approx&\big\langle  0|
e^{-i\theta\frac{b_0}{a_0}\int_0^\ell dx\ \cos\big(\sqrt{\pi }
\phi(x)\big)}|0\big\rangle\ ,
\label{FTGF}
\eea
where $|0\rangle$ is the Fock vacuum. The representation \fr{FTGF}
reveals that $G^z_\ell(\theta)$ maps onto a simple vertex operator
two-point function in the free boson theory, whereas
$F^\alpha_\ell(\theta)$ correspond to expectation values of non-local
operators. The alert reader will have noted that we did not provide
a bosonized expression for $G^x_\ell(\theta)$. The reason is that 
the field theory calculation of $G^x_\ell(\theta)$ is easier in a
somewhat different setup and we return to this issue in
section~\ref{ssec:Gx}. 
\section{Generating functions for the staggered subsystem magnetization} 
\label{sec:stagg}
We start by considering the probability distributions of the staggered
subsystem magnetizations $P^\alpha_N(m,\ell)$ and the corresponding
generating functions $F^\alpha_\ell(\theta)$. We first present
analytic results in certain limits and then compare these to numerical ones.

\subsection{The XX point \texorpdfstring{$\Delta=0$}{Lg}}
At the XX point the Heisenberg model can be mapped to non-interacting
spinless fermions by means of a Jordan-Wigner transformation. Using
standard techniques\cite{LSM} we can derive the following determinant
representation for the longitudinal generating function
$F^z_\ell(\theta)$ 
\bea
F^z_\ell(\theta) &=& \det\left[{\mathbb B} \right]\ ,\qquad
\label{FzD0}
{\mathbb B}_{nm}= e^{i\frac{\theta}{2}(-1)^n}\delta_{n,m}+2\sin(\theta/2)
(-i)^{(-1)^n+(-1)^m}\mathbb{C}_{n,m}\ .
\eea
Here $\mathbb{C}_{nm}$ is the correlation matrix of the free fermion
chain obtained by the Jordan Wigner transformation
\be
{\mathbb C}_{nm}= \frac{\sin\big(\frac{\pi}{2} (m-n)\big)}{\pi (m-n)}\ .
\ee
The matrix $\mathbb{B}_{nm}$ is a Toeplitz matrix\cite{remark1}. Its properties have
been analyzed in great detail in the context of entanglement entropies
in Ref.~\onlinecite{ce-10}. For large values of the subsystem size
$\ell$ one obtains the following asymptotic expansion\cite{ce-10}
\be
F^z_\ell(\theta)=\big(\cos(\theta)\big)^\frac{\ell}{2}
\sum_{m=-\infty}^\infty
(-1)^{m\ell}(2\ell)^{-2\big(m+\beta(\theta)\big)^2}
G^2\big(m+1+b(\theta)\big)G^2\big(1-m-b(\theta)\big)
\left[1+\frac{{c}_2\big(m+b(\theta)\big)}{\ell^2}
+\dots\right]
\label{Ffree}
\ee
where 
\bea
b(\theta)&=&\frac{1}{2\pi
i}\ln\left[\frac{\cos(\theta/2)-\sin(\theta/2)}
{\cos(\theta/2)+\sin(\theta/2)}\right]\ ,\qquad
{c}_2(\beta)=-\frac{\beta^2(1+8\beta^2)}{6}\ .
\eea
It follows from \fr{Ffree} that $F^z_\ell(\theta)$ is very small except in the
vicinities of $\theta\approx 0,\pi$. To analyze the behaviour in these
regions it is useful to define the following scaling limits: 
\begin{itemize}
\item[S1:]
$\theta\to 0$, $\ell\to\infty$, while keeping $z=\theta\ell^{1/2}$ fixed.

In this regime the generating function reduces to a simple Gaussian in
the scaling variable
\be
F^z_\ell(\theta)\sim e^{-z^2/4}\ .
\ee
\item[S2:]
$\theta\to \pi$, $\ell\to\infty$, while keeping
$y=(\pi-\theta)\ell^{1/2}$ fixed. 

In this regime the behaviour depends on the parity of the subsystem
size
\be\label{eq:Fz_scaling_pi}
F^z_\ell(\theta)\sim 
\sqrt{2} \, G^2(1/2)G^2(3/2) \,
e^{-y^2/4} \times
\left\{
\begin{array}{ll}
(-1)^{\ell/2} \ell^{-1/2} \,, & \quad\quad \ell {\rm\quad even} \\
& \\
(-1)^{(\ell-1)/2}\,\ell^{-1} [c+\log(2\ell)]\; y / \pi \,, & \quad\quad \ell {\rm\quad odd}
\end{array}
\right.
\ee
where $c = 2\log(2)+ \gamma_{E}$ and $\gamma_{E}\approx 0.577216$.
\end{itemize}
We will see in the following that the two limits S1 and S2 are useful
for analyzing numerical results for $F^z_\ell(\theta)$.

\subsection{Field theory approach}
\label{ssec:BSG}
In the field theory approach we are tasked with evaluating the
expressions \fr{FTGF} for $F^\alpha_\ell(\theta)$. This can be done by
following the analysis of Refs~\onlinecite{gritsev,IGD06,IGD08}, which
considered generating function for phase fluctuations in Luttinger
liquids. Expanding in powers of $\theta$ we obtain
\bea
F^x_\ell(\theta)&\approx&\sum_{n=0}^\infty\left(\frac{i\theta
b_0}{2a_0}\right)^{2n}\frac{1}{(2n)!}
\int_0^\ell dx_1\dots
\int_0^\ell dx_{2n}\ \langle2\cos\big(\sqrt{\pi}\theta(x_1)\big)\dots
2\cos\big(\sqrt{\pi}\theta(x_{2n})\big)\rangle\nn
&=&\sum_{n=0}^\infty\frac{1}{(n)!}\left(\frac{i\theta\ell
b_0}{4\pi a_0}\right)^{2n}\
Z^{(0)}_{2n}(L,\ell/L,K)\ ,
\eea
where
\bea
Z^{(0)}_{2n}(L,z,K)&=&\int_0^{2\pi} \frac{du_1}{2\pi}\dots
\frac{du_n}{2\pi}\int_0^{2\pi}\frac{dv_1}{2\pi}\dots\frac{dv_n}{2\pi}\frac{\prod_{j,k=1}^n
G(u_j-v_k)}{\prod_{j<k}^n G(u_j-u_k)G(v_j-v_k)}\ ,\nn
G(u)&=&\left(\frac{L}{\pi a_0}\sin\big(\frac{uz}{2}\big)\right)^{-\frac{1}{2K}}\ .
\eea
As the leading singularities of the integrand occur when $u_j\approx
v_k$ we now make the further approximation
\be
Z^{(0)}_{2n}(L,\ell/L,K)\approx Z^{(0)}_{2n}(\ell,1,K)\ .
\ee
This results in
\be
F^x_\ell(\theta)\approx\sum_{n=0}^\infty\frac{1}{(n!)^2}
\left(\frac{i\theta\ell b_0}{2 a_0}\right)^{2n}
Z^{(0)}_{2n}(\ell,1,K).
\label{BSG}
\ee
The right hand side of \fr{BSG} is equal to the partition function of
a boundary sine-Gordon model, for which exact results are
available in the literature\cite{FLS1,FLS2,FS,BLZ1,DT99,BLZ2}. Using a result of
Ref.~\onlinecite{BLZ2} for the right hand side of \fr{BSG} one has
\be
F^x_\ell(\theta)\approx A^{({\rm
vac})}(\lambda)\ ,\qquad
\lambda=\sin\big(\frac{\pi}{4K}\big)b_0
\left(\frac{\ell}{2\pi a_0}\right)^{1-1/4K}\theta\ .
\label{USF1}
\ee
The function $A^{({\rm vac})}(\lambda)$ can be computed very
efficiently from the solution of the single-particle Schr\"odinger equation
\be
-\partial_x^2\Psi(x)+\left[x^{4K-2}-\frac{1}{4x^2}\right]\Psi(x)=E\Psi(x)\ .
\label{SE}
\ee
Denoting by $\Psi^+(x,E)$ and $\chi^+(x,E)$ the solutions to \fr{SE}
with asymptotics
\bea
\Psi^+(x,E)&\sim&\sqrt{\frac{\pi x}{2K}}\qquad \text{for }x\to 0\ ,\nn
\chi^+(x,E)&\sim&x^{-2K-\frac{1}{2}}e^{-x^{4K}/4K}\qquad\text{for
}x\to\infty\ ,
\eea
we have
\be
A^{({\rm vac})}(\lambda)=\frac{1}{2}
W[\chi^+,\Psi^+]\Bigg|_{E=\rho\lambda^2}\ ,\quad
\rho= (8K)^{2-1/2K}\Gamma^2(1-1/4K),
\label{USF2}
\ee
where $W[f,g]=f\partial_x g-g\partial_x f$ denotes the Wronskian.

Let us now turn to the longitudinal generating function
$F^z_\ell(\theta)$. It has an integral representation
\bea
F^z_\ell(\theta)&\approx&\sum_{n=0}^\infty\left(\frac{i\theta
c_1}{2a_0}\right)^{2n}\frac{1}{(2n)!}
\int_0^\ell dx_1\dots
\int_0^\ell dx_{2n}\ \langle2\sin\big(\sqrt{4\pi}\phi(x_1)\big)\dots
2\sin\big(\sqrt{4\pi}\phi(x_{2n})\big)\rangle\ .
\label{FzSG}
\eea
This expression needs to be regularized because
\be
\langle e^{i\sqrt{4\pi}\phi(x)}\ e^{-i\sqrt{4\pi}\phi(x)}\rangle=
\left(\frac{L}{\pi a_0}\sin\big(\frac{\pi x}{L}\big)\right)^{-2K}\ ,
\ee
and $K$ ranges from $1/2$ at the isotropic point $\Delta=1$ to
infinity when the ferromagnetic point is approached ($\Delta\to
-1$). The right-hand side of \fr{FzSG} can again be related to the
partition function of a boundary sine-Gordon model, but the boundary
interaction for $\Delta<1$ is now irrelevant. This suggests that
$F^z_\ell(\theta)$ will be independent of $\Delta$ and equal to the
result at $\Delta=0$, i.e.
\be
F^z_\ell(\theta)\approx e^{-z^2/4}\ ,\quad z=\theta\ell^{1/2}.
\label{FzSG2}
\ee
Eqns \fr{USF1}, \fr{USF2} and \fr{FzSG2} provide us with explicit
expressions for the generating functions $F^\alpha_\ell(\theta)$ that
now can be compared to numerical results for the lattice model.

\subsection{Numerical Method}
Our numerical approach is based on the iTEBD algorithm\cite{iTEBD,iTEBD2}.
A translationally invariant MPS representation of the ground state of
the XXZ Hamiltonian is obtained as follows. We initialize the system
in the simple product state  
$\bigotimes_{j\in\mathbb{Z}} (|\uparrow\rangle_{j}+|\downarrow\rangle_{j})/\sqrt{2}$,
which admits an MPS representation with auxiliary dimension
$\chi=1$. We then evolve the state in imaginary time by the operator
$\exp(-\tau H)$ by means of a second order Suzuki-Trotter
decomposition with imaginary time-step $\tau J= 10^{-3}$. During
imaginary time evolution the MPS loses its canonical form, which we then
restore before taking expectation values of operators.

In order to control the convergence of the imaginary time algorithm
we keep track of the energy density. In practice we run the algorithm
until the energy density becomes stationary (within machine precision). 
We repeat this procedure with auxiliary dimensions of up to $\chi = 256$. 
In Table \ref{tab_E} we present our best estimates for the
ground-state energy densities for different values of $\Delta$ and
compare them to the known exact result
\be
E_{GS} = \frac{1}{4}\int_{-\infty}^{\infty} d\lambda \, 
\frac{1}{2\pi\cosh\left(\frac{\lambda}{1-\eta}\right)}
\frac{4(\Delta^2-1)}{\cosh(2\lambda)-\Delta}
+ \frac{\Delta}{4}.
\label{eq_EGS}
\ee

\begin{table}[ht]
\center
\caption{Energy densities from iTEBD and exact formula (\ref{eq_EGS}) 
for several values of $\Delta$.}\label{tab_E}
\begin{tabular}{c c c c}
\hline\hline
$\Delta$ & $E_{\rm MPS}$ & $E_{\rm GS}$ & $\delta E (\times 10^{-8})$\\
\hline
$-0.8$ & $-0.256339667$ & $-0.256339677$ & $0.96$\\

$-0.6$ & $-0.267640618$ & $-0.267640628$ & $1.03$\\

$-0.4$ & $-0.282089887$ & $-0.282089903$ & $1.62$\\

$-0.2$ & $-0.299086657$ & $-0.299086680$ & $2.23$\\

$0$ & $-0.318309858$ & $-0.318309886$ & $2.84$\\

$0.2$ & $-0.339564266$ & $-0.339564304$ & $3.84$\\

$0.4$ & $-0.362727187$ & $-0.362727227$ & $4.06$\\

$0.6$ & $-0.387725863$ & $-0.387725910$ & $4.71$\\

$0.8$ & $-0.414528779$ & $-0.414528832$ & $5.35$\\
\hline\hline
\end{tabular} 
\end{table}

Our numerical results for the energy densities differ from the exact
values by $O(10^{-8})$, which is quite satisfactory given that the
model is gapless. Once we have obtained the MPS description of the
ground state, we can straightforwardly evaluate the generating
functions $F^{\alpha}_{\ell}(\theta)$ and $G^{\alpha}_{\ell}(\theta)$ with
a computational cost that scales as $O(\ell\chi^3)$. A useful check on
the numerical accuracy of our results can be obtained by considering
the noninteracting case $\Delta=0$, where the exact determinant formula
(\ref{FzD0}) for the generating function of the longitudinal staggered
magnetization is available. The discrepancy between the iTEBD data and
the exact result increases as expected with the subsystem size
$\ell$. However up to subsystem sizes of $\ell = 200$ the relative
error of our iTEBD result is less than $0.1\%$.

\subsection{Numerical results for the transverse generating function
\texorpdfstring{$F^x_\ell(\theta)$}{Lg}} 
\begin{figure}[ht]
\begin{center}
\includegraphics[width=0.3\textwidth]{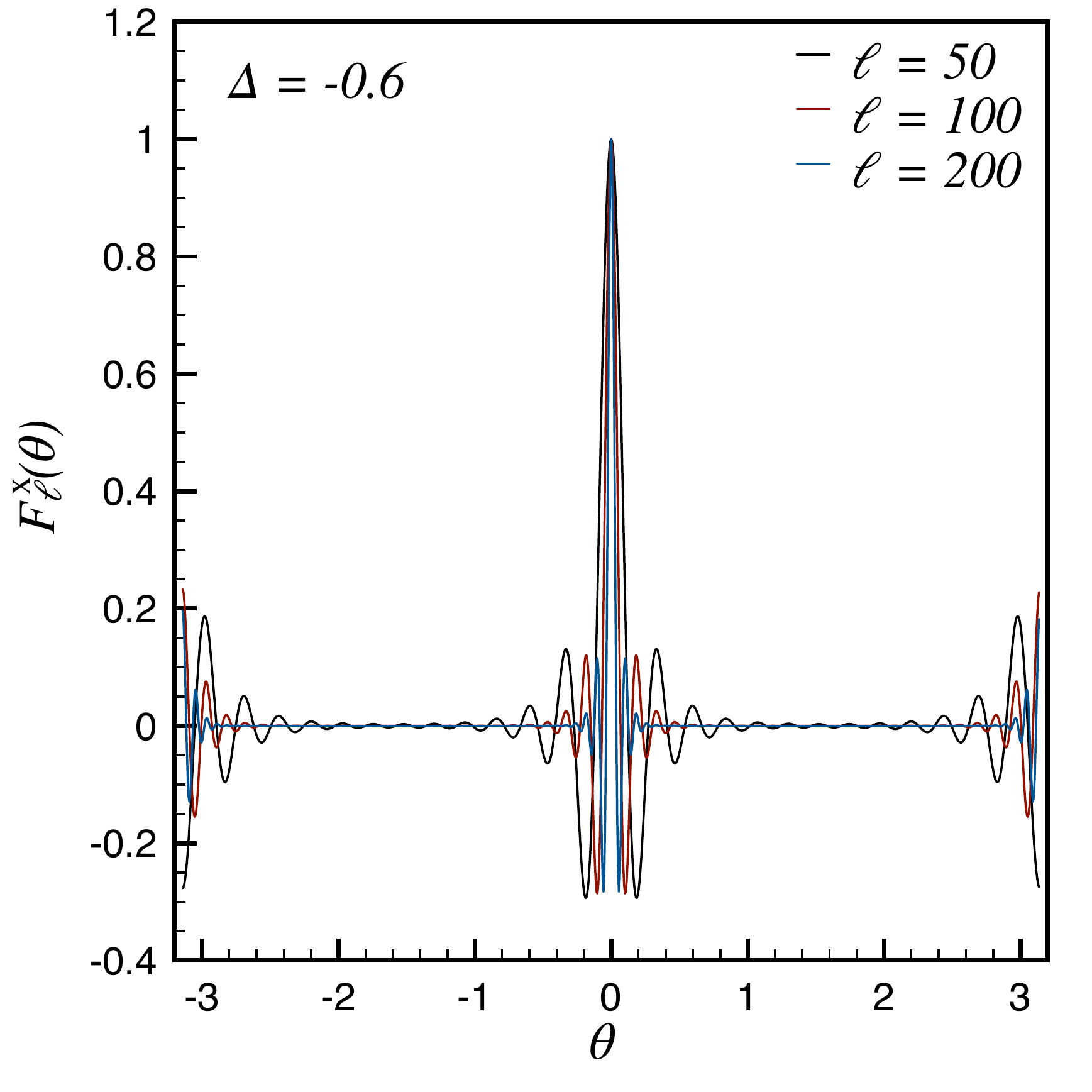}
\qquad
\includegraphics[width=0.3\textwidth]{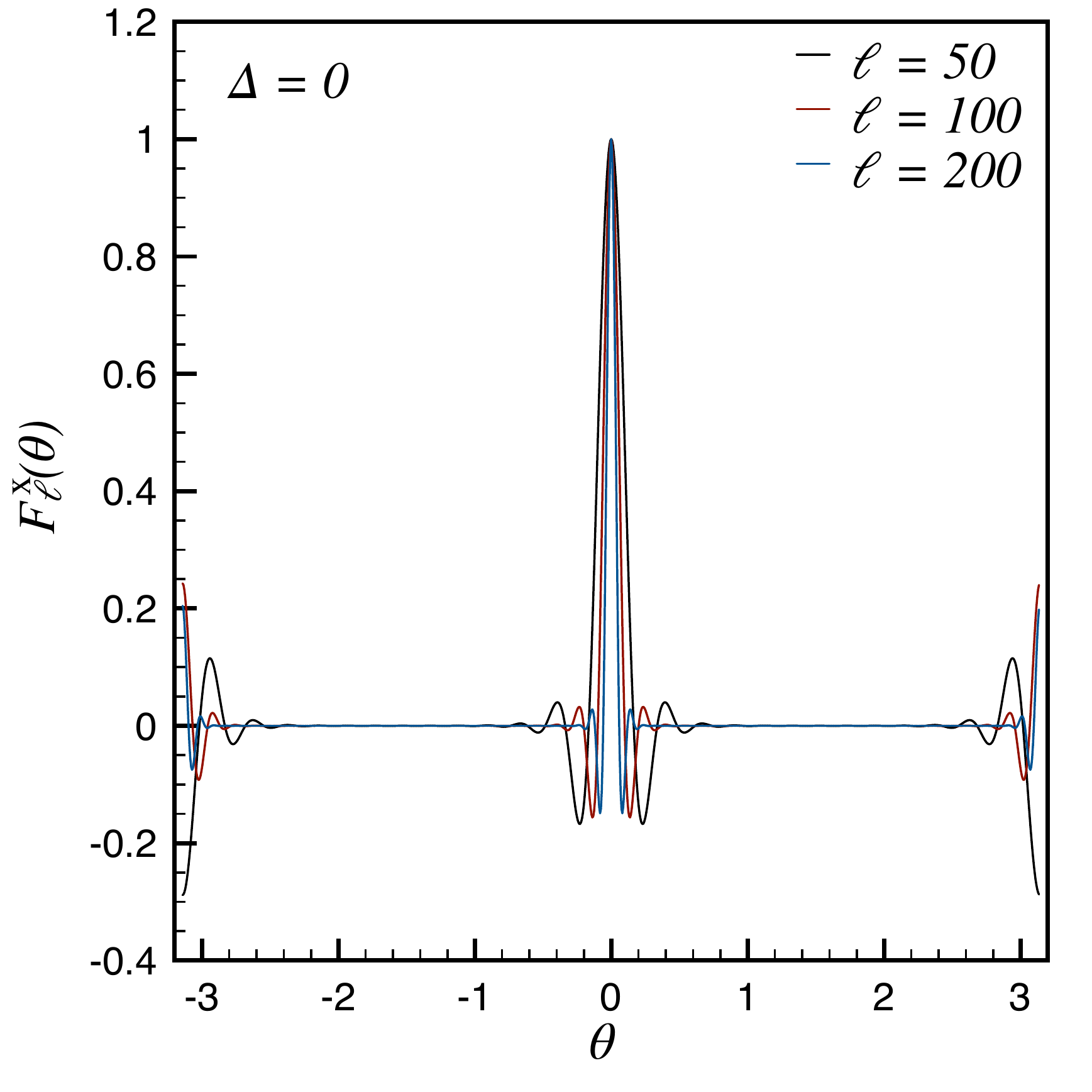}
\qquad
\includegraphics[width=0.3\textwidth]{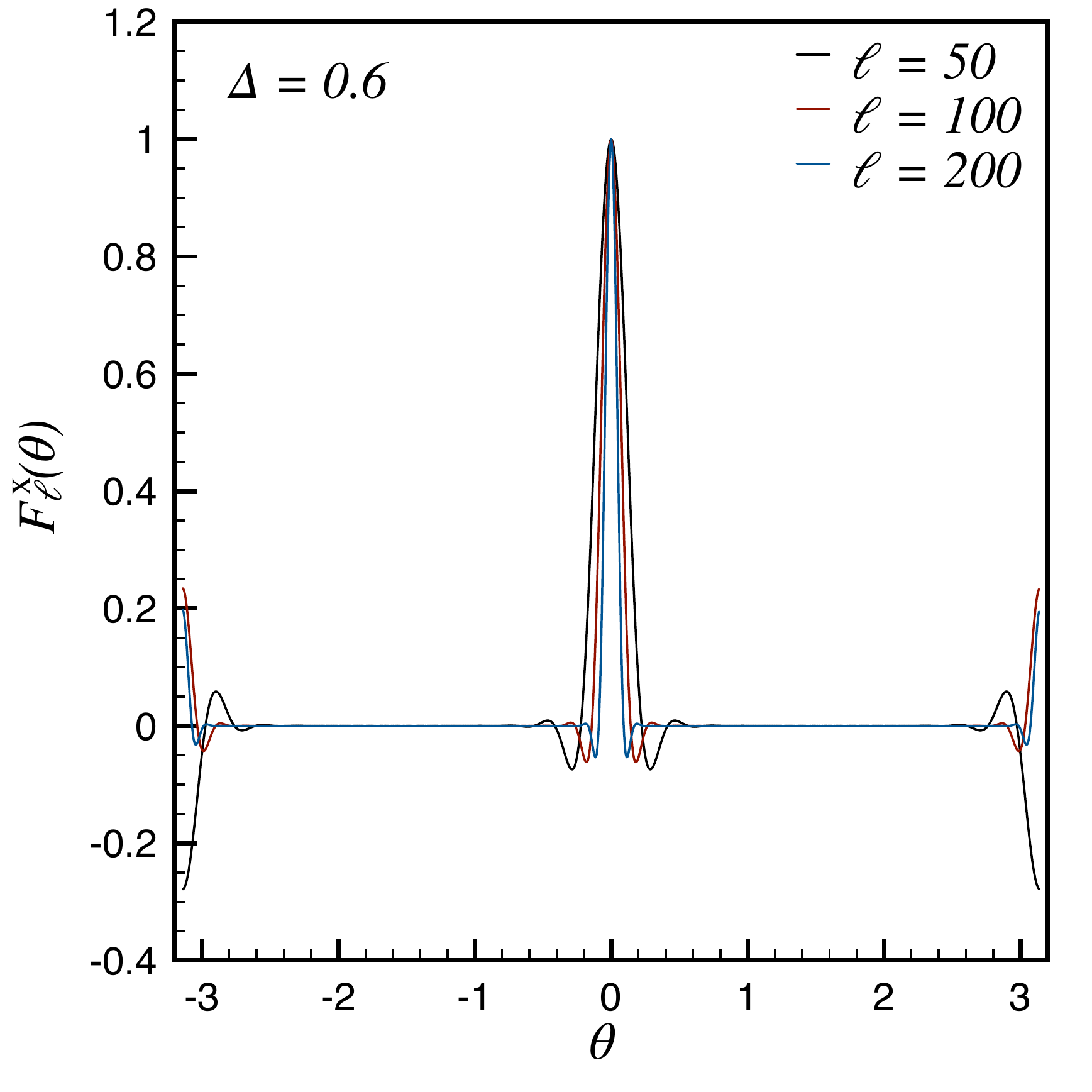}
\caption{\label{fig:Fx}
Staggered transverse generating function $F^{x}_{\ell}(\theta)$, 
for representative values of $\Delta$ and $\ell$.}
\end{center}
\end{figure}

Numerical results for $F^x_\ell(\theta)$ as a function of $\theta$ for
several values of the subsystem size $\ell$ are shown in
Fig.~\ref{fig:Fx}. 
We see that the generating function is very small
everywhere except in the vicinities of $\theta=0,\pi$. We also observe
that the oscillatory behaviour as a function of $\theta$ becomes more
pronounced in the attractive regime $\Delta<0$.

\begin{figure}[ht]
\begin{center}
\includegraphics[width=0.3\textwidth]{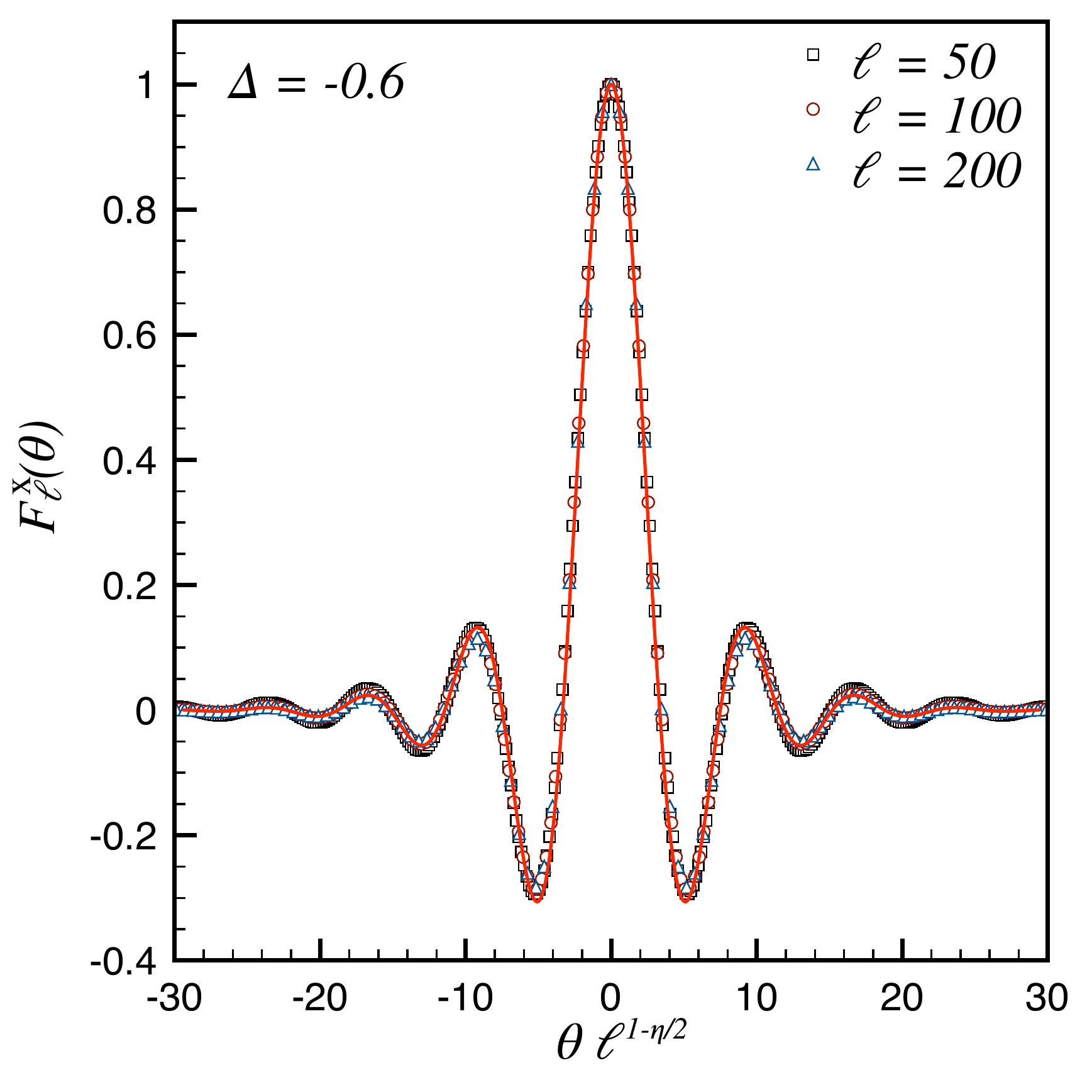}
\qquad
\includegraphics[width=0.3\textwidth]{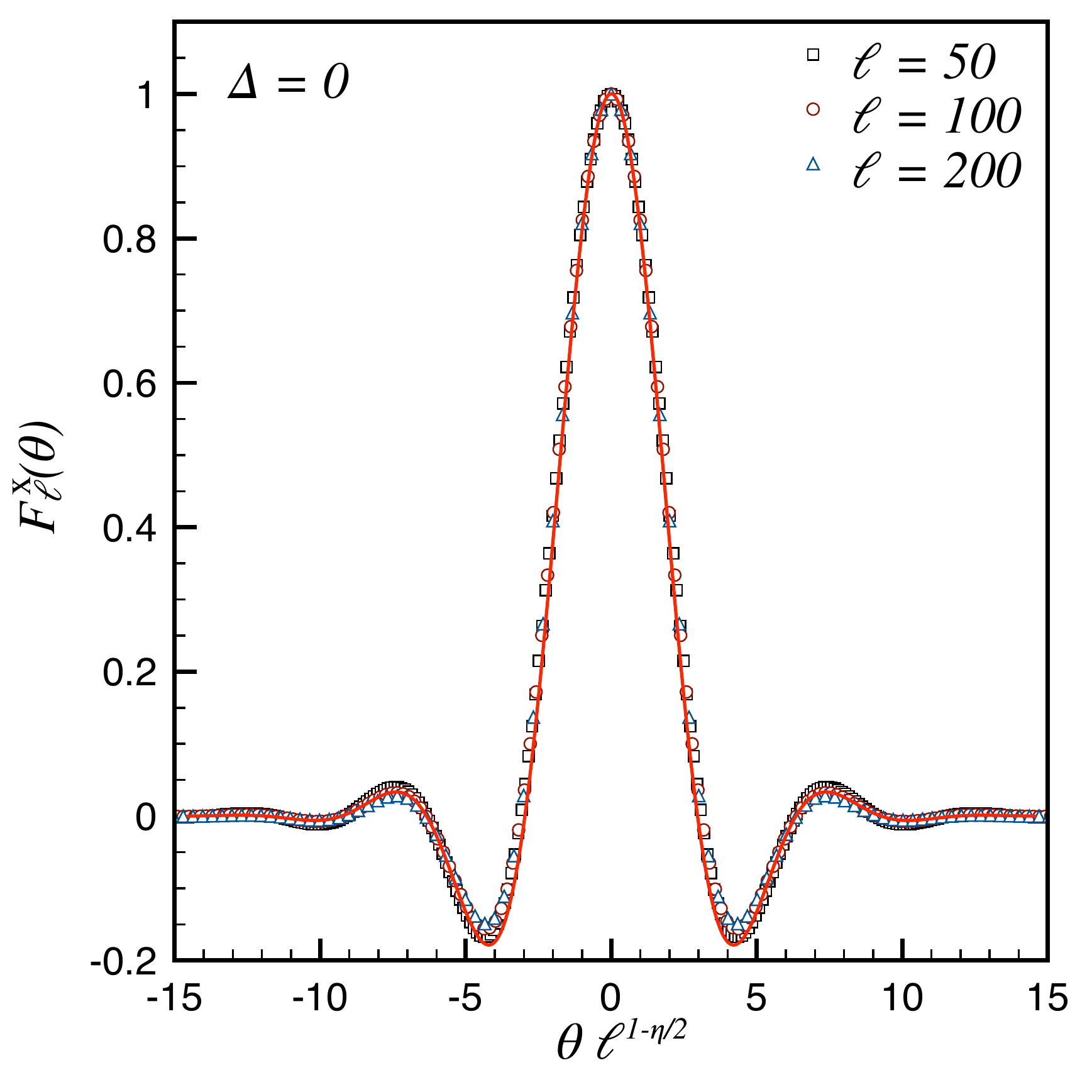}
\qquad
\includegraphics[width=0.3\textwidth]{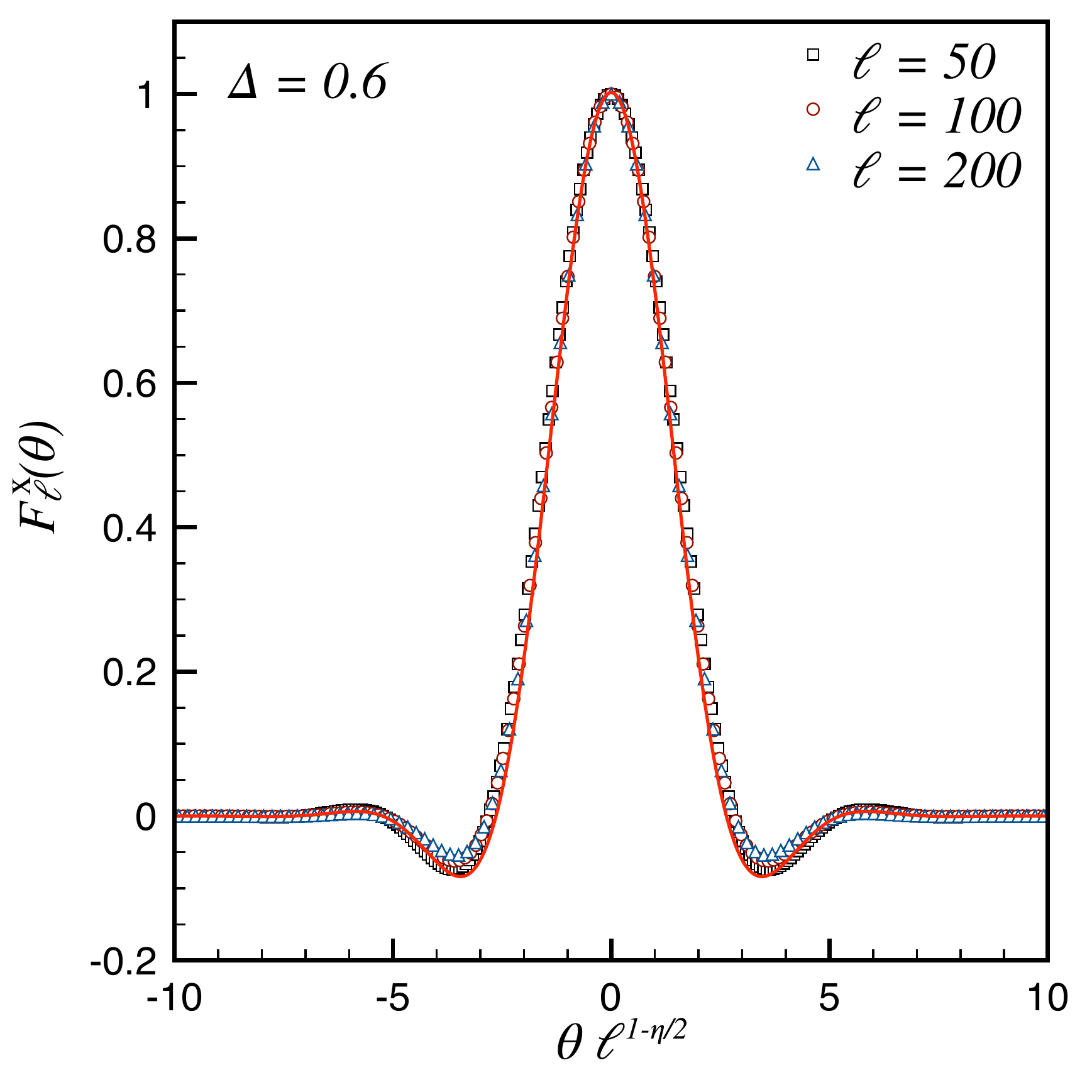}
\caption{Staggered transverse generating function $F^{x}_{\ell}(\theta)$
for several values of $\Delta$. The numerical results (symbols) are
seen to exhibit scaling collapse in the variable
$\theta \ell^{1-2\eta}$ and are well described by the universal
scaling function \fr{USF1}, \fr{USF2} calculated from the boundary
sine-Gordon model (red line).} 
\label{fig:Fx1}
\end{center}
\end{figure}

Based on the field theory analysis of section \ref{ssec:BSG} we expect
the $\theta\approx 0$ regime to exhibit scaling with a universal
scaling function given by \fr{USF1}, \fr{USF2}
\be
F^{x}_{\ell}(\theta\approx 0) =
\mathcal{F}^{x}_0(z)\ ,\quad
z=\theta \ell^{1-\eta/2}\ .
\label{scalingansatz1}
\ee
Here ${\cal F}^{x}_0(z)$ is related to the function $A^{({\rm
vac})}(\lambda)$ in \fr{USF2} by
\be
\mathcal{F}^{x}_0(z)=
A^{({\rm vac})}(cz)\ ,
\label{FA}
\ee
where $c$ is a non-universal $\Delta$-dependent constant that arises
from the fact that while the ratios \fr{moments} are universal, the
second moment itself is not, \emph{cf.} the discussion preceding
eqn \fr{rescaling}. In practice we determine $c$ by carrying out a best
fit of our numerical data to \fr{FA}. In Fig.~\ref{fig:Fx1} we present
a comparison of our numerical results for $F^x_\ell(\theta)$ to the
field theory prediction \fr{USF1}, \fr{USF2}. We see that the numerical data
exhibits scaling collapse and the agreement with the theoretical
scaling function is clearly very good. This holds for all values of
$\Delta$ we have considered in the critical regime $-1<\Delta\leq
1$. We again see that in the attractive regime $\Delta<0$ the oscillatory
behaviour away from $\theta=0$ becomes more pronounced.  

\begin{figure}[ht]
\begin{center}
\includegraphics[width=0.3\textwidth]{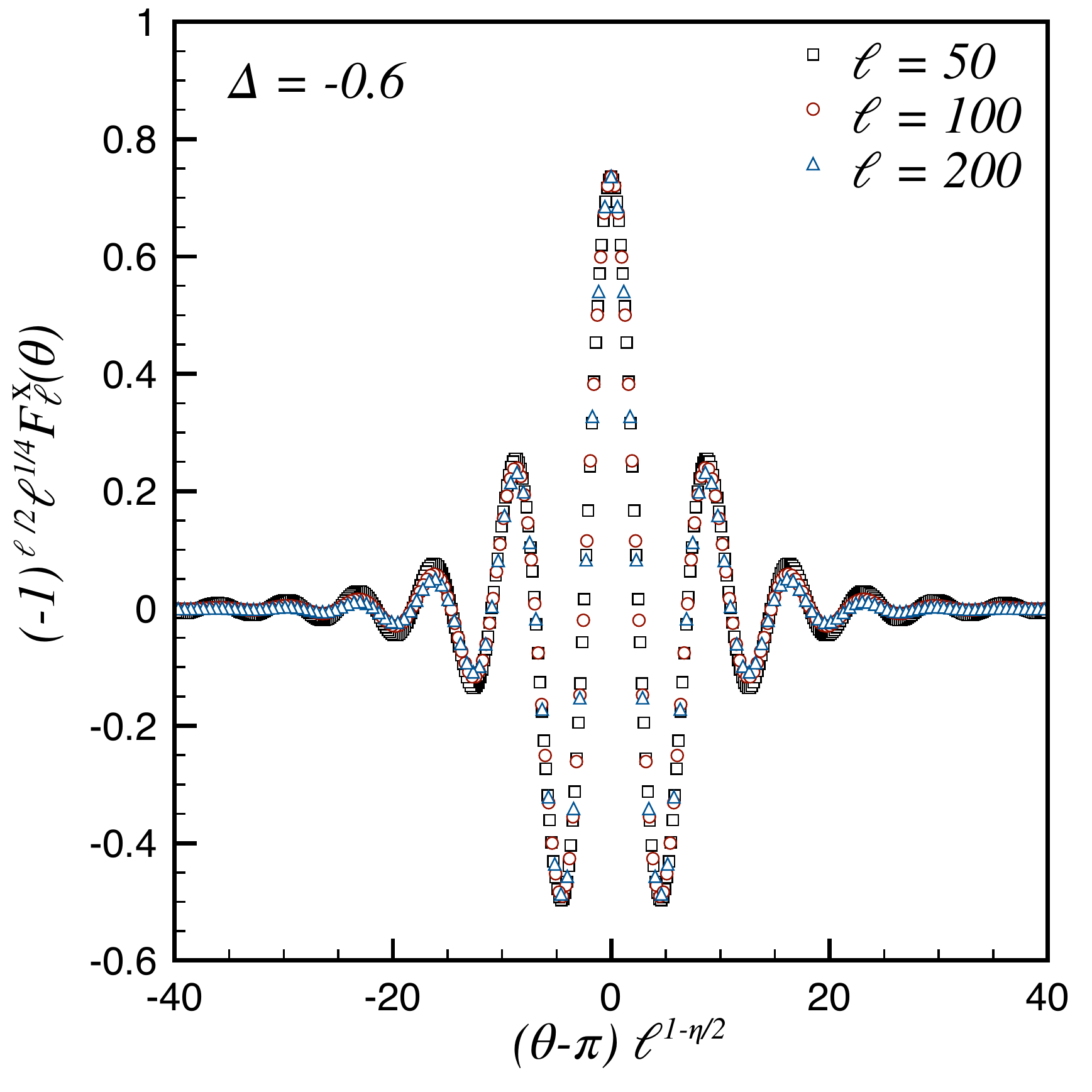}
\qquad
\includegraphics[width=0.3\textwidth]{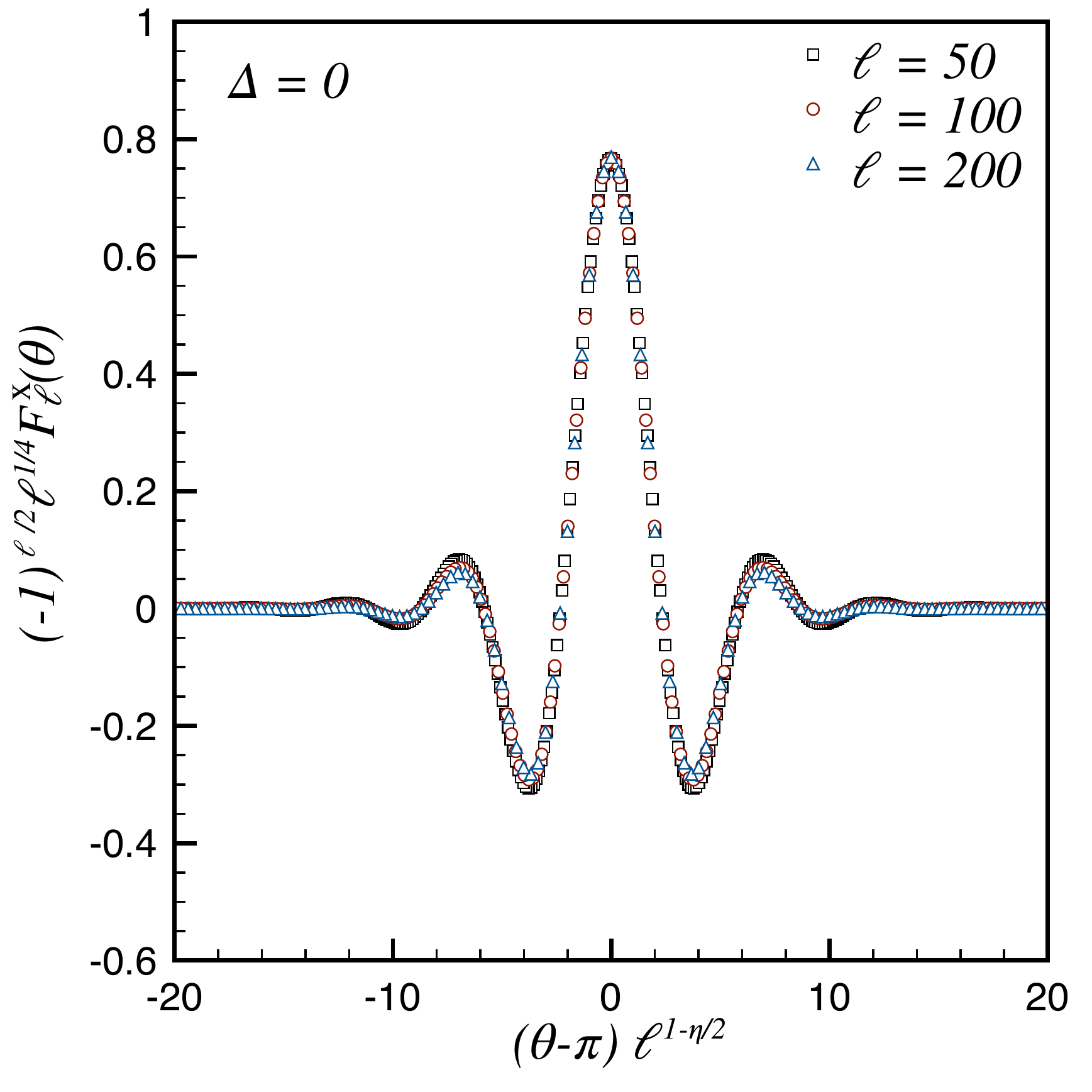}
\qquad
\includegraphics[width=0.3\textwidth]{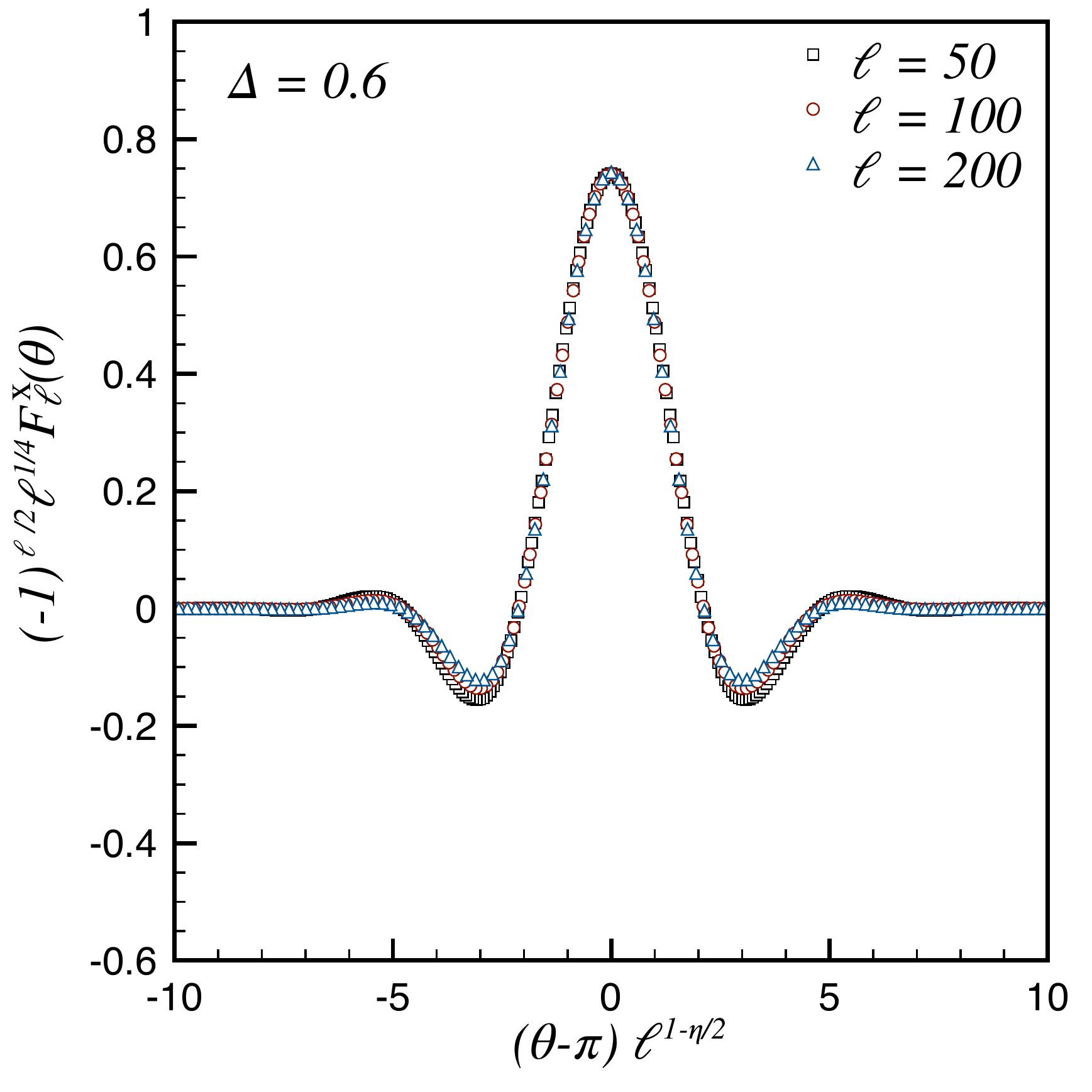}
\caption{\label{fig:Fx_pi_even}
Scaling behaviour of the staggered transverse generating function
$F^{x}_{\ell}(\theta)$ for $\theta\approx\pi$, even subsystem sizes
$\ell$ and several values of exchange anisotropy $\Delta$.}
\end{center}
\end{figure}

\begin{figure}[ht]
\begin{center}
\includegraphics[width=0.3\textwidth]{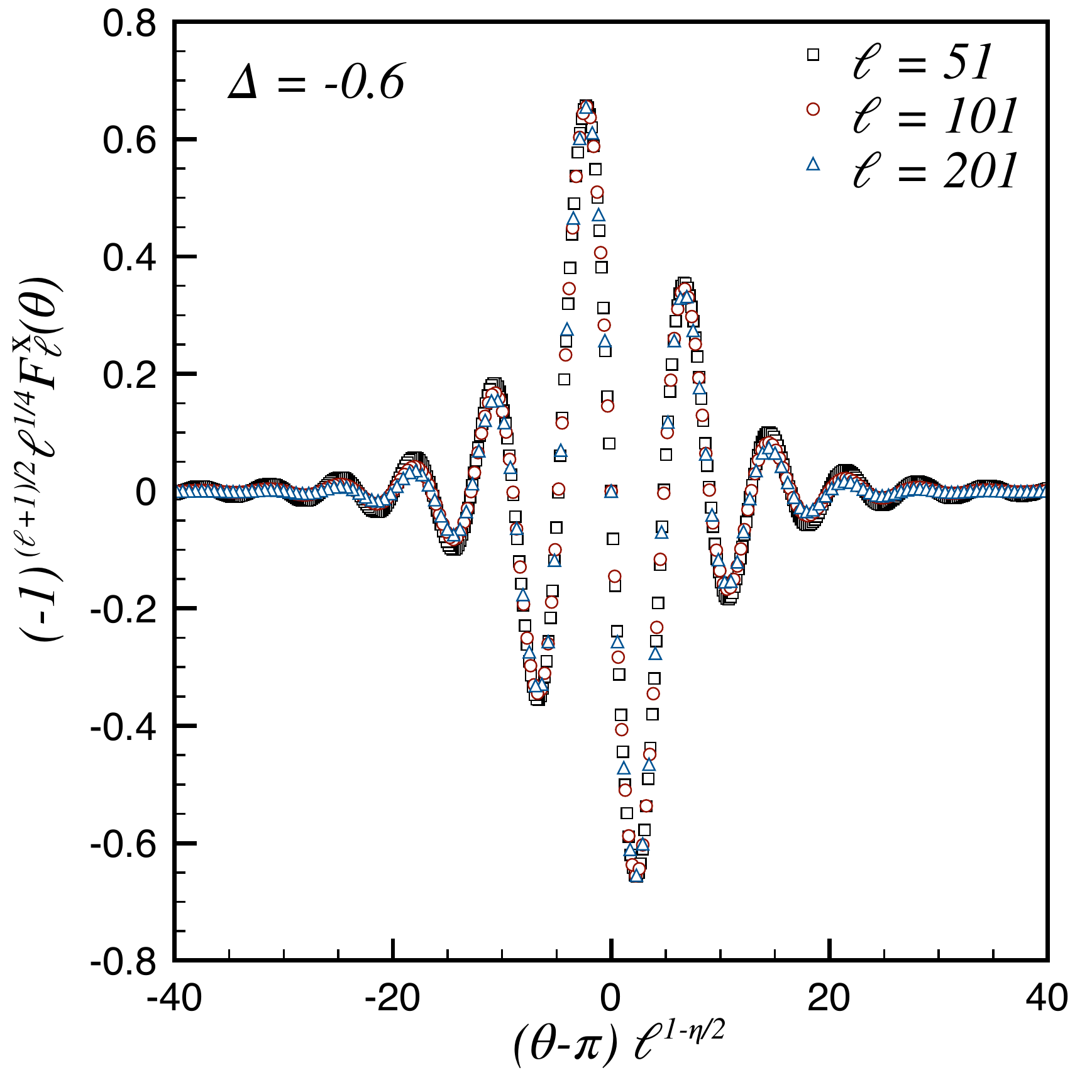}
\qquad
\includegraphics[width=0.3\textwidth]{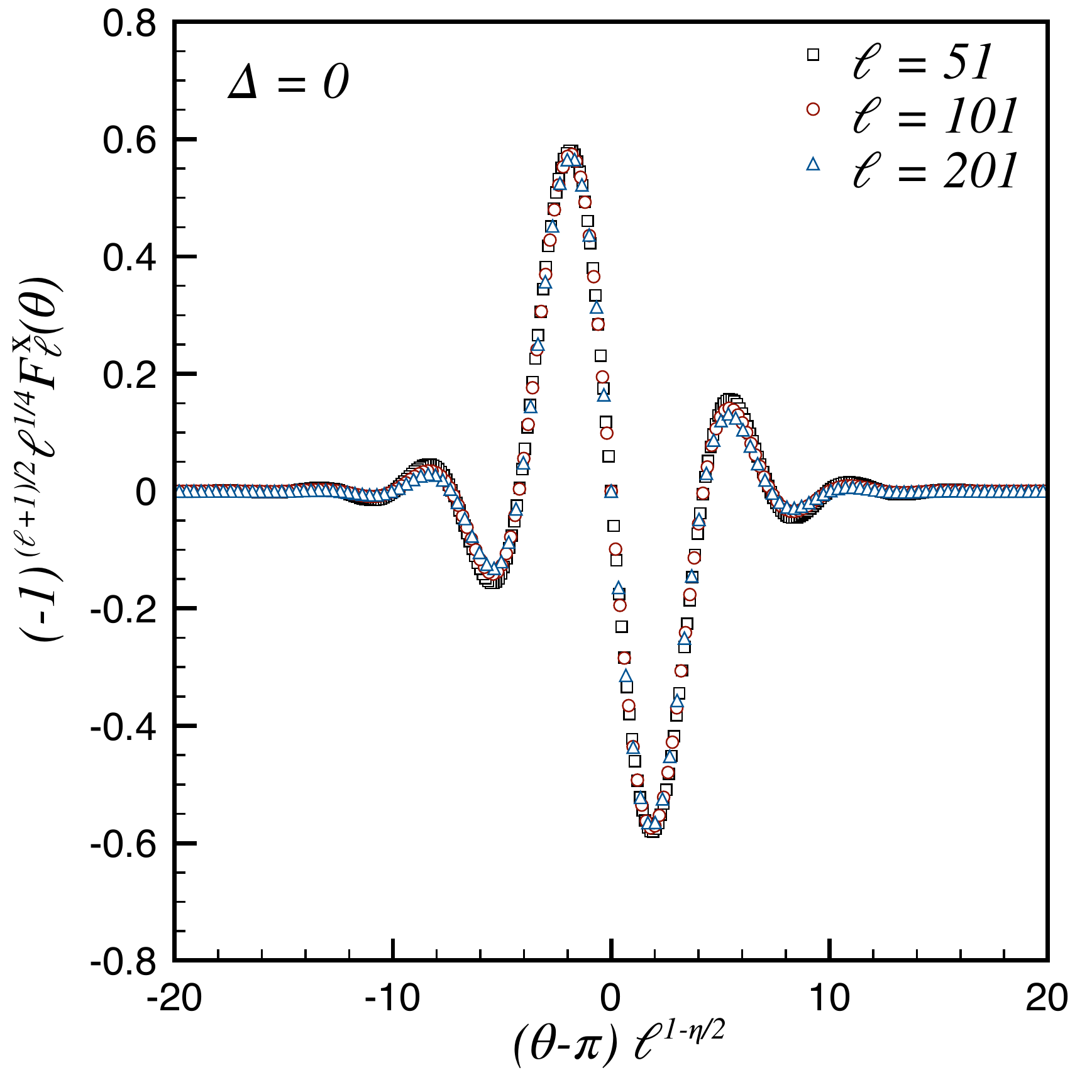}
\qquad
\includegraphics[width=0.3\textwidth]{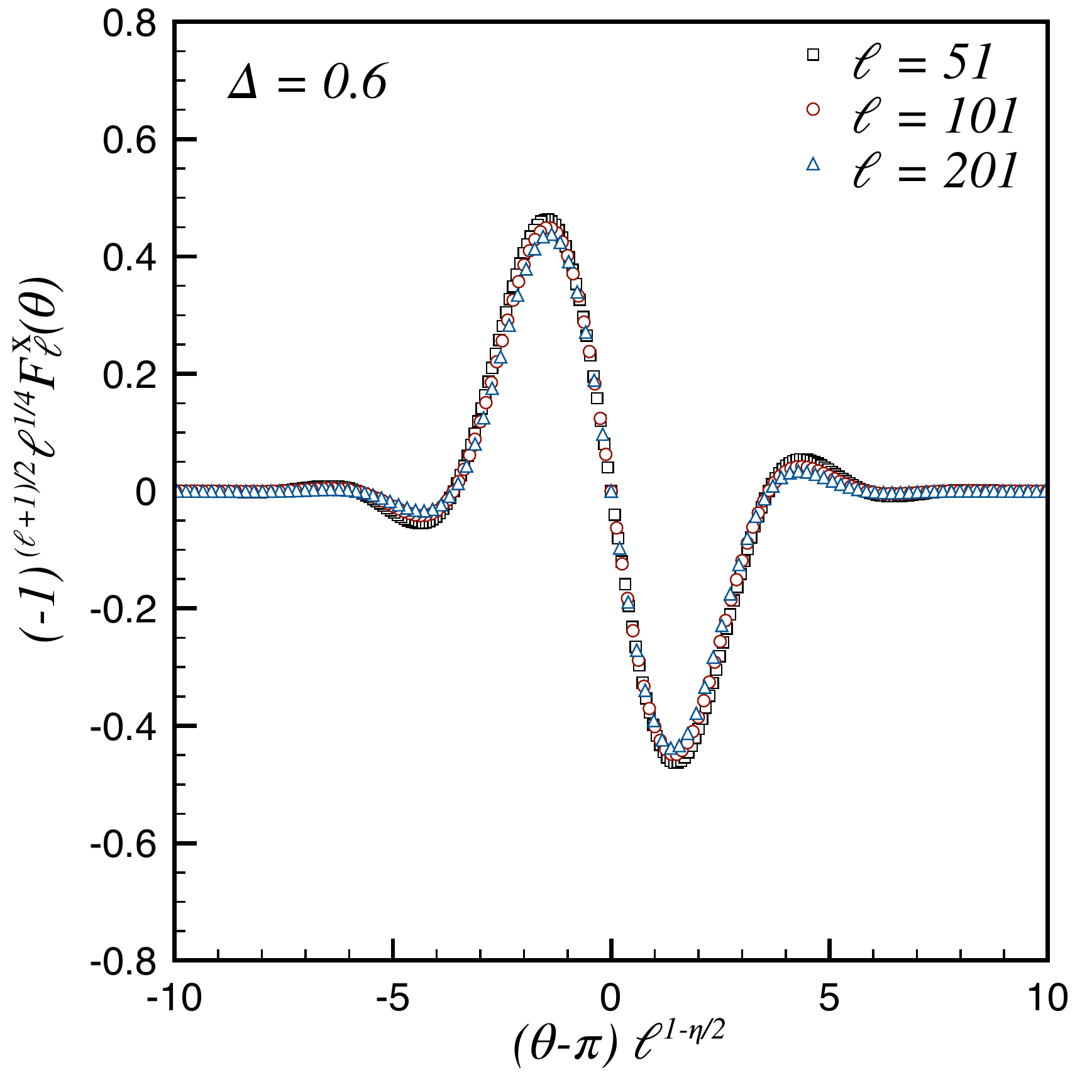}
\caption{\label{fig:Fx_pi_odd}
Same as Fig.~\ref{fig:Fx_pi_even} for odd $\ell$.}
\end{center}
\end{figure}

We now turn to the other region in which $F^x_\ell(\theta)$ is
sizeable, namely $\theta\approx\pi$. Interestingly, as shown in
Figs~\ref{fig:Fx_pi_even} and \ref{fig:Fx_pi_odd}, we observe
scaling behaviour here as well. There is a strong parity effect in
the subsystem size $\ell$ which requires us to consider even and odd
$\ell$ separately.
Our numerical data in the vicinity of $\theta=\pi$ is well described
by the scaling ansatz 
\be
F^{x}_{\ell}(\theta\approx\pi) 
\simeq (-1)^{\lfloor \ell /
2 \rfloor} \ell^{-1/4}\mathcal{F}^{x}_{e/o}(z)\ ,\quad
z=(\theta-\pi) \ell^{1-\eta/2}\ ,
\label{scalingansatz}
\ee
where $e/o$ refers to even and odd subsystem size $\ell$
respectively. Inspection of Figs~\ref{fig:Fx_pi_even} and 
\ref{fig:Fx_pi_odd} shows that the ansatz is in excellent agreement
with the data. We note that at $\theta = \pi$ the numerical data (for
$\ell$ even) exhibit a  perfect algebraic decay $\sim \ell^{-1/4}$,
independent of the value of the interaction $\Delta$. The
form \fr{scalingansatz} suggests that for very large subsystem sizes
in the thermodynamic limit the feature at
$F^{x}_{\ell}(\theta\approx\pi)$ becomes less and less important
compared to $F^{x}_{\ell}(\theta\approx 0)$. At present no analytic
results on $F^{x}_{\ell}(\theta\approx\pi)$ are known. It should in
principle be possible to calculate $F^x_{e/o}(z)$ using field theory
methods.  

\subsection{Probability distribution \texo{$P^x_N(m,\ell)$} of the
transverse, staggered subsystem magnetization}
\begin{figure}[ht!]
\begin{center}
\includegraphics[width=0.3\textwidth]{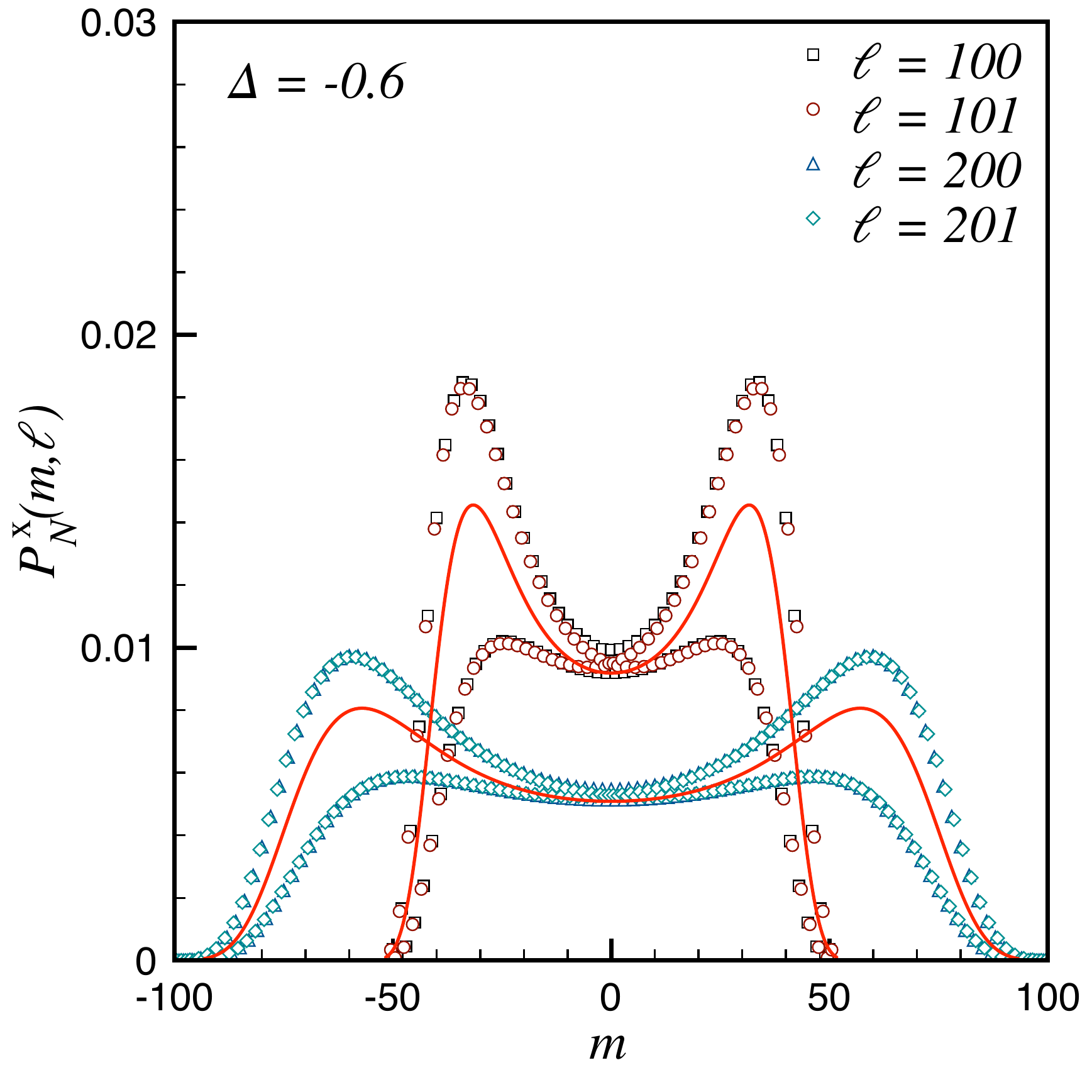}
\qquad
\includegraphics[width=0.3\textwidth]{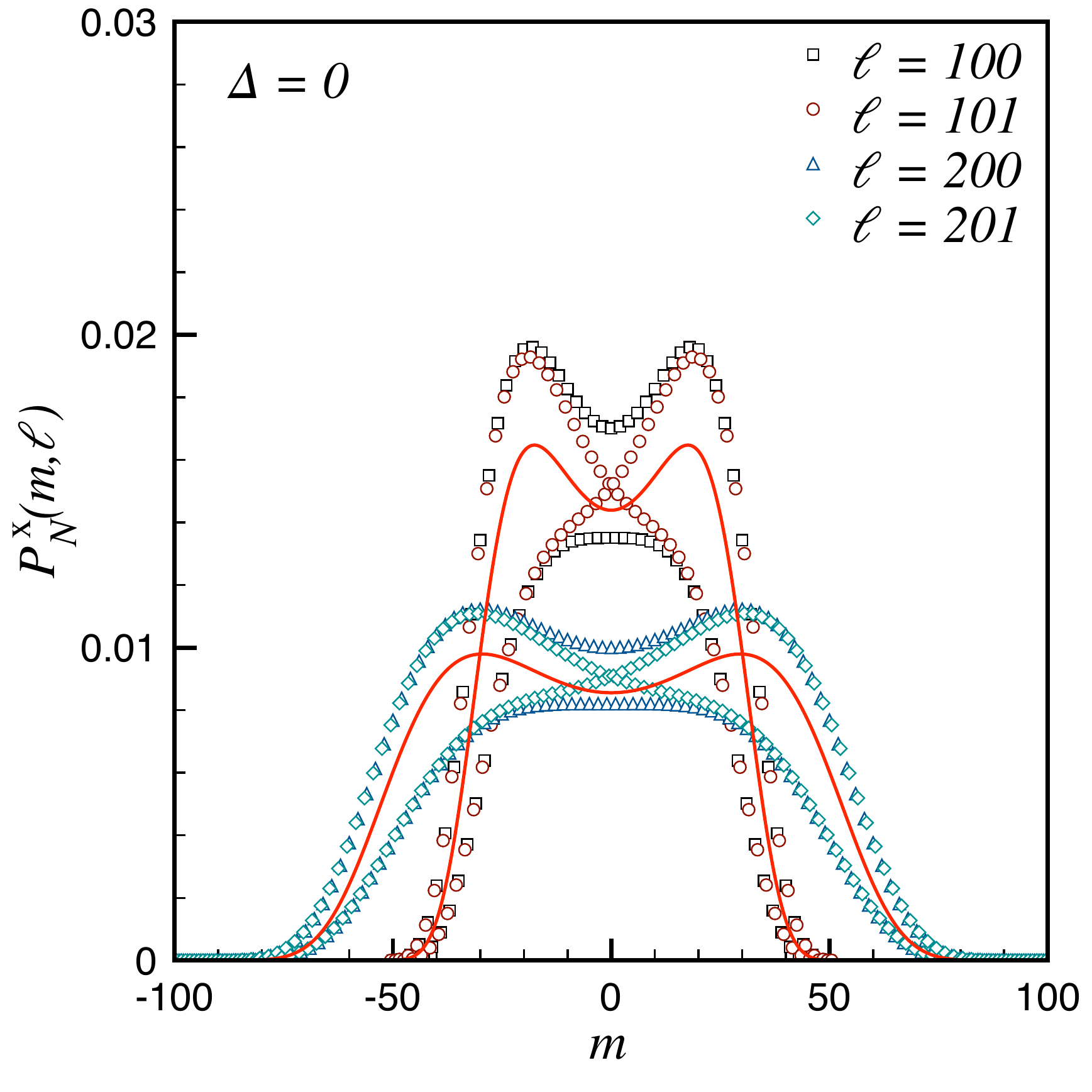}
\qquad
\includegraphics[width=0.3\textwidth]{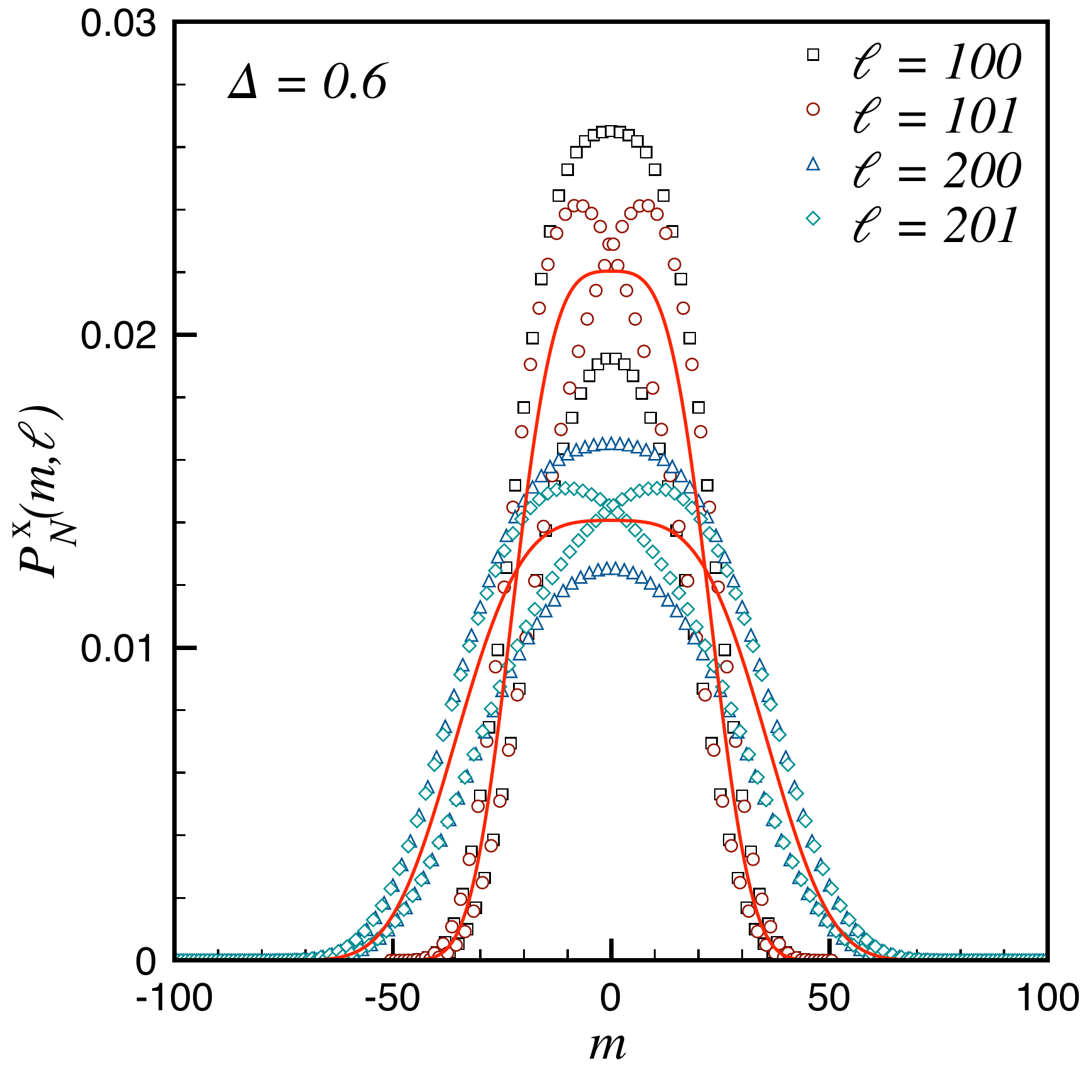}
\caption{Probability distribution functions $P_N^x(m,\ell)$ for
$\Delta=-0.6$, $\Delta=0$ and $\Delta=0.6$. As $P^{x}_{N}(m,\ell)$ is
a sum over $\delta$-functions we plot the corresponding weights at the
appropriate values of $m$. The solid red lines show the field theory
result, which becomes exact in the large-$\ell$ limit.} 
\label{fig:Px}
\end{center}
\end{figure}

We are now in a position to determine the probability distribution
$P^x_N(m,\ell)$ from the generating function $F^x_\ell(\theta)$ using
eqn \fr{PxN}. In Fig.~\ref{fig:Px} we show results for $P^x_N(m,\ell)$
as a function of $m$ for several values of the  exchange anisotropy
$\Delta$ and subsystem sizes $\ell$. As $P^{x}_{N}(m,\ell)$ is a sum
over $\delta$-functions, \emph{cf.} eqn \fr{PxN}, we plot the
corresponding weights $\widetilde{F}_\ell^x$ at the appropriate values
of $m$.
We observe that
\begin{enumerate}
\item{} There is a strong even/odd effect in $m$. The results for even
and odd $m$ follow different smooth curves. The separations between
even and odd curves slowly tend to zero as $\ell^{-1/4}$ as the
subsystem size $\ell$ is increased.
\item{} There is a weaker even/odd effect in the subsystem size
$\ell$. This effect remains visible even for the large subsystem sizes
we consider here. The magnitude of this effect grows with $\Delta$ and is
strongest for $\Delta\to 1$, i.e. when we approach the isotropic
antiferromagnet. 
\item{} The probability distributions are quite broad, implying strong
quantum fluctuations in the staggered transverse subsystem
magnetization.
\item{} The width of $P^{x}_{N}(m,\ell)$ increases as the interaction
becomes more attractive and the distribution flattens.
\item{} The distribution for attractive and moderately repulsive
interactions is bi-modal, while for $\Delta\approx 1$ it displays a
single maximum.
\end{enumerate}

\begin{figure}[ht]
\begin{center}
\includegraphics[width=0.4\textwidth]{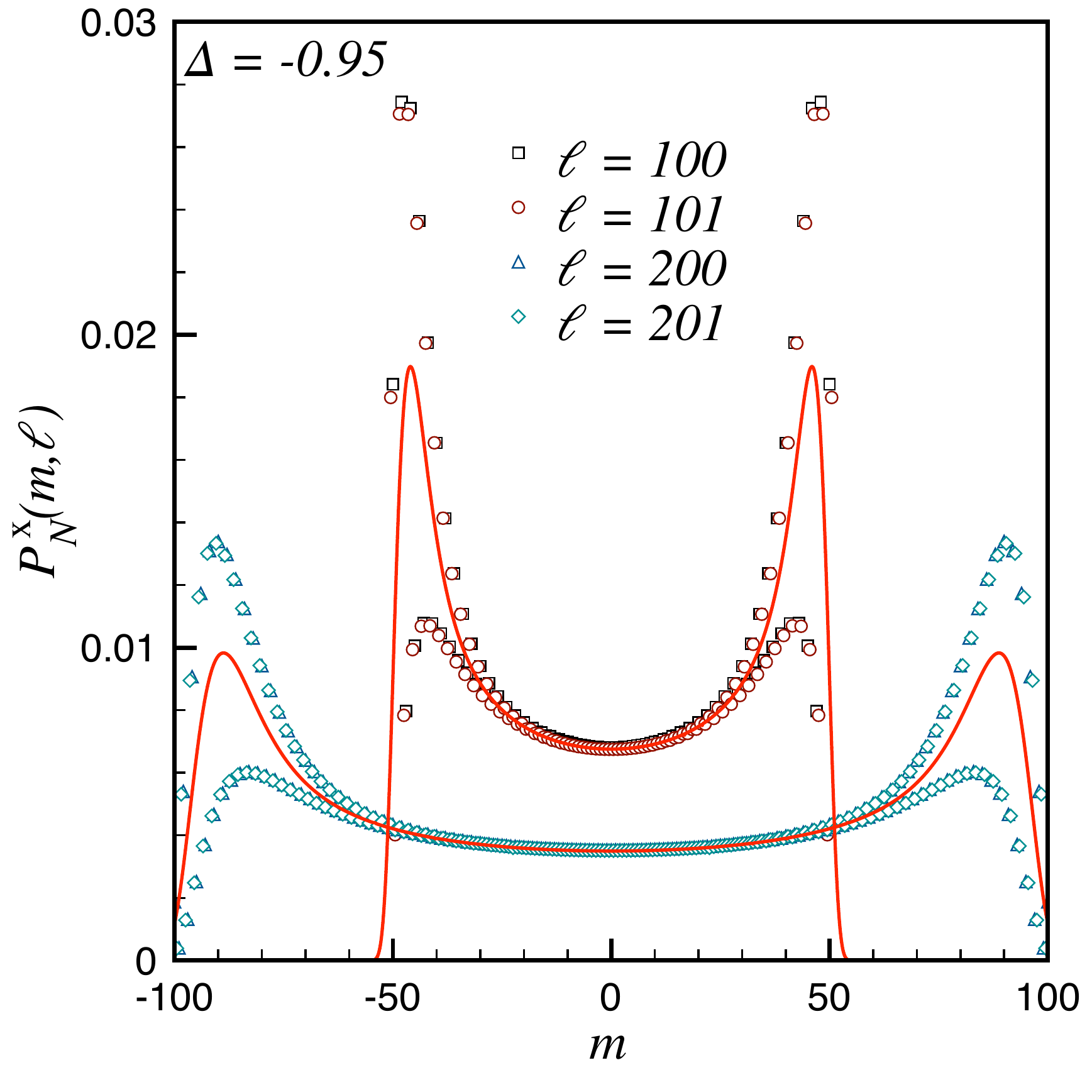}
\caption{Probability distribution functions $P_N^x(m,\ell)$ for
$\Delta=-0.95$. As the system become more and more "ferromagnetic" 
the probability distribution for the staggered  subsystem magnetisation 
tends to become broader and flat.}
\label{fig:Px-095}
\end{center}
\end{figure}

These observations can be understood in terms of our scaling analysis
of the generating function $F^x_\ell(\theta)$. The probability
distribution $P_N(m,\ell)$ is dominated by the behaviour of
$F^x_\ell(\theta)$ in the regions $\theta=0,\pi$, and exploiting the
observed scaling behaviour of these contributions we conclude that
\be
P^{x}_{N}(m,\ell) \simeq
\ell^{\frac{\eta}{2}-1} \mathcal{\widetilde F}^{x}_{0}(m/\ell^{1-\eta/2}) 
+ (-1)^{\lfloor \ell / 2 \rfloor + \lfloor
m\rfloor} \ell^{\frac{\eta}{2}-5/4}\mathcal{\widetilde
F}^{x}_{e/o}(m/\ell^{1-\eta/2}), 
\ee
where $\mathcal{\widetilde F}^{x}_{0}$ and $\mathcal{\widetilde
F}^{x}_{e/o}$ are obtained by Fourier transforming the functions
${\cal F}^x_0(z)$ and ${\cal F}^x_{e/o}(z)$ that describe the scaling
behaviour of the generating function around $\theta=0$ and
$\theta=\pi$ respectively. For large subsystem sizes $\ell$ the
even/odd effect in $m$ disappears and we are left with
$\ell^{\frac{\eta}{2}-1} \mathcal{\widetilde
F}^{x}_{0}(m/\ell^{1-\eta/2})$, which can be calculated exactly using
the boundary sine-Gordon mapping. The corresponding contribution is
shown by a solid red line in Figs~\ref{fig:Px}
and \ref{fig:Px-095}. We see that for attractive and moderately strong
repulsive interactions there is an enhanced probability to form a
large positive or negative staggered moment in the xy-plane. However,
this enhancement is not particularly pronounced. The effect is
strongest close to the ferromagnet at $\Delta=-1$ as can be seen in
Fig.~\ref{fig:Px-095}, which presents results for $P^x_N(m,\ell)$ at
$\Delta=-0.95$. We observe that for large subsystem sizes $\ell$ the
probability distribution becomes fairly flat 
over most of allowed range of staggered magnetizations $-\ell/2\leq
m\leq \ell/2$ except for an enhancement close to the maximal possible
values $m\approx\pm\ell/2$. 

\subsection{Longitudinal generating function \texo{$F^z_\ell(\theta)$}
and probability distribution \texo{$P^z_N(m,\ell)$}}
\begin{figure}[ht]
\includegraphics[width=0.4\textwidth]{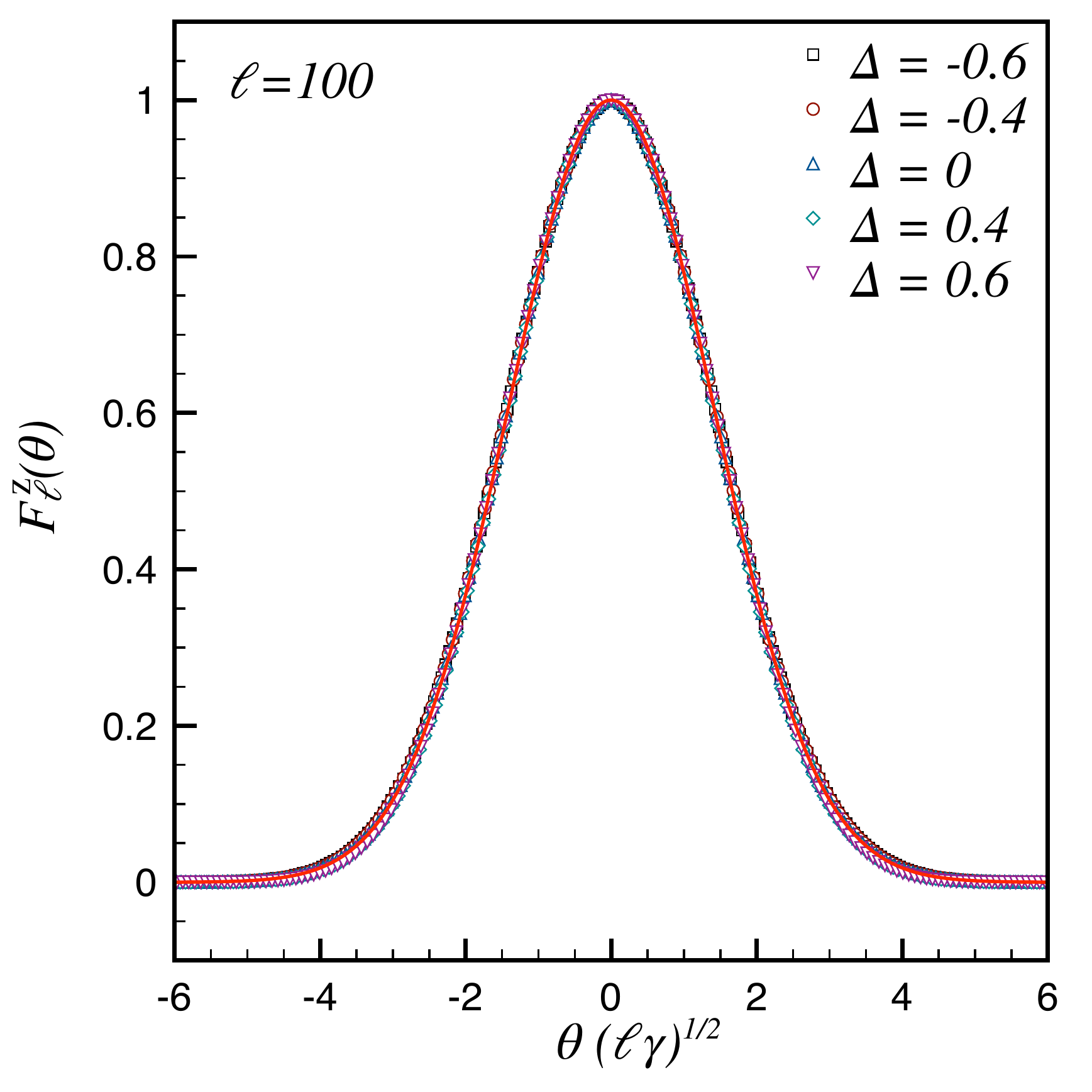}
\caption{\label{fig:Fz1}
Staggered longitudinal generating function $F^{z}_{\ell}(\theta\approx
0)$ for several values of $\Delta$. The numerical data exhibit scaling
collapse that is in excellent agreement with the universal scaling
function $\exp(-\gamma z^2/4)$ (red line).
}
\end{figure}

\begin{figure}[ht]
\begin{center}
\includegraphics[width=0.3\textwidth]{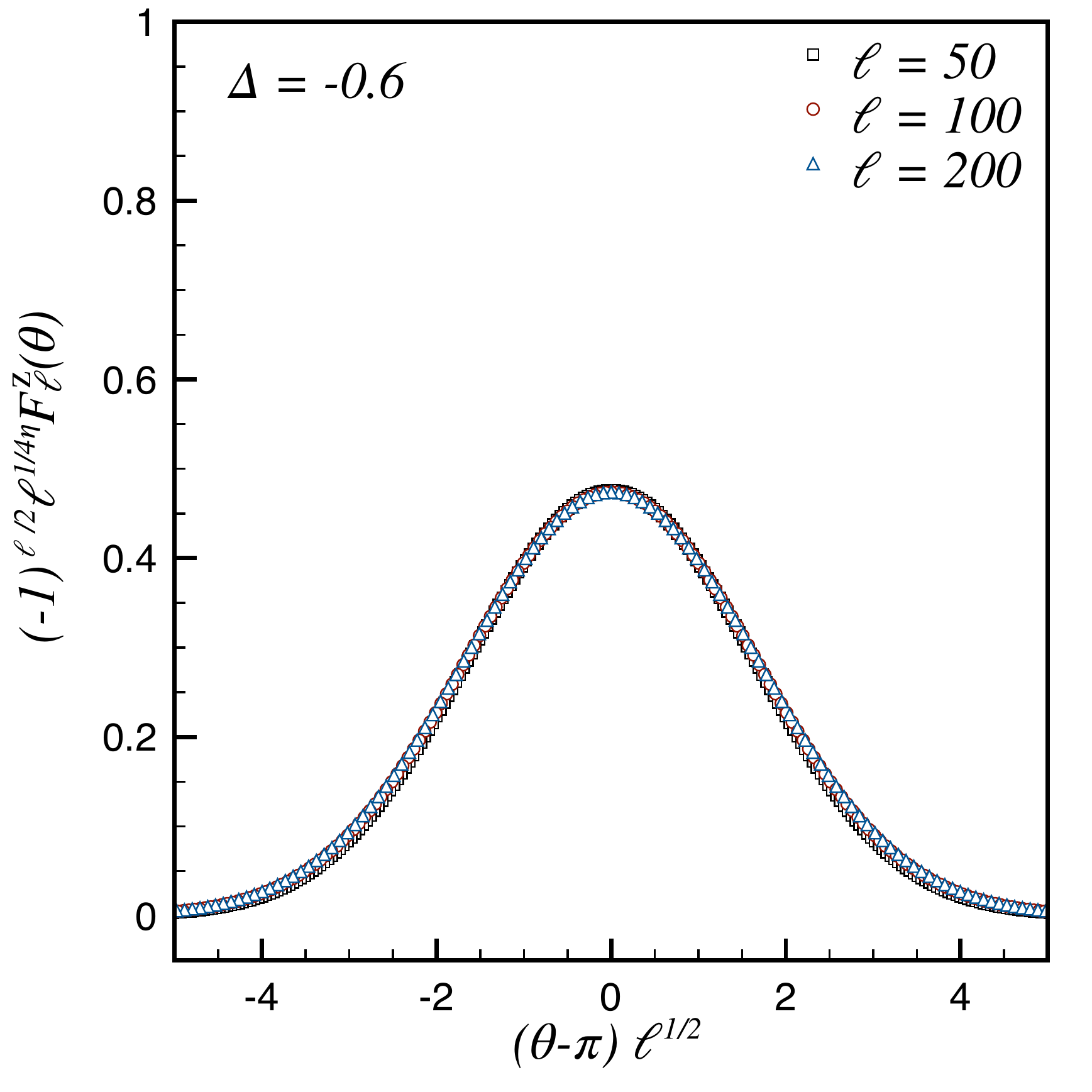}
\qquad
\includegraphics[width=0.3\textwidth]{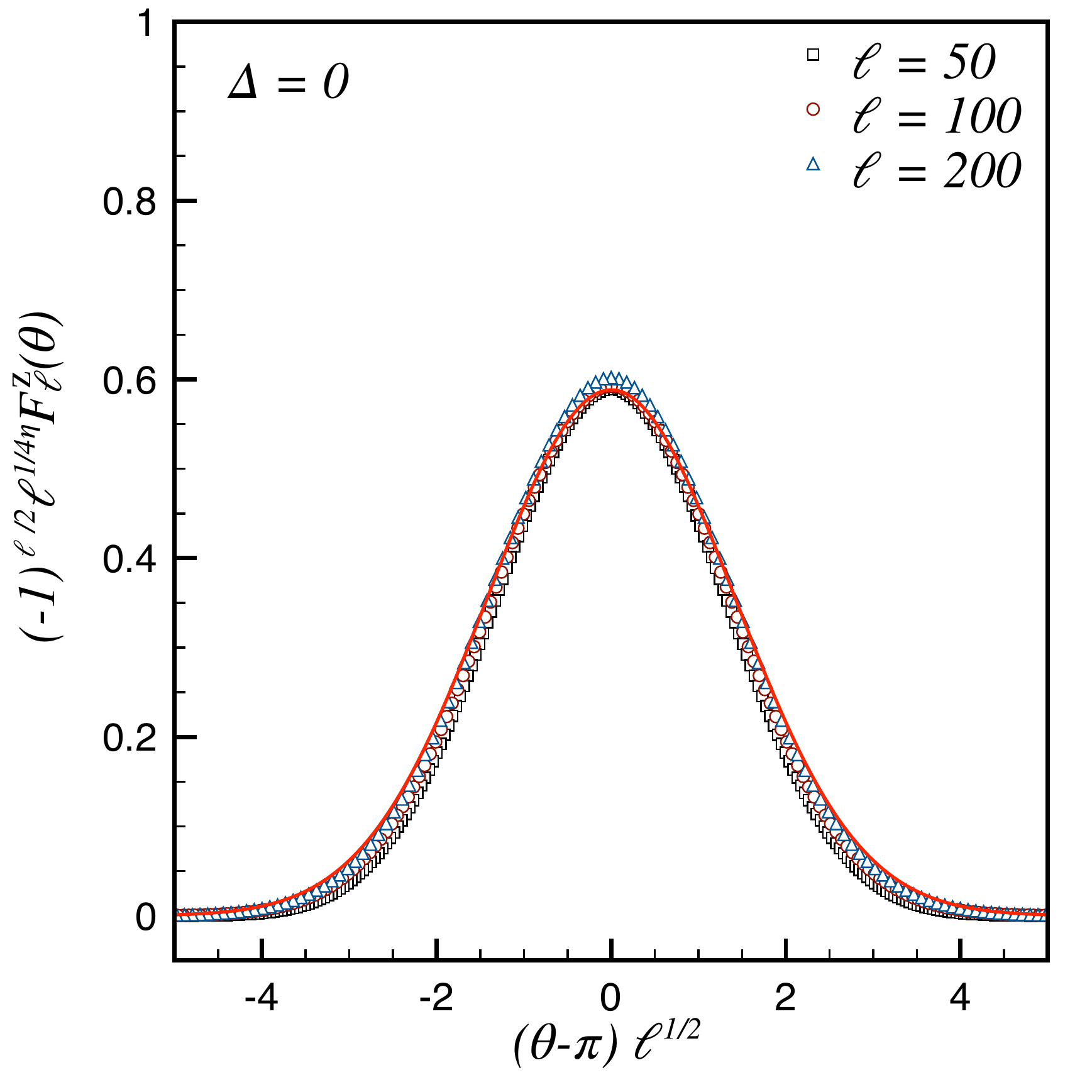}
\qquad
\includegraphics[width=0.3\textwidth]{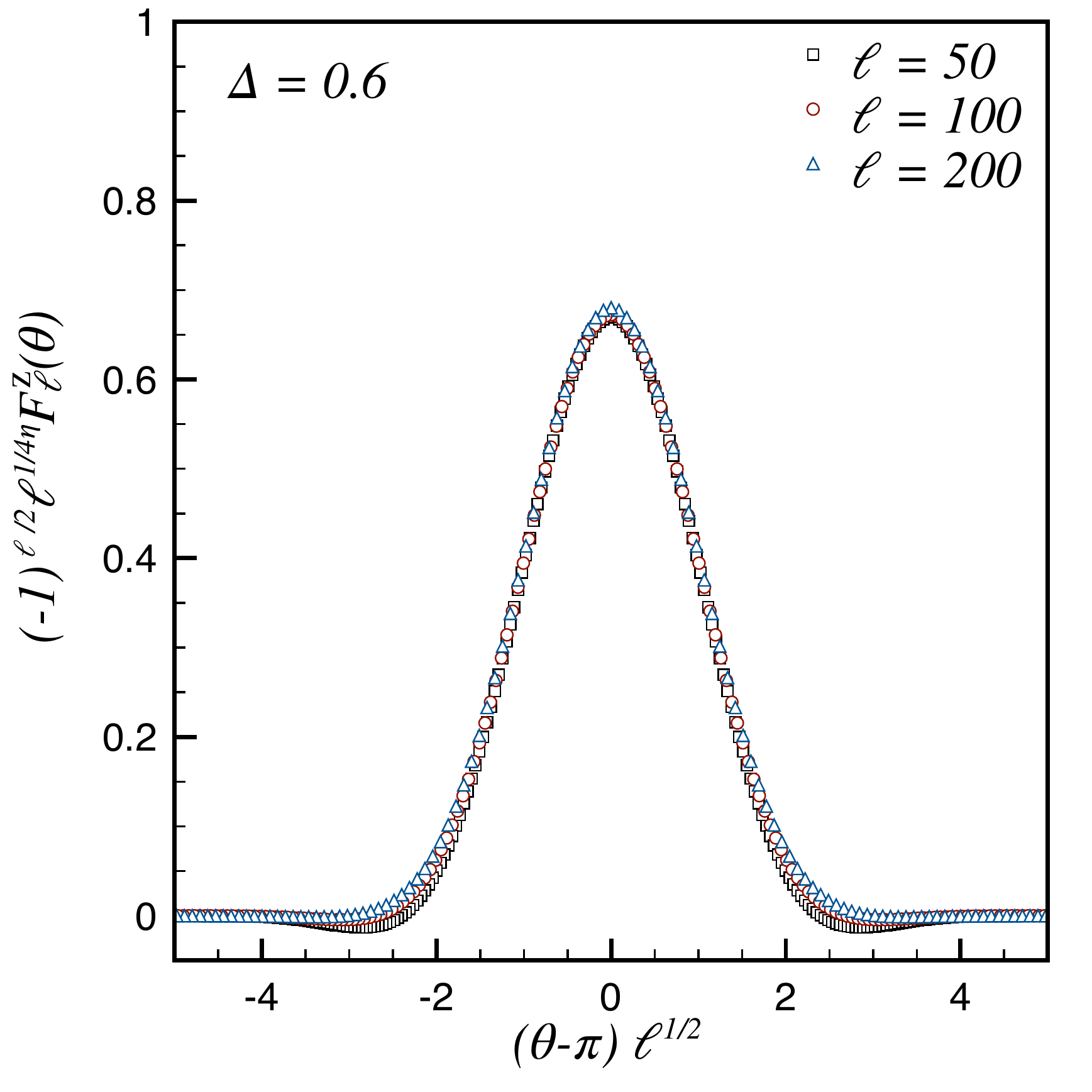}
\caption{\label{fig:Fz_pi_even}
Scaling behaviour of the staggered longitudinal generating function
$F^{z}_{\ell}(\theta\approx\pi)$ for even subsystem sizes $\ell$ and
several values of $\Delta$. For the $XX$ point the full red line
represents the analytical scaling function $0.588353 \exp(-z^2/4)$.} 
\end{center}
\end{figure}

\begin{figure}[ht]
\begin{center}
\includegraphics[width=0.3\textwidth]{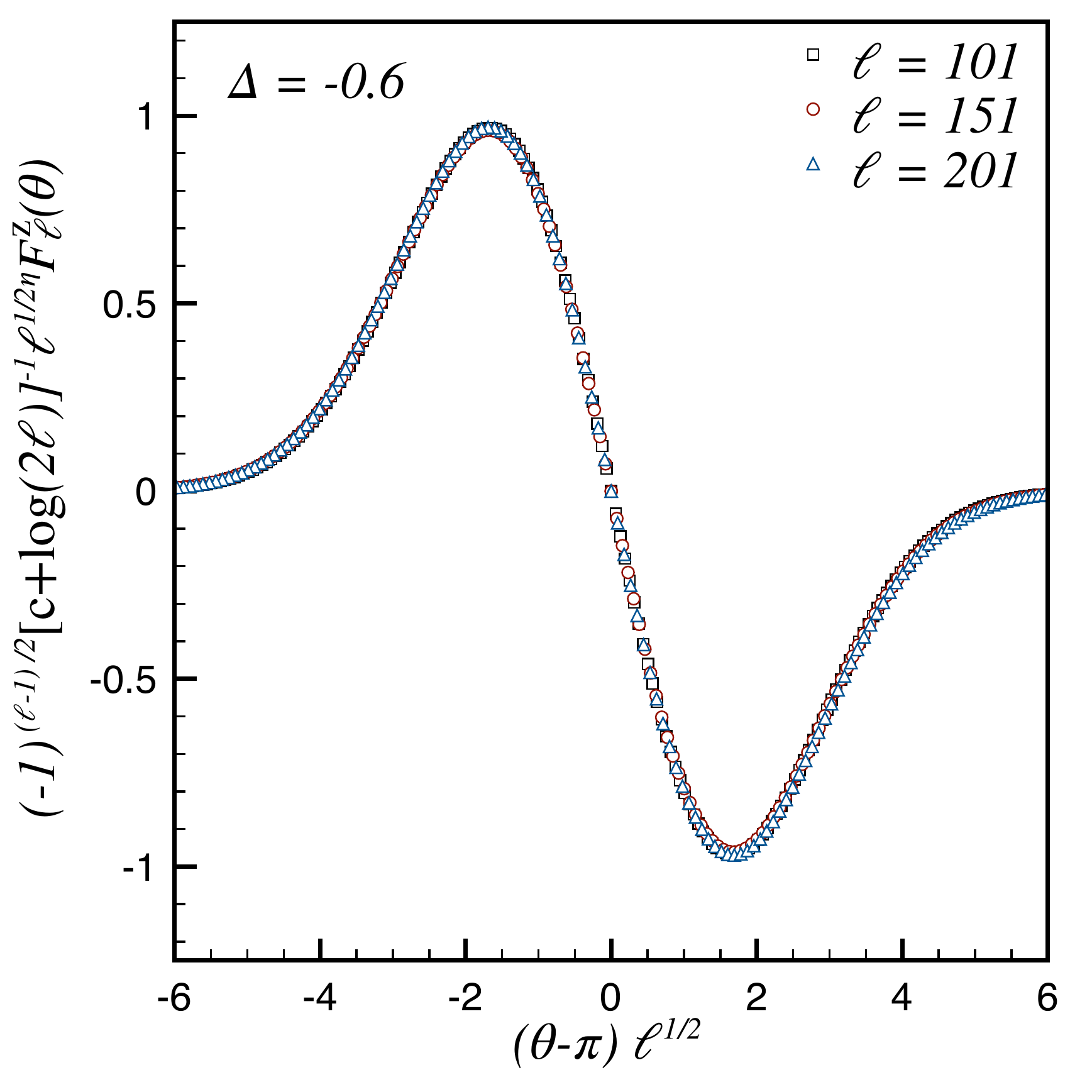}
\qquad
\includegraphics[width=0.3\textwidth]{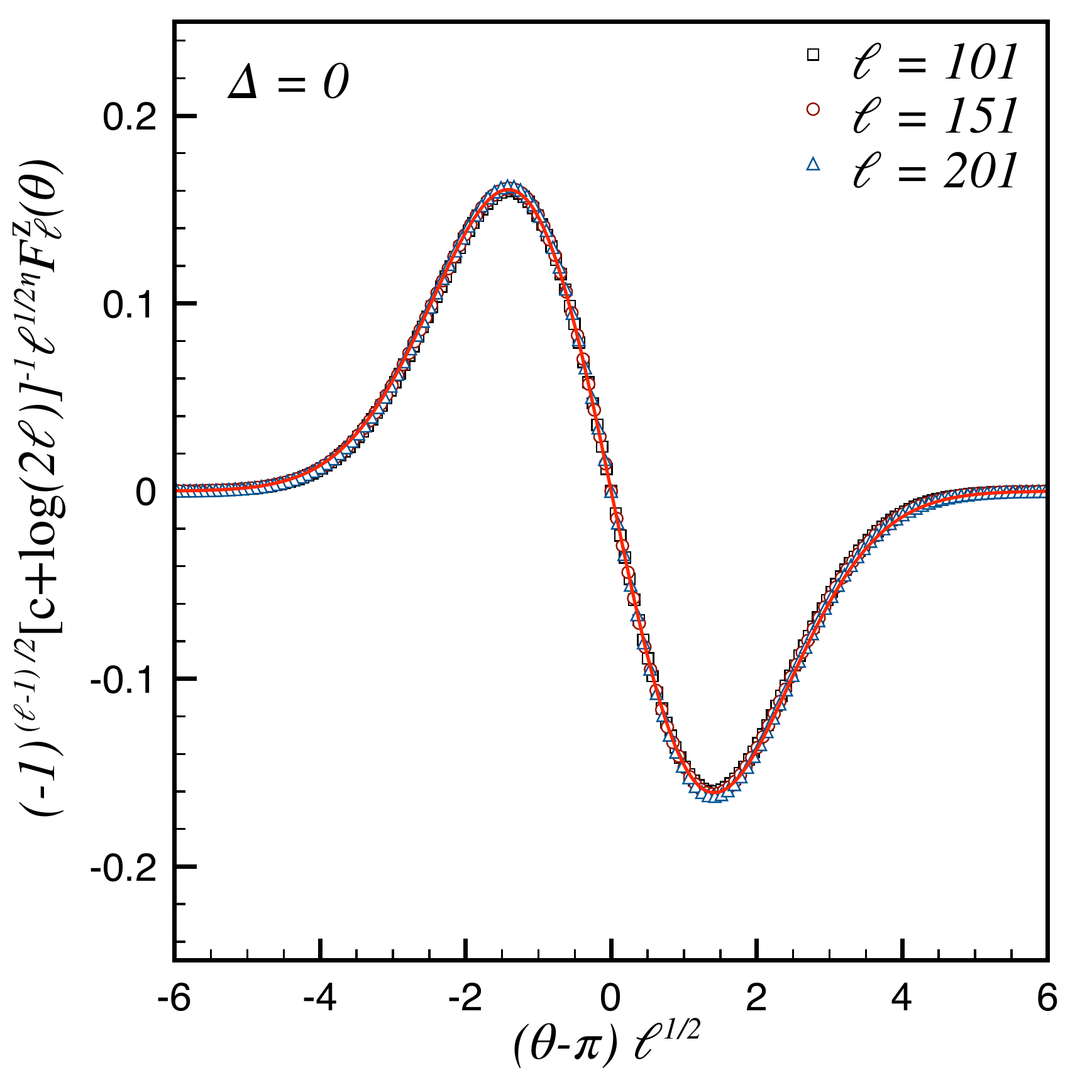}
\qquad
\includegraphics[width=0.3\textwidth]{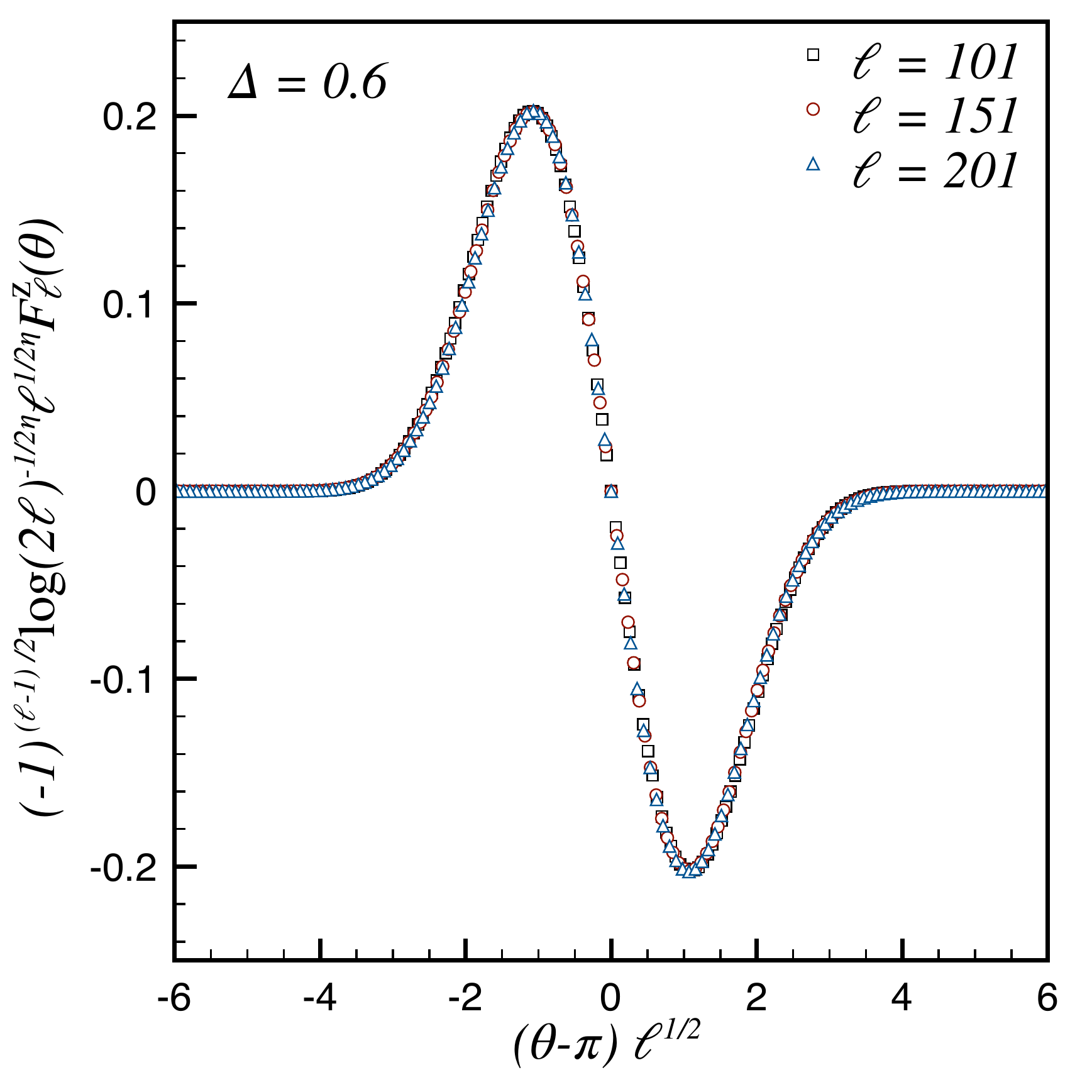}
\caption{\label{fig:Fz_pi_odd}
Scaling behaviour of the staggered longitudinal generating function
$F^{z}_{\ell}(\theta\approx\pi)$ for odd subsystem sizes $\ell$ and
several values of $\Delta$. For the $XX$ case, the  red line represent
the analytical scaling function in Eq. (\ref{eq:Fz_scaling_pi}). The
constant $c$ equals $2\log(2)+\gamma_{E}$ for $\Delta=0$ and has been
set to $c\simeq -2.8$ for $\Delta=-0.6$ and $c\simeq 0.5$ for
$\Delta=0.6$ respectively.}
\end{center}
\end{figure}

\begin{figure}[ht]
\begin{center}
\includegraphics[width=0.3\textwidth]{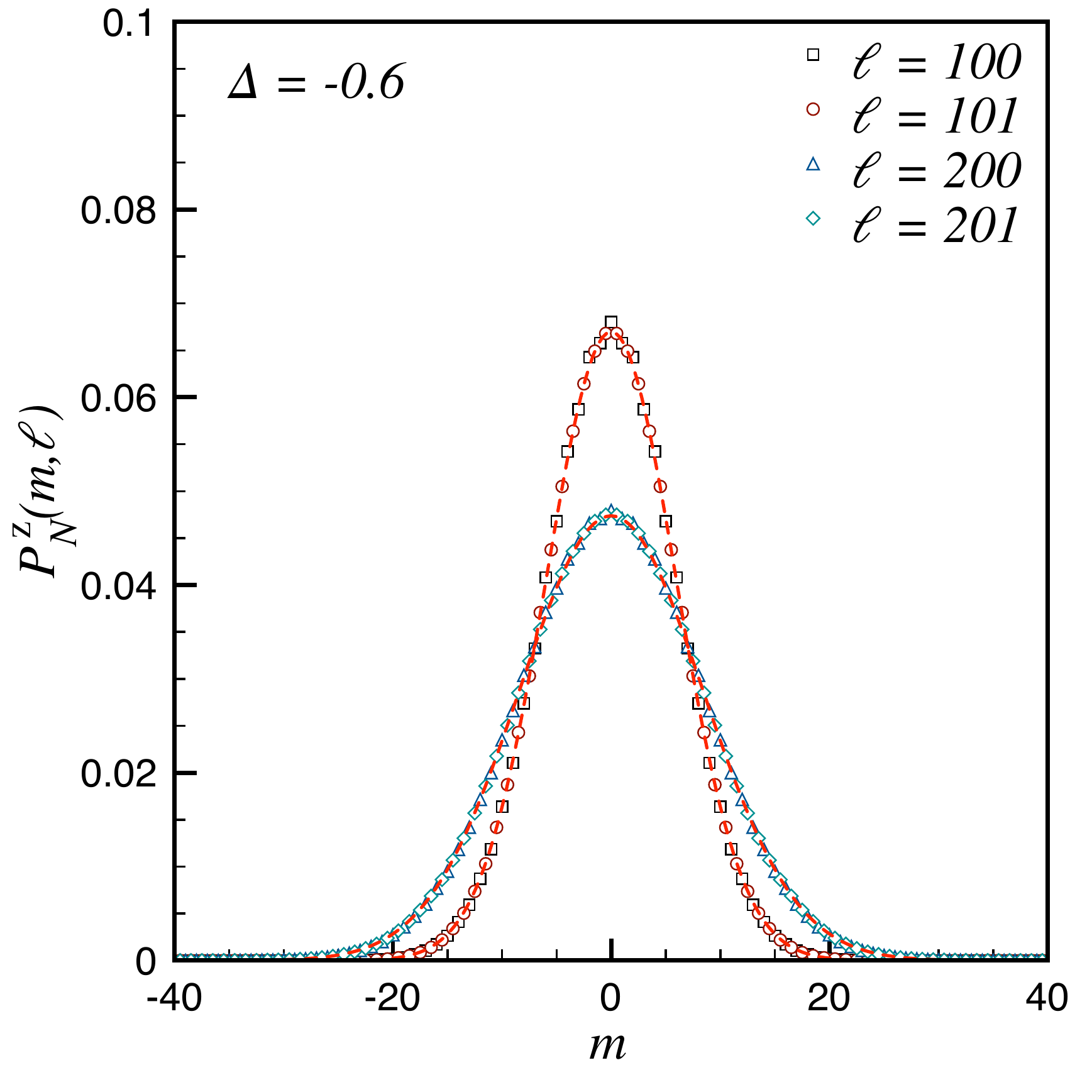}
\qquad
\includegraphics[width=0.3\textwidth]{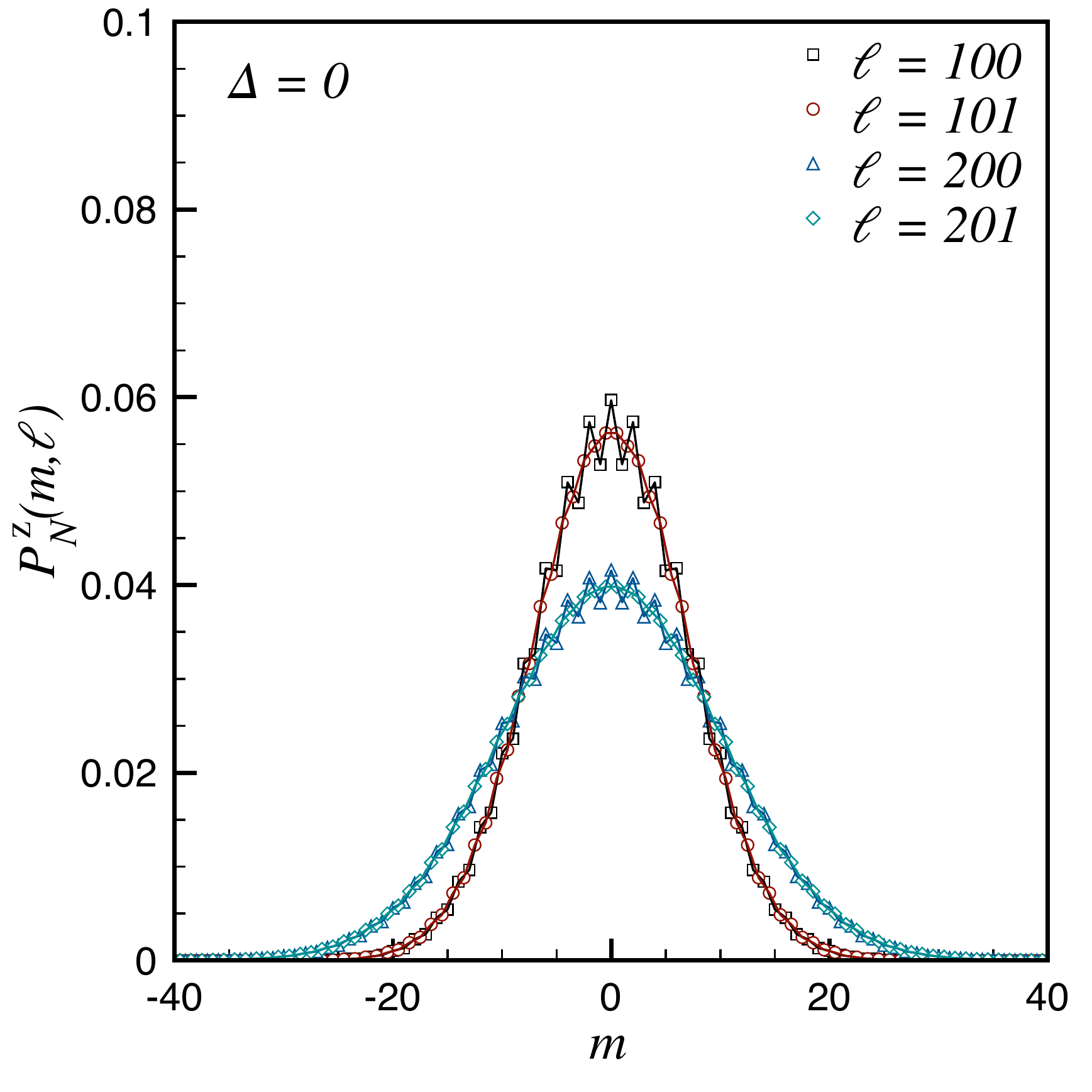}
\qquad
\includegraphics[width=0.3\textwidth]{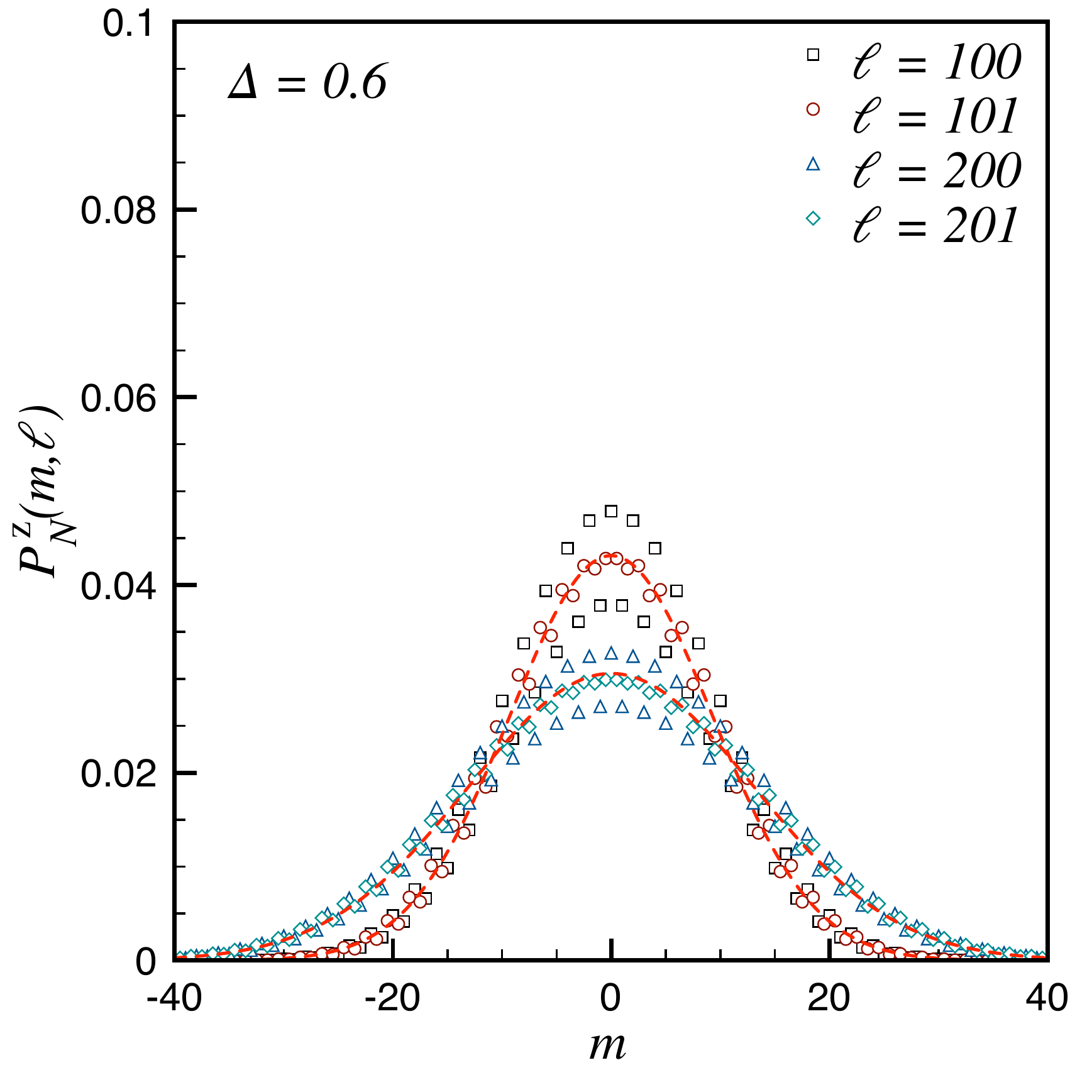}
\caption{Probability distribution functions $P_N^z(m,\ell)$ for
$\Delta=-0.6$, $\Delta=0$ and $\Delta=0.6$. 
In the noninteracting case the full lines are
the exact results obtained using the determinant formula.
In the interacting cases, the red dashed lines are the leading smooth contribution
coming from the scaling behaviour of the generating function in the vicinity of $\theta = 0$.
Notice in particular that, in the ferromagnetic regime, the sub-leading staggered corrections are almost 
invisible for the sizes considered here. Otherwise, as the antiferromagnetic regime is approached,
the sub-leading parity effects become more significant.}
\label{fig:Pz}
\end{center}
\end{figure}

We now turn to the longitudinal generating function
$F^z_\ell(\theta)$. As we will see, its behaviour is rather different
from $F^x_\ell(\theta)$. According to the field theory approach
discussed in section~\ref{ssec:BSG} we expect $F^z_\ell(\theta)$ to be
described by the scaling function
\be
F^z_\ell(\theta\approx 0)=e^{-\gamma z^2/4}\ ,\quad z=\theta\ell^{1/2}\ ,
\label{fz1}
\ee
where $\gamma$ is a $\Delta$-dependent constant that encodes the fact that
appropriate ratios of moments are universal, while the second moment
itself is not, \emph{cf.} eqn \fr{rescaling}. Numerical results for
$F^z_\ell(\theta)$ are shown in Fig.~\ref{fig:Fz1} and are seen to be
in excellent agreement with the scaling form \fr{fz1}. 
The coefficient $\gamma$ is found to be consistent with
\be
\gamma=\frac{1}{2-2\eta}.
\ee
We conclude that the fluctuations of $N^z(\ell)$ have a very simple
form: the second moment is 
\be
\langle{\rm GS}|\big(N^z(\ell)\big)^2|{\rm GS}\rangle=
\frac{\ell}{8-8\eta}+o(\ell)\ ,
\ee
while all higher cumulants vanish.
We now turn to the behaviour of $F^z_\ell(\theta\approx\pi)$. Guided
by the exact result \fr{eq:Fz_scaling_pi} for $\Delta=0$ we have
attempted to describe our numerical data by the ansatz
\be
F^z_\ell(\theta\approx\pi)=
\begin{cases}
(-1)^{\ell/2}A\ \ell^\alpha e^{-\gamma z^2/4} & \text{$\ell$ even}\\
(-1)^{(\ell-1)/2} \frac{B}{c+\log(2\ell)}\ell^\beta\ z e^{-\gamma z^2/4}
& \text{$\ell$ odd}
\end{cases}\ ,\qquad
z=(\pi-\theta)\ell^{1/2}.
\label{Fzansatz1}
\ee
Here $A$, $B$, $\alpha$, $\beta$ and $c$ are $\Delta$-dependent parameters that
we fix by considering the $\ell$-dependencies of $F^z_\ell(\pi)$ (for
$\ell$ even) and $\partial_\theta F^z_\ell(\theta)$ (for $\ell$ odd).
Our numerical results suggest that
\be
\alpha=\frac{1}{4\eta}\ ,\qquad \beta=\frac{1}{2\eta}\ .
\label{Fzansatz2}
\ee
In Figs~\ref{fig:Fz_pi_even} and \ref{fig:Fz_pi_odd} we compare
numerical results for $F^z_\ell(\theta\approx\pi)$ for several values
of $\Delta$ and $\ell$ to the scaling ansatz \fr{Fzansatz1}, \fr{Fzansatz2}.
The agreement is seen to be quite satisfactory in all cases.
Having determined the generating function $F^z_\ell(\theta)$ we can
now use it to obtain the probability distribution of the longitudinal
staggered subsystem magnetization $P_N^z(m,\ell)$ by Fourier
transform. Results for several values of $\Delta$ and $\ell$ are
presented in Fig.~\ref{fig:Pz}.

The probability distribution is again a sum over delta-functions that
fix the allowed values of $m$ and we plot the corresponding weights.
We observe that in all cases $P_N^z(m,\ell)$ exhibits a single maximum
centred at $m=0$ and is significantly narrower than its transverse
counterpart $P_N^x(m,\ell)$. There is again an even/odd effect in $m$
that increases in magnitude as $\Delta$ approaches $1$, but it is
generally weaker than in the transverse case. There also is an
even/odd effect in the subsystem size $\ell$.

\section{Generating functions for the subsystem magnetization} 
We now turn to the probability distribution of the (smooth) subsystem 
magnetization. We first consider the longitudinal generating function
$G^z_\ell(\theta)$, as analytic results are readily available for it. 
\subsection{Longitudinal generating function 
\texorpdfstring{$G^z_\ell(\theta)$}{Lg}}
For large subsystem sizes the longitudinal generating function can be
determined by standard methods: at $\Delta=0$ free fermion techniques
apply, while for general values of $\Delta$ Luttinger liquid methods
provide detailed predictions.
\subsubsection{The XX point \texorpdfstring{$\Delta=0$}{Lg}}
As it is straightforward to take into account a magnetic field along
the z-direction in this case, we present results for the subsystem
magnetization in the ground state of the Hamiltonian 
\be
\label{H_XX}
H = J \sum_{j=1}^{L} \left( S^{x}_{j}S^{x}_{j+1} +
S^{y}_{j}S^{y}_{j+1}\right ) -h\sum_{j=1}^L
S^{z}_{j} \ ,\quad 0<h<1.
\ee
Here a simple determinant formula for $G^z_\ell(\theta)$ is
known \cite{Colomo93} 
\bea
G^z_\ell(\theta) &=& e^{i\frac{\theta\ell}{2}}\det\left[ \mathbb{I} +  (
e^{-i  \theta}-1)  {\mathbb C} \right]\ ,\nn
\label{RhoD0}
{\mathbb C}_{nm}&=& \frac{\sin(k_F (m-n))}{\pi (m-n)}\ ,\qquad
k_F={\rm arccos}(h/J)\ .
\eea
The Toeplitz determinant \fr{RhoD0} is related to a determinant
that has been analyzed in great detail in the
context of entanglement entropies\cite{ce-10}. Using the results of
Ref.~\onlinecite{ce-10} the large-$\ell$ asymptotics of 
$G^z_\ell(\theta)$ can be expressed in the form
\be
G^z_\ell(\theta)=e^{i\ell\theta/2}\sum_{j=-\infty}^\infty \rho_\ell(j+\frac{\theta}{2\pi})\,
\label{Gfree}
\ee
where 
\bea
\rho_\ell(\beta)=e^{- 2i \beta k_F \ell} \big(2\ell\sin(k_F)\big)^{-2\beta^2}G^2(1+\beta)G^2(1-\beta)
 \left[1+ \frac{c_1(\beta)}{\ell}+ \frac{c_2(\beta)}{\ell^2} + \dots\right].
\eea
Here ${G(z)}$ is the Barnes $G$-function and
\bea
\label{finalcoef}
c_1(\beta)&=&2i  \beta^3 \cot(k_F)\,,\\
c_2(\beta)&=& \frac{\beta^2}6 (-1 + 7 \beta^2 + 12 \beta^4
- 3 \beta^2 (5 + 4 \beta^2) \csc^2 k_F) \ .
\eea
The leading terms in \fr{Gfree} correspond to $j=0,1$ and have been
considered previously in Ref.~\onlinecite{bss-07}. The constant
$c_1(\beta)$ has been conjectured in Ref.~\onlinecite{aiq-11}.
At zero magnetic field we have $k_F=\pi/2$ and
\be
c_1(\beta)=0\ ,\qquad c_2(\beta)=-\frac16 \beta^2 (1 + 8 \beta^2).
\ee
We note that at zero magnetic field the generating function is
real. This is because all odd cumulants vanish as a consequence of
particle-hole symmetry. This ceases to be the case at finite magnetic
fields, but odd cumulants still vanish in the large $\ell$ limit (as
for the gas \cite{CMV-12}). 

\subsubsection{Luttinger liquid description}
The large-$\ell$ behaviour of the expression \fr{FTGF} for
$G^z_\ell(\theta)$ has been previously determined by bosonization
methods in Refs~\onlinecite{bss-07,aem-07}. In zero magnetic field this gives
a power-law decay 
\be\label{GzLL}
G^{z}_\ell(\theta) \simeq \sum_{j=-\infty}^\infty e^{-i \pi \ell j} 
D_j(\theta)\ \ell^{-\nu(\theta+2\pi j)}\ , 
\ee
where $\nu(\theta)= \frac{\theta^2}{4\pi^2\eta}$ and $\eta=1-\frac{1}{\pi}{\rm
arccos}(\Delta)$. An analytic expression for the amplitudes of the
leading terms in \fr{GzLL} was conjectured in
Ref.~\onlinecite{bss-07} 
\bea
D_0(\theta)&=& \left[\frac{\Gamma\left(\frac{\eta}{2-2\eta}\right)}{2\sqrt{\pi}
\Gamma\left(\frac{1}{2-2 \eta}\right)}\right]^{\theta^2/(4\eta\pi^2)}
\exp\Big[-\int_0^\infty\frac{d t}{t}
\Big(\frac{\sinh^2\frac{\theta}{2\pi} t}{\sinh t \,\cosh(1-\eta) t\,\sinh\eta t}
- \frac{\theta^2 e^{-2 t}}{4\eta \pi^2}\Big)\Big],\quad
\eea
and $D_{-1}(\theta)=D_0(|\theta|-2\pi)$.
\subsubsection{Comparison to iTEBD results}
\begin{figure}[ht]
\includegraphics[width=0.3\textwidth]{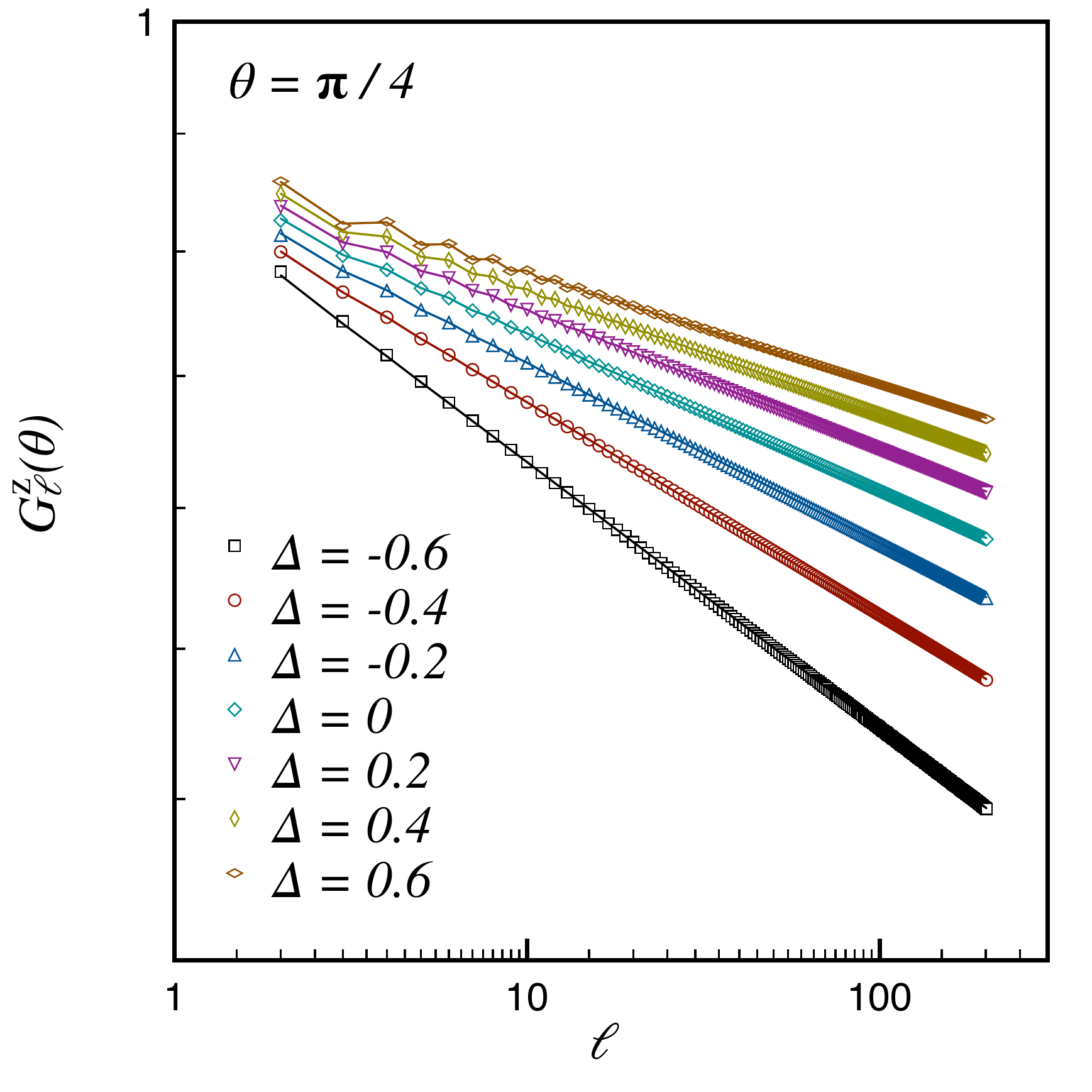}
\qquad
\includegraphics[width=0.3\textwidth]{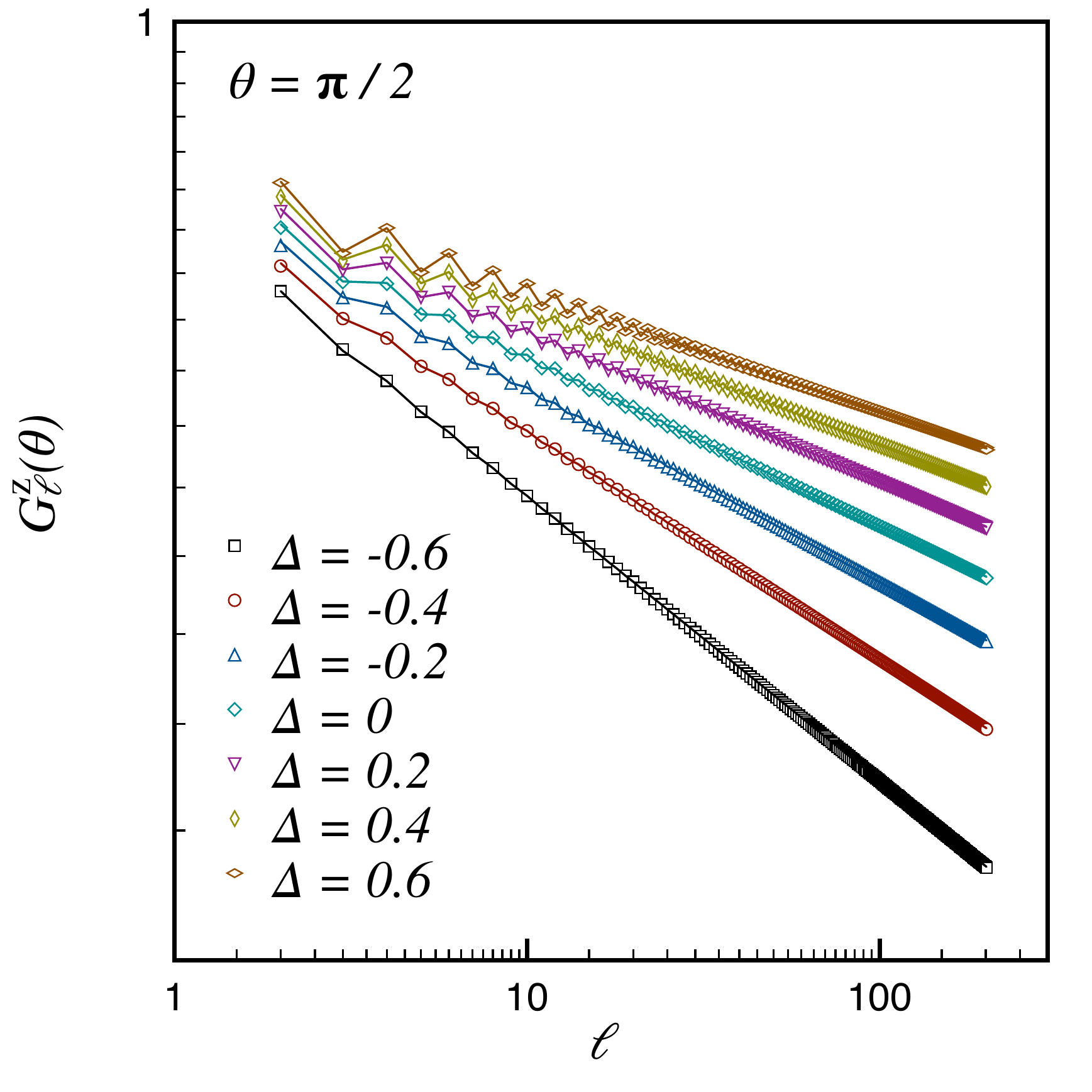}
\qquad
\includegraphics[width=0.3\textwidth]{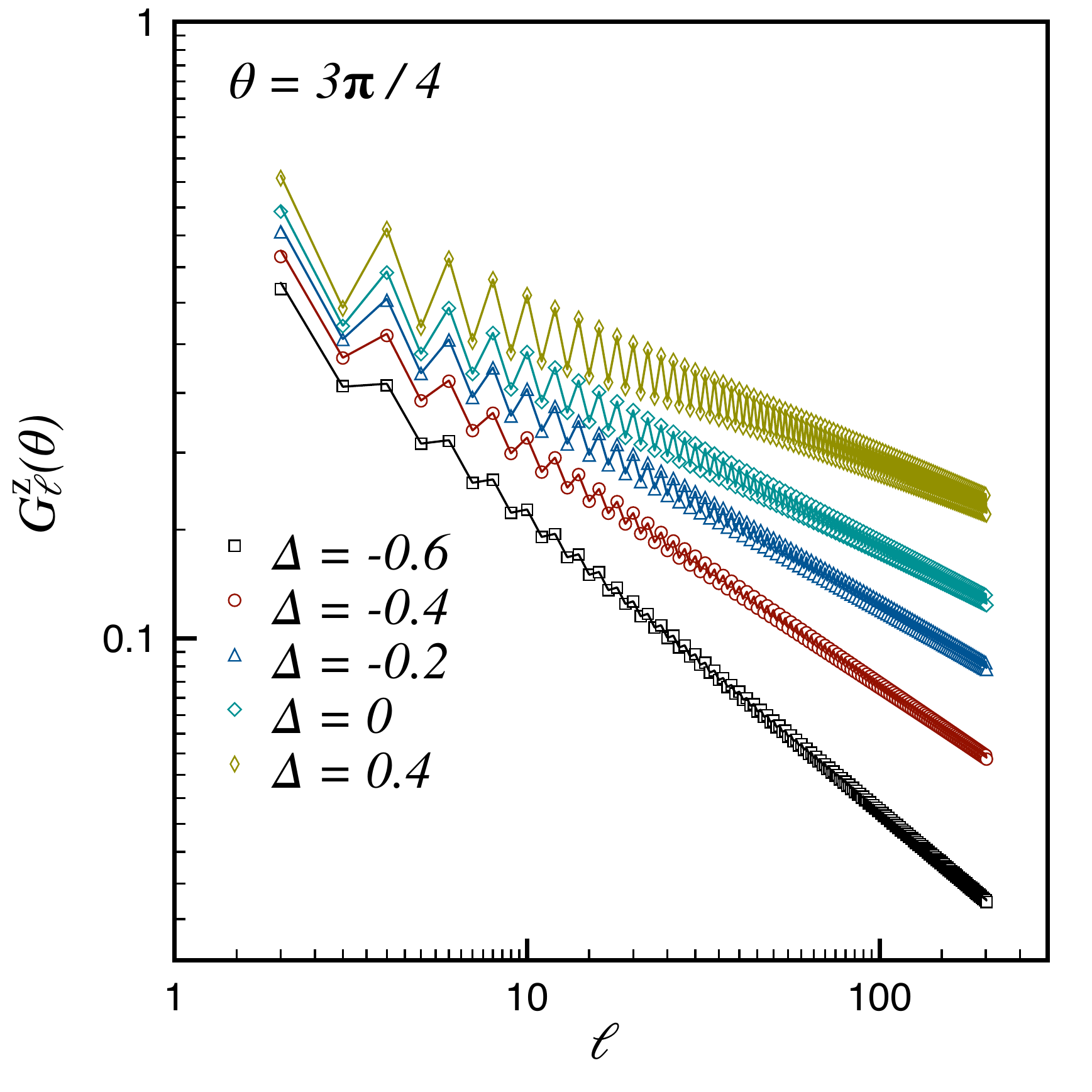}
\caption{\label{fig_Gz}
$G^{z}_{\ell}(\theta)$ as a function of subsystem size $\ell$ 
for different values of the anisotropy $\Delta$ and $\theta$. iTEBD
results (symbols) are compared to the Luttinger liquid prediction
(\ref{GzLL}) (full lines), apart from XX case ($\Delta=0$) where the
exact formula (\ref{RhoD0}) has been used.
}
\end{figure}

In Fig.~\ref{fig_Gz} we show the power law decay of $G^z_\ell(\theta)$
with subsystem size $\ell$ for several fixed values of $\theta$. In
all cases the agreement with the Luttinger liquid prediction \fr{GzLL}
is seen to be excellent. 
A comparison with the exact results at $\Delta=0$  provides a useful
accuracy check for our iTEBD data. As expected, the discrepancy
grows with increasing subsystem size $\ell$. Moreover, it
also depends on $\theta$ and grows as $\theta$ approaches $\pm \pi$. 
However up to subsystem sizes of $\ell = 200$ the relative error of our
iTEBD result is less than $0.1\%$.

\begin{figure}[ht]
\begin{center}
\includegraphics[width=0.3\textwidth]{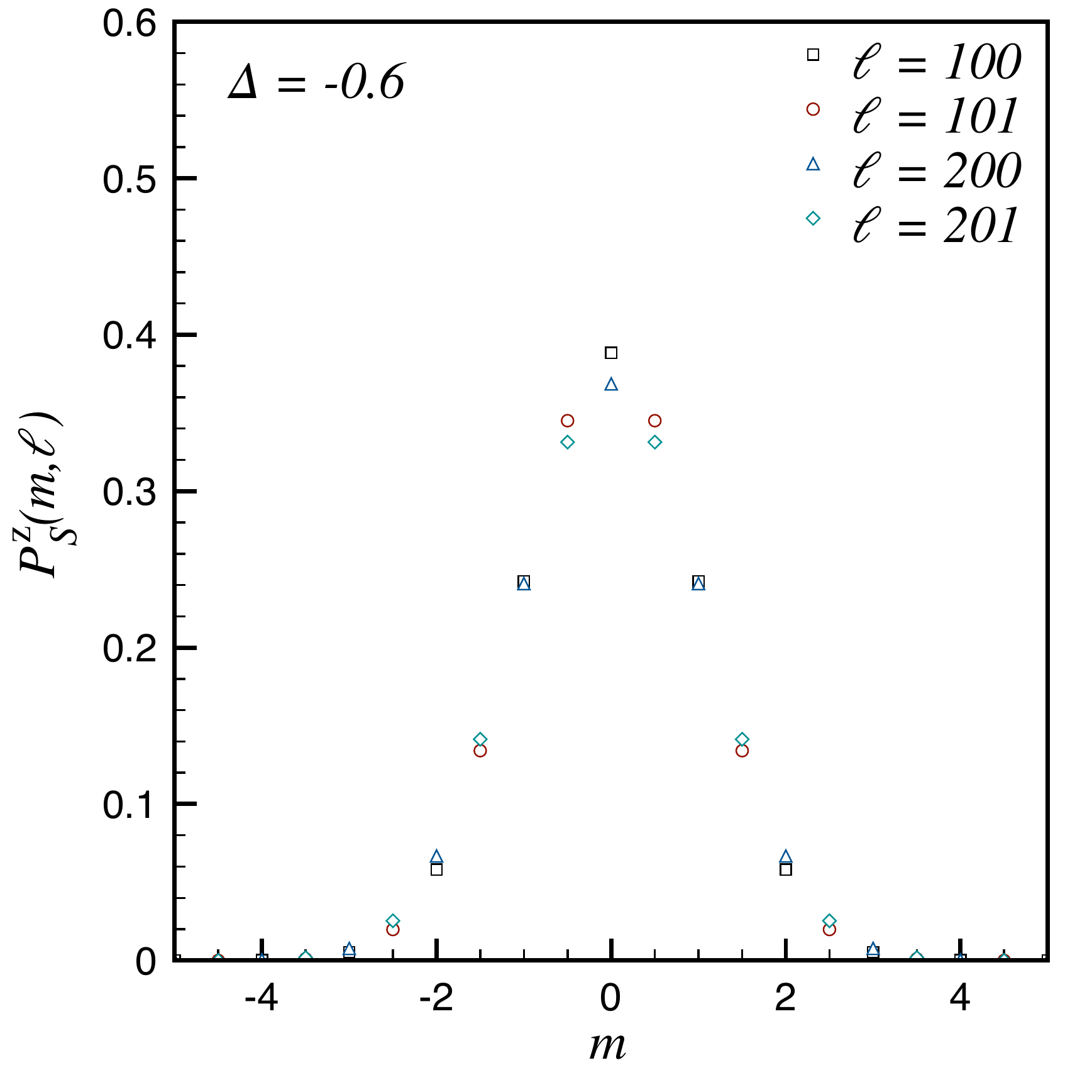}
\qquad
\includegraphics[width=0.3\textwidth]{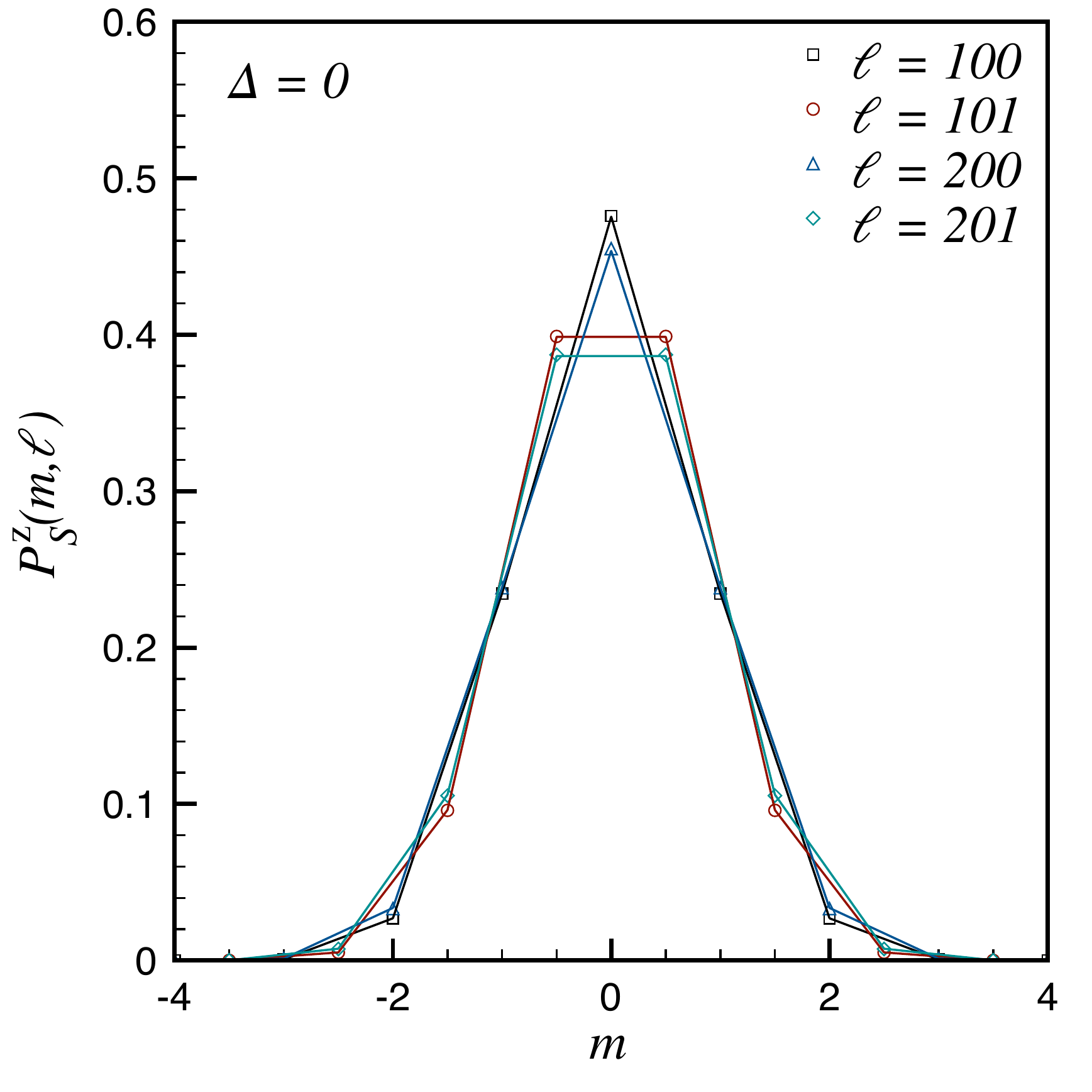}
\qquad
\includegraphics[width=0.3\textwidth]{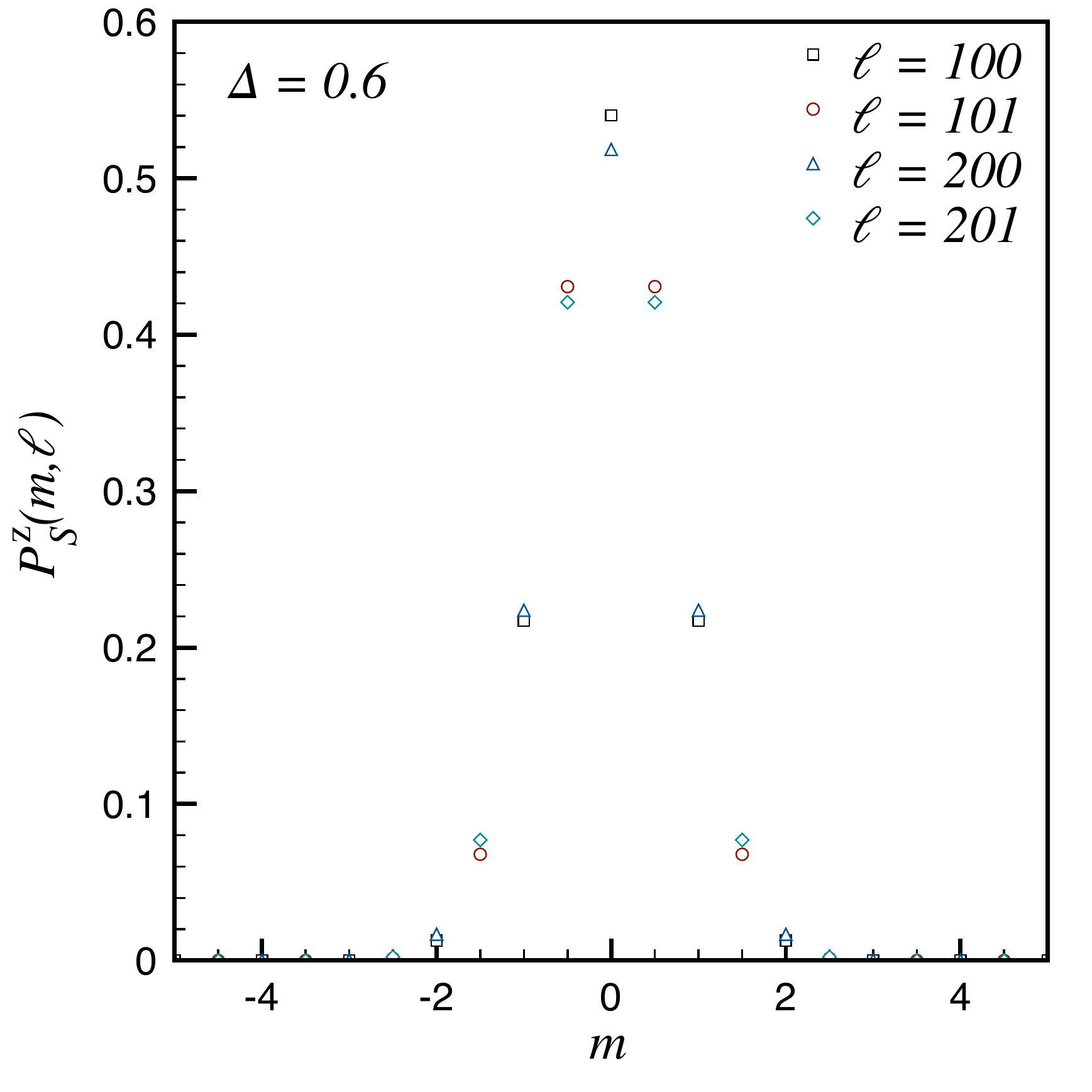}
\caption{\label{fig:PSz}
Probability distribution functions $P_S^z(m,\ell)$ for
$\Delta=-0.6$, $\Delta=0$ and $\Delta=0.6$. 
In the noninteracting case the full lines are
the exact results obtained integrating the determinant formula.
}
\end{center}
\end{figure}

The probability distribution $P_S^z(m,\ell)$ is then readily obtained
by Fourier transforming $G^{z}_{\ell}(\theta)$. Plotting again the
weights of the delta-functions that fix the possible values of $m$
gives the results shown in Fig.~\ref{fig:PSz}.
The probability distribution is centred around $m=0$ and is very
narrow for all anisotropies $\Delta$. Moreover, there is very little
subsystem size dependence for the large values of $\ell$ considered.

\subsection{Transverse generating function \texo{$G^x_\ell(\theta)$}}
\begin{figure}[ht]
\includegraphics[width=0.3\textwidth]{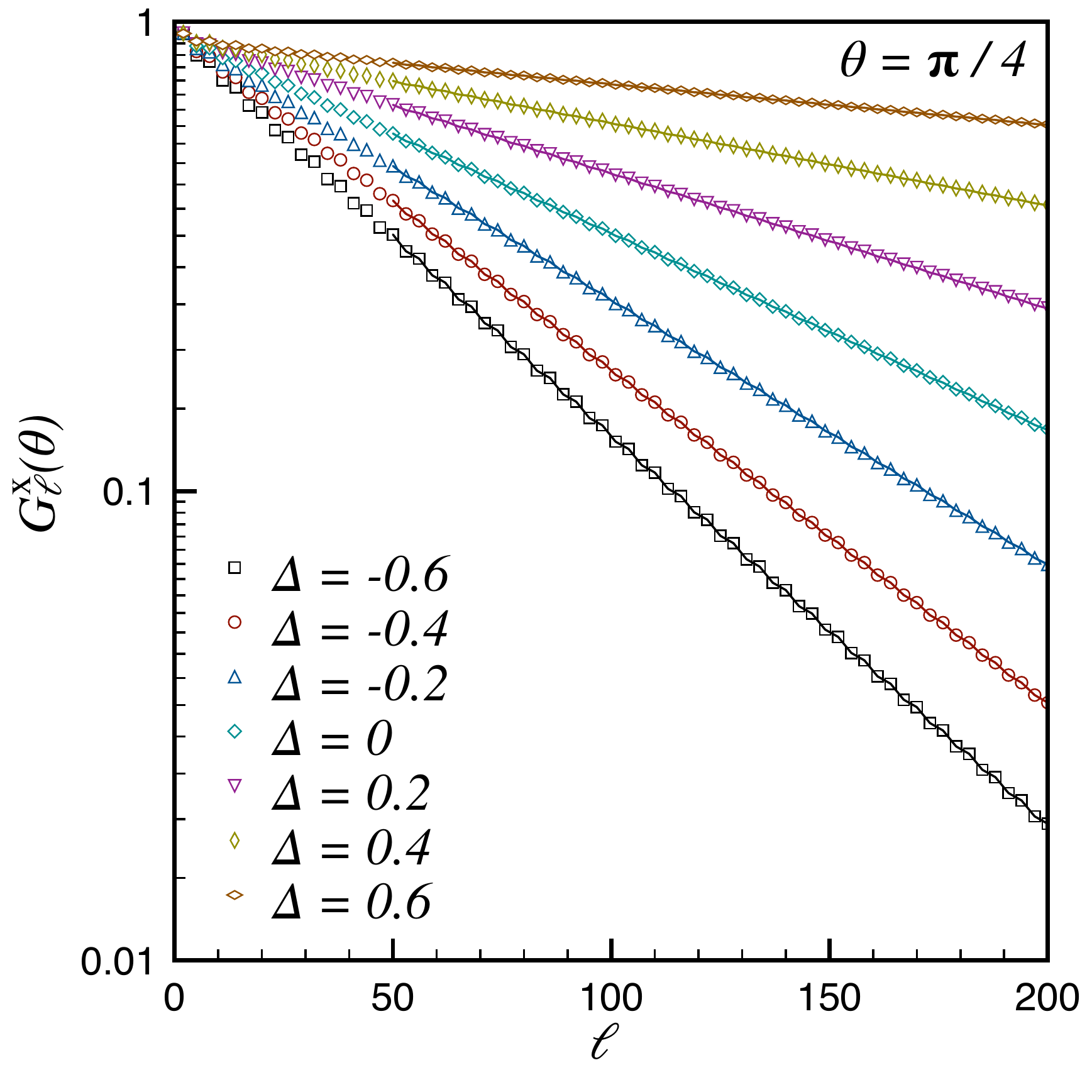}
\qquad
\includegraphics[width=0.3\textwidth]{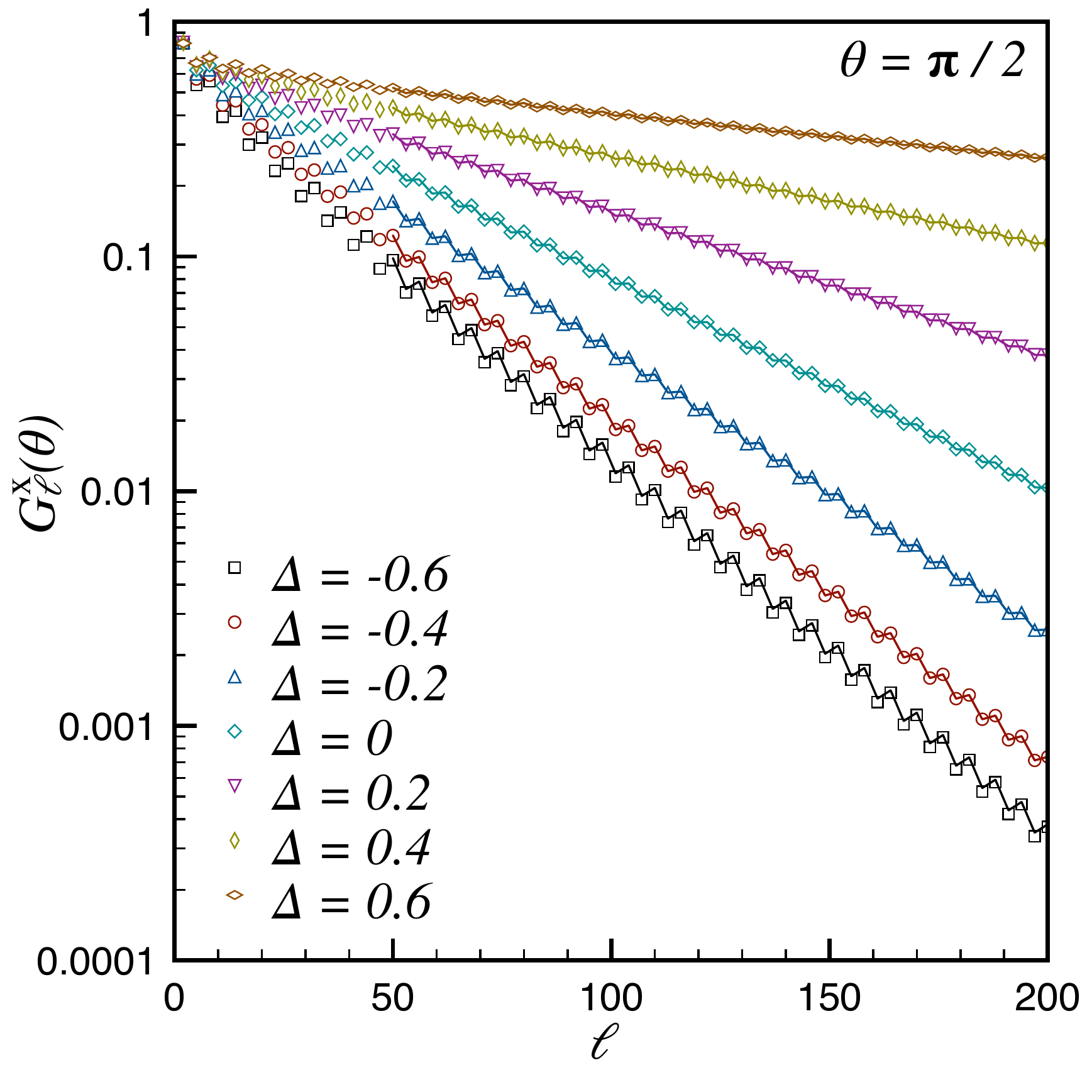}
\qquad
\includegraphics[width=0.3\textwidth]{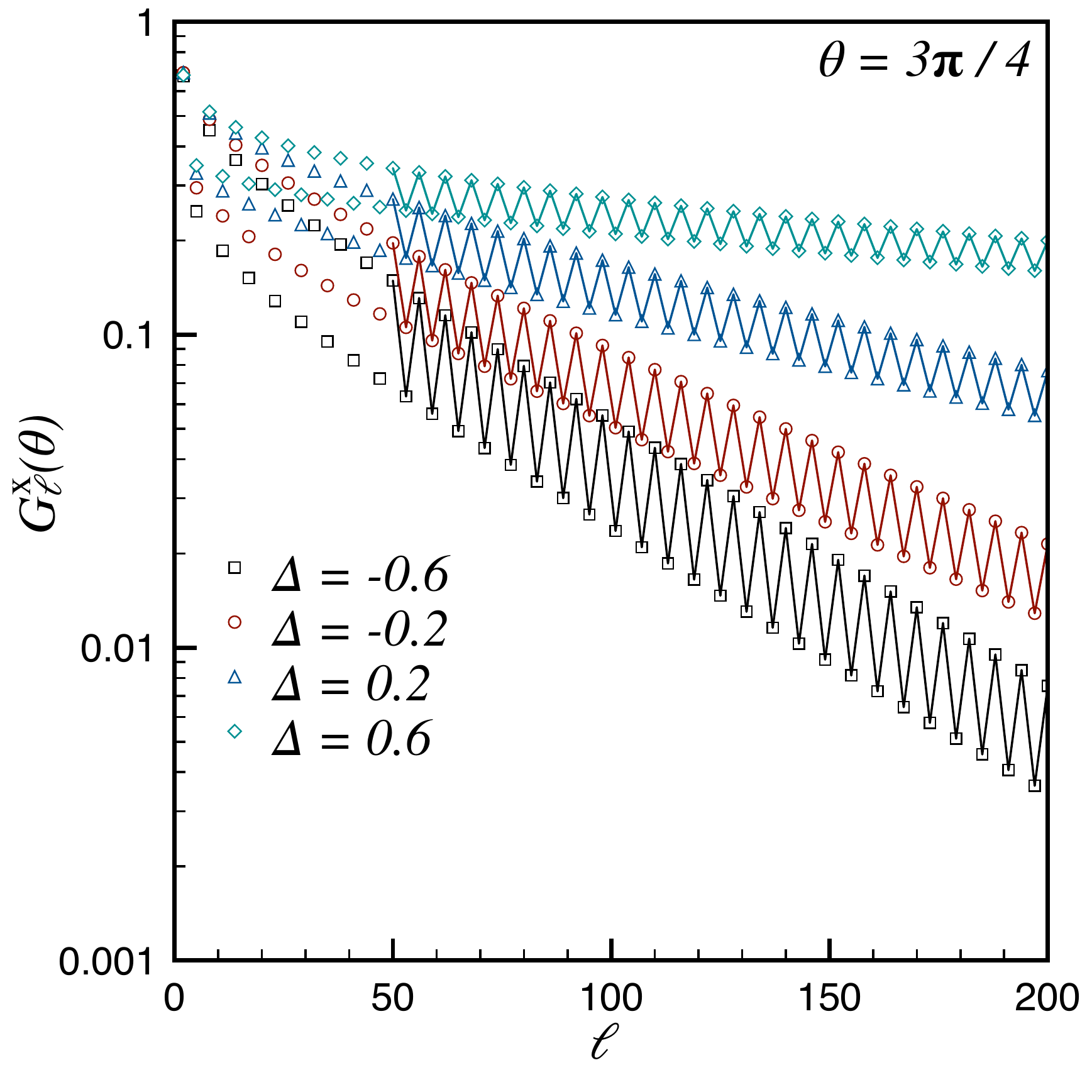}
\caption{Transverse generating function $G^{x}_{\ell}(\theta)$ vs
$\ell$ for different values of the anisotropy $\Delta$ and 
parameter $\theta$. The full lines are best fits to the functional
form \ref{fit_func}. 
}
\label{fig_Gx}
\end{figure}

\begin{figure}[ht]
\includegraphics[width=0.4\textwidth]{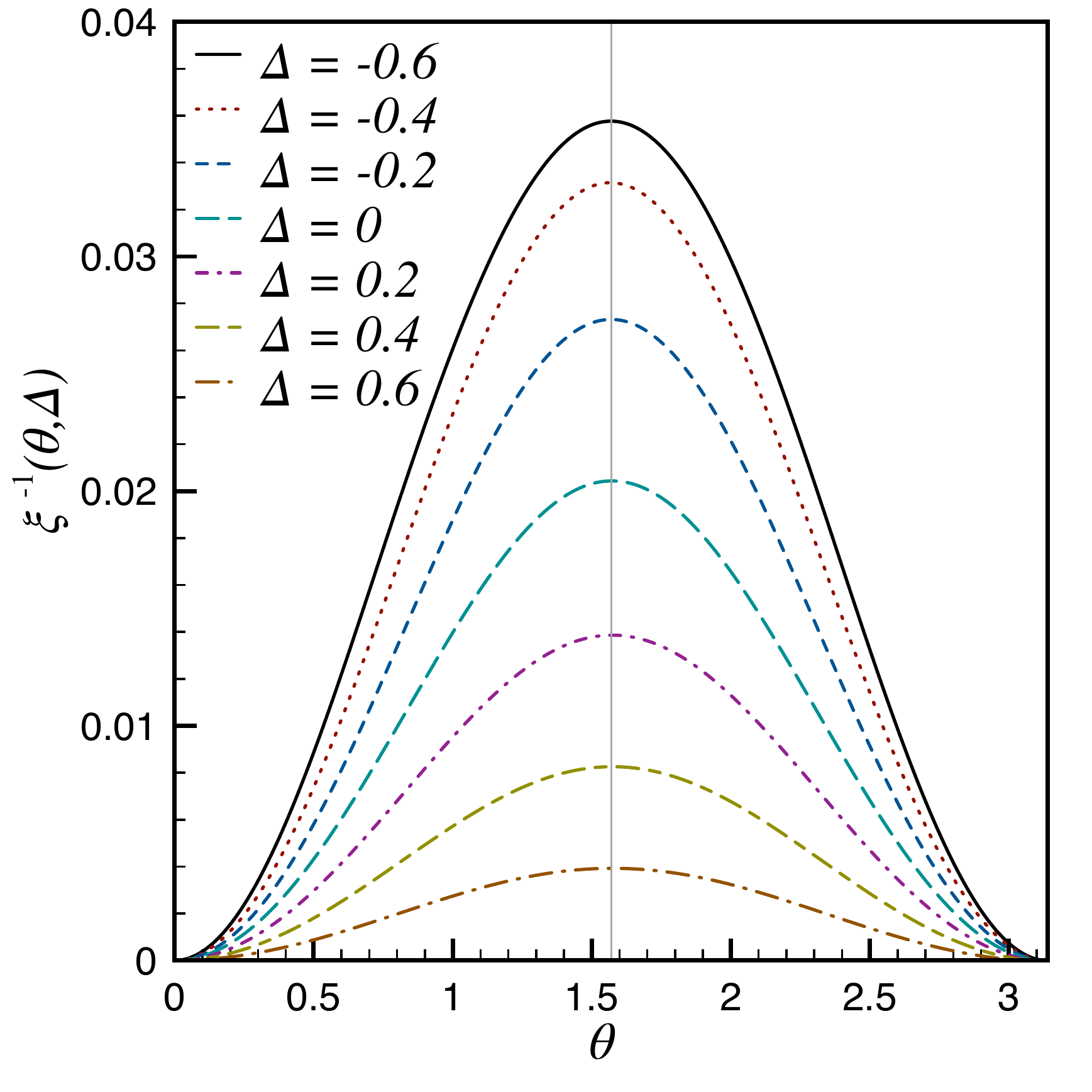}
\qquad
\includegraphics[width=0.4\textwidth]{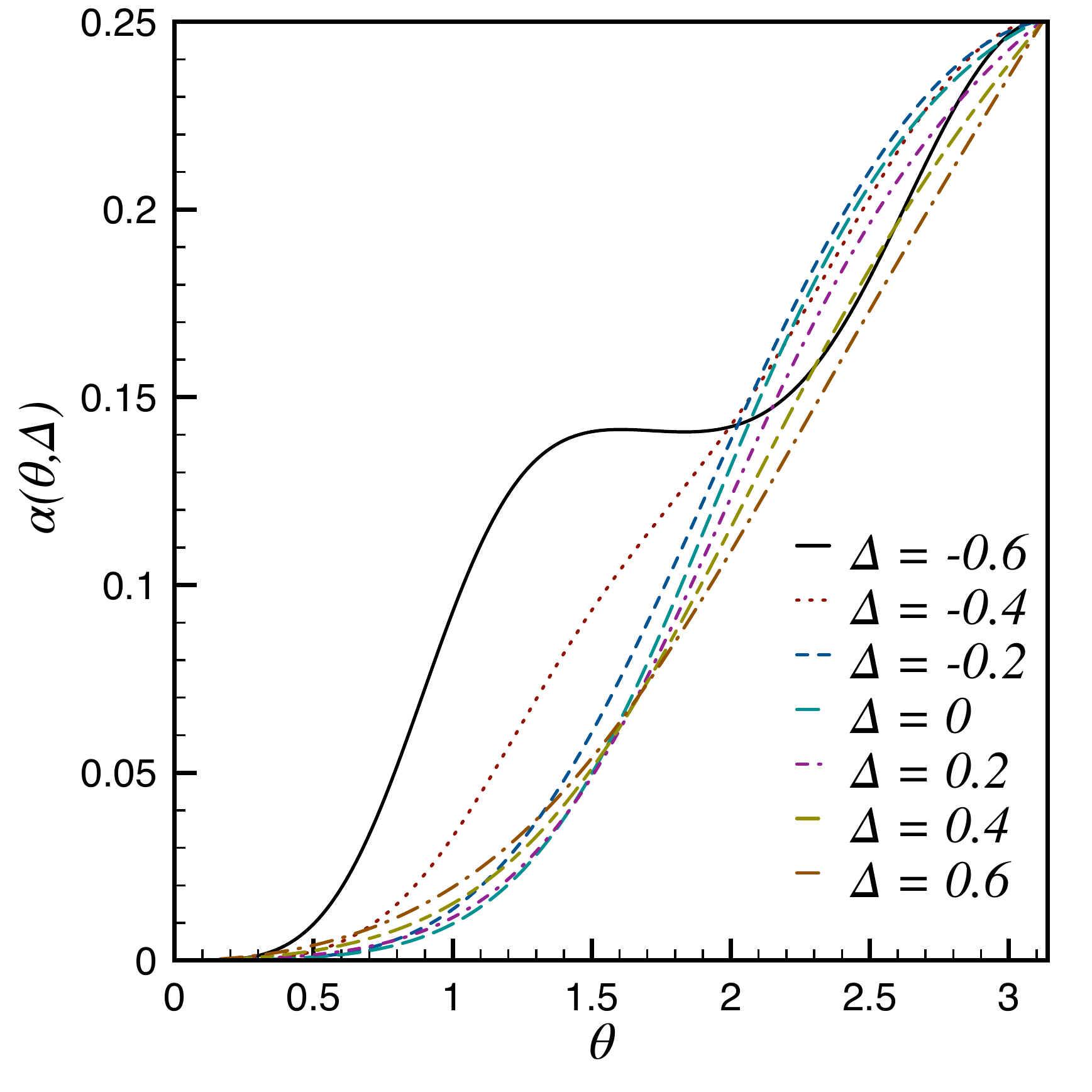}
\caption{Inverse decay length $1/\xi$ (right) and power-law exponent $\alpha$
of $G^{x}_{\ell}(\theta)$ as functions of $\theta$ for several
values of the interaction strength $\Delta$. The best fit parameters
have been obtained using the function defined in (\ref{fit_func}) and
considering only even values of $\ell$.} 
\label{fig1}
\end{figure}

The generating function $G^x_\ell(\theta)$ cannot be easily analyzed
by either free fermion or bosonization methods. As we will see in
section \ref{ssec:Gx} it is however possible to determine it by field
theory methods in particular limits. Our numerical results indicate
that $G^x_{\ell}(\theta)$ decays exponentially in the subsystem size
$\ell$ for all values of $\theta$ except $\theta=\pm\pi$, where it
displays a power-law decay for even subsystem sizes (and vanishes for
odd $\ell$) 
\be
G^x_{2\ell}(\pm\pi)\propto (2\ell)^{-1/4}\ .
\ee
In order to analyze our numerical data for other values of $\theta$ we
have carried out fits to the following functional form
\be\label{fit_func}
G^x_\ell(\theta) \simeq A(\theta,\Delta)
\frac{{\rm
e}^{-\ell/\xi(\theta,\Delta)}}{\ell^{\alpha(\theta,\Delta)}}\left[
1+B(\theta,\Delta)\frac{(-1)^\ell}{\ell^{\beta(\theta,\Delta)}}
\right] .
\ee
The resulting fits to our numerical data are shown in
Figure \ref{fig_Gx} for some representative values of $\theta$.
The agreement between the fits and the data is seen to be very good.
In Figure \ref{fig1} we show the functions $\xi^{-1}(\theta,\Delta)$
and $\alpha(\theta,\Delta)$ resulting from our fits for several values
of the anisotropy parameter $\Delta$.  We checked the stability of the
results against the ``fit window'' $[\ell_{min},\ell_{max}]$ of
subsystem sizes used by suitably varying $\ell_{min}$ and $\ell_{max}$
in the interval $[50,200]$. 
The inverse decay length $\xi^{-1}(\theta,\Delta)$ is seen to take its
maximum around $\theta=\frac{\pi}{2}$ and is generally quite small. As
the anisotropy $\Delta$ approaches unity $\xi^{-1}(\theta,\Delta)$ is
seen to approach zero. This is expected because for $\Delta=1$ the
spin rotational symmetry imposes $G^x_\ell(\theta)=G^z_\ell(\theta)$
and as we have seen the latter decays as a power law in
$\ell$. Similarly, for the correlation length diverges for
$\theta\rightarrow\pm\pi$, which indicates power-law behaviour in
$\ell$ for these values of $\theta$. The exponent
$\alpha(\theta,\Delta)$ of the power-law factor in \fr{fit_func}
appears to be a monotonically increasing function of
$\theta\in[0,\pi]$. For small $\theta$ it behaves as $\alpha(\theta\ll
1,\Delta) \sim \theta^{2}$. 

\begin{figure}[ht]
\begin{center}
\includegraphics[width=0.3\textwidth]{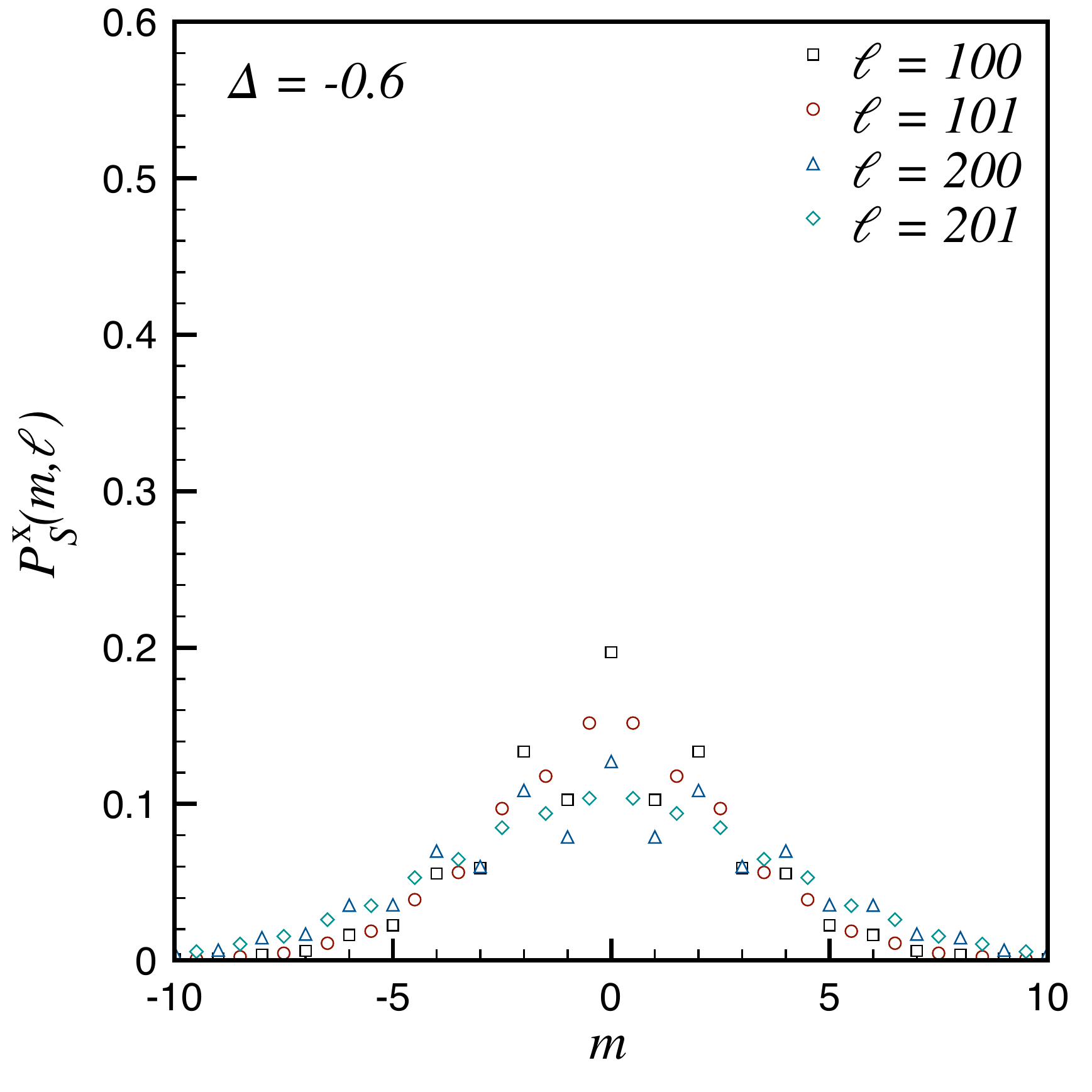}
\qquad
\includegraphics[width=0.3\textwidth]{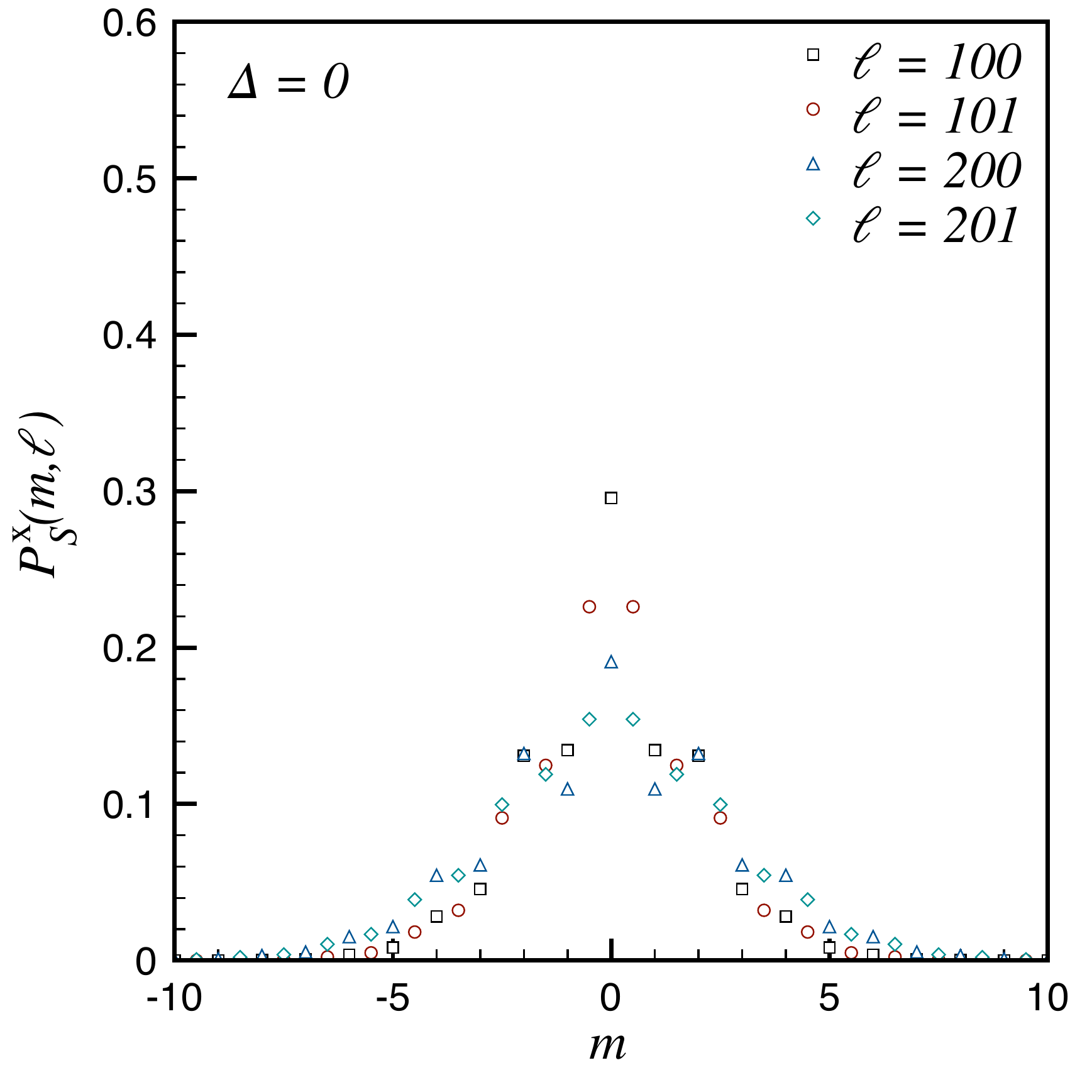}
\qquad
\includegraphics[width=0.3\textwidth]{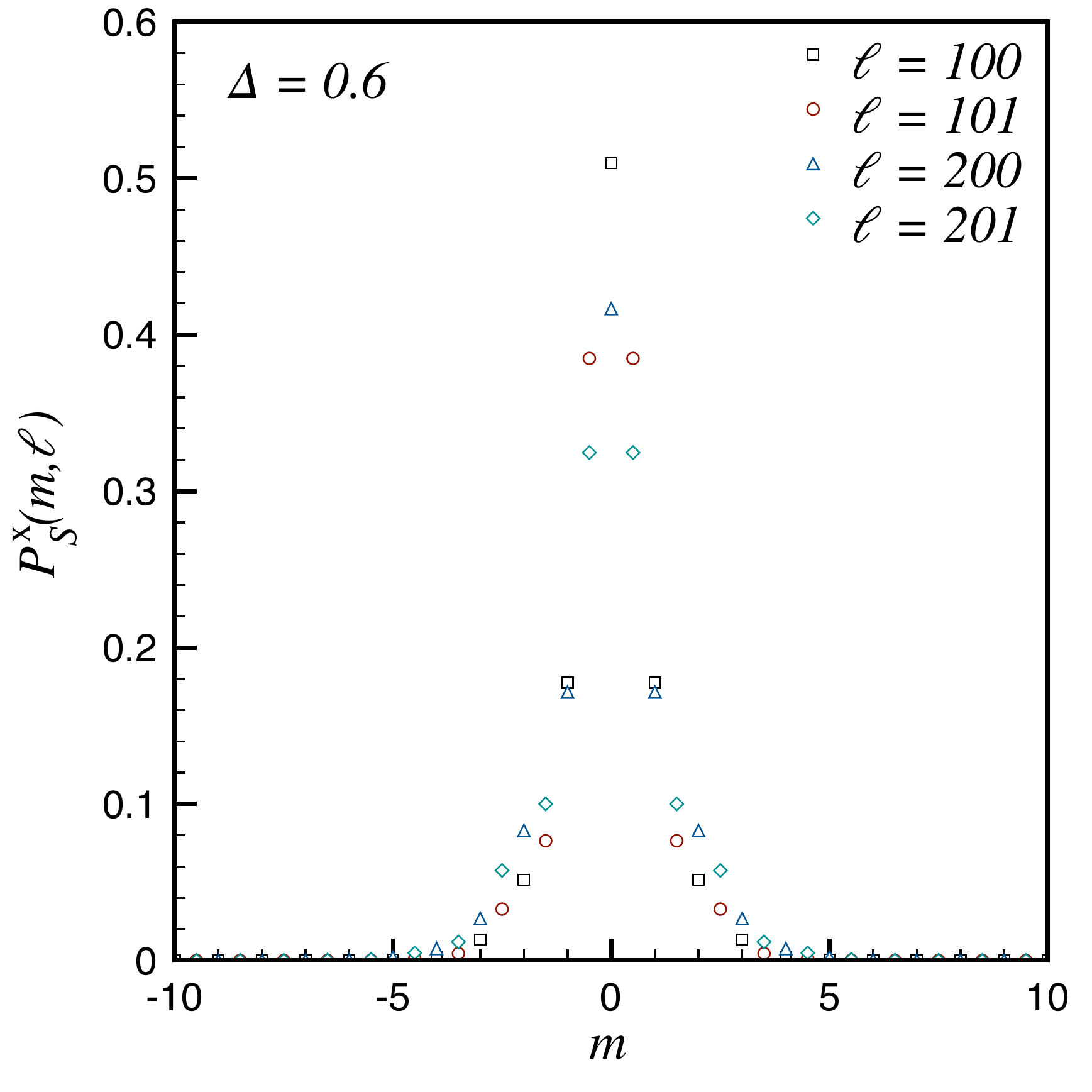}
\caption{\label{fig:PSx}
Probability distribution functions $P_S^x(m,\ell)$ for
$\Delta=-0.6$, $\Delta=0$ and $\Delta=0.6$. The even/odd effect
in $m$ is more pronounced for even $\ell$ and negative values of
$\Delta$. 
}
\end{center}
\end{figure}

Finally we present results for the probability distribution
$P_S^x(m,\ell)$ of the smooth, transverse subsystem magnetization in
Fig.~\ref{fig:PSx}.
We see that the probability distribution has a single maximum at $m=0$
and is generally quite narrow (albeit not as narrow as
$P^z_S(m,\ell)$). Its width increases with diminishing $\Delta$ and is
largest when $\Delta\to -1$. 
\subsubsection{Perturbed Luttinger liquid approach to the transverse
generating function} 
\label{ssec:Gx}
\begin{figure}[ht]
\begin{center}
\includegraphics[width=0.45\textwidth]{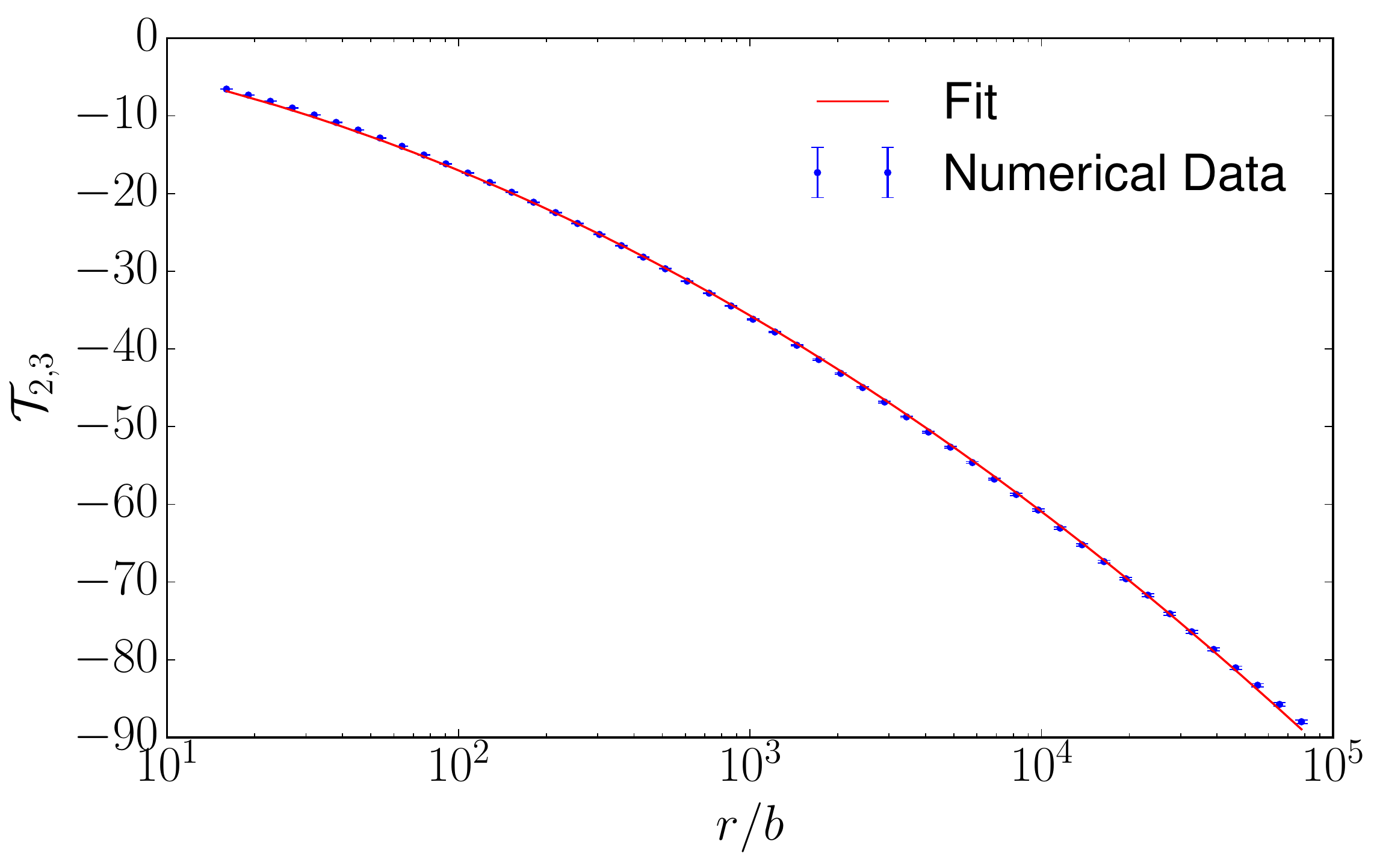}
\qquad
\includegraphics[width=0.45\textwidth]{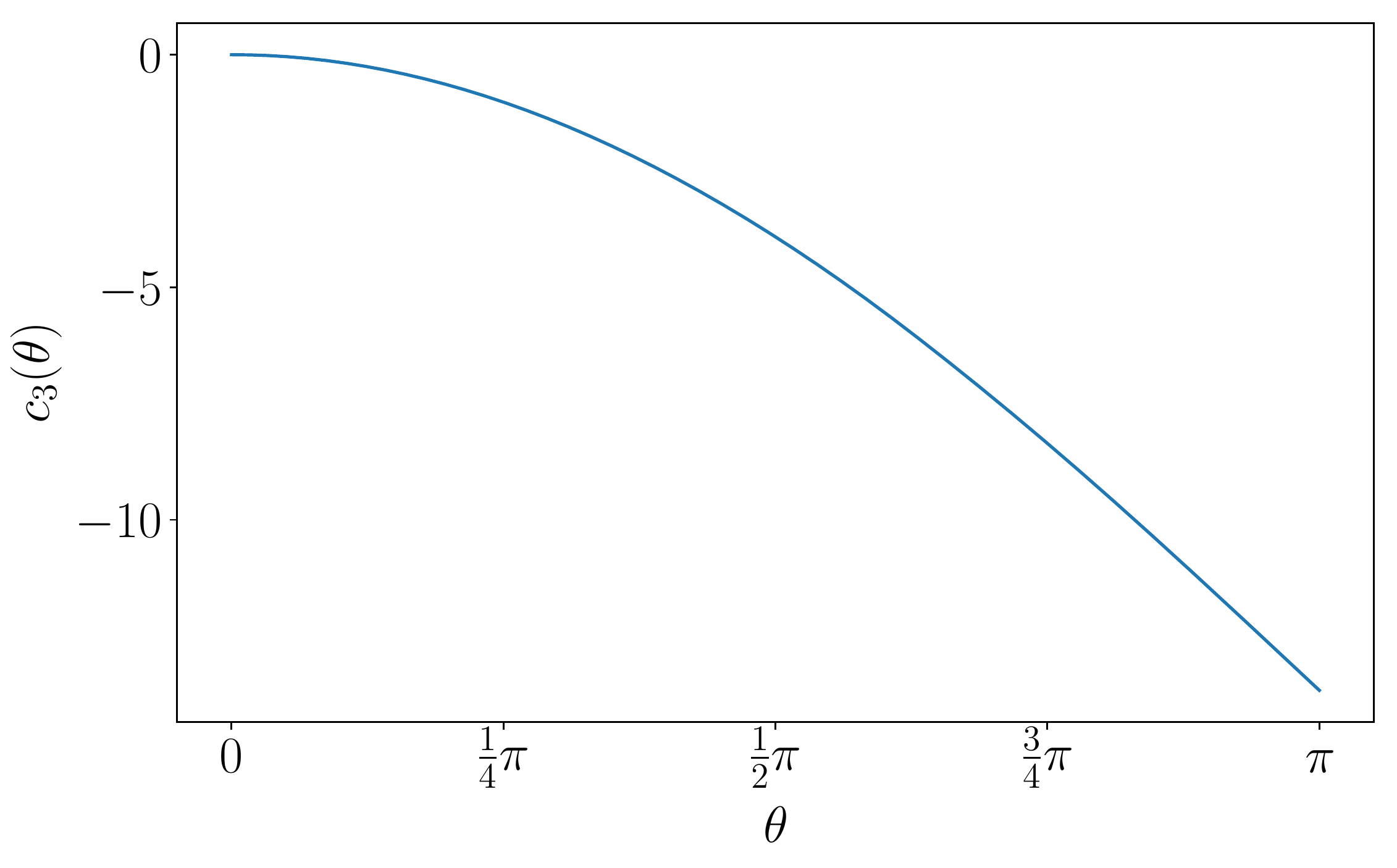}
\caption{\label{fig:FitTermT3}
(Left) Best fit of numerical results for the integral \fr{T230} to the
functional form \fr{T23}. The agreement is seen to be very good.
(Right) Function $c_3(\theta)$ extracted from the numerical fit.}
\end{center}
\end{figure}

In order to make progress, we consider the vicinity of the XXX point
$\Delta=1$ and choose our anisotropy axis to lie along the $x$ direction, i.e. 
\begin{align}
H(\Delta)=J\sum_{j=1}^L \tS_j^y \tS_{j+1}^x+\tS_j^y \tS_{j+1}^y+\tS_j^z
\tS_{j+1}^z+(\Delta-1)\tS_j^x \tS_{j+1}^x\ .
\end{align}
We now bosonize at the XXX point and then take the anisotropy into
account as a perturbation. In the low energy limit the Hamiltonian can
be written in the form
\begin{align}
\mathcal{H}(\Delta)&=\frac{v}{2}\int\mathrm{d}
x\; \left(\partial_x \Theta\right)^2+\left(\partial_x\Phi\right)^2
+\int\mathrm{d}x\; \sum_a \tilde g_a J^{a}\bar J^{a}\ ,
\label{HPTLL}
\end{align}
where $\tilde{g}_a\propto (\Delta-1)$ and $J^a$ and $\bar J^a$ are the right
and left chiral currents respectively. Defining chiral fields
by
$\Phi(x,t)=\phi_R(vt-x)+\phi_L(vt+x)=\bar\varphi(\bar z)+\varphi(z)$
where $z=v\tau+ix$, we have the following expressions
for the currents
\bea
J^3&=&\frac{i}{\sqrt{2\pi}}:\partial\varphi(z):\ ,\quad
\bar J^3 =-\frac{i}{\sqrt{2\pi}} :\bar\partial\bar\varphi(\bar z):\ ,\nn
J^+&=&\frac{1}{2\pi a_0} :\exp{i\sqrt{8\pi}\varphi(z)}:\ ,\quad
\bar J^+=\frac{1}{2\pi a_0}:\exp{-i\sqrt{8\pi}\bar\varphi(\bar z)}:\ .
\eea
By virtue of the global $U(1)$ symmetry of $H(\Delta)$ we have only two
independent coupling constants and $g_3=g_2$. The couplings fulfill
the Kosterlitz-Thouless RG-equations \cite{GNT}
\bea
\frac{d\tilde g_1}{d\log(L/b)}&=&\frac{1}{2\pi v}\tilde g_2^2 \ ,\qquad
\frac{d\tilde g_2}{d\log(L/b)}=\frac{1}{2\pi v}\tilde g_1 \tilde
g_2 \ ,
\label{eq:runningcoupl} 
\eea
where $L$ and $b$ are hard long and short-distance cutoffs. To proceed
it is convenient to define new couplings by
\be
g_a=-\frac{1}{2\pi v}\tilde{g_a}\ .
\ee
The combination $g_1^2-g_2^2=\mu^2$ is an RG invariant and can be
obtained by matching to Bethe-Ansatz calculations\cite{gia} 
\begin{align}
  \mu = 2\Big(1-\frac{\pi}{2\ \text{arccos}(\Delta-1)}\Big).
\end{align}
The low-energy projection of the spin operator in the $z$-direction
becomes  
\begin{align}
  \tS^z_j \approx m-\frac{a_0}{\sqrt{2\pi}}\partial_x \Phi(x) + (-1)^j
  a_1 \sin{\sqrt{2\pi}\Phi(x)}+\dots,
\end{align}
where $a_0$ is the lattice spacing and the amplitude $a_1$ is known
exactly\cite{lukyanovterras}.
If we now make the assumption that the staggered piece of the spin
operator can be neglected (as was the case for the longitudinal
generating function considered above), we have  
\begin{align}
G^x_l(\theta)=\langle{\rm GS}|\exp\Big(i\theta\sum_{j=1}^\ell\tS^z_j
\Big)|{\rm GS}\rangle
\approx \langle{\rm GS}|e^{-i\frac{\theta}{\sqrt{2\pi}}(\Phi(l
a_0)-\Phi(0))}|{\rm GS}\rangle\ .
\label{GxFT}
\end{align}
We will now determine \fr{GxFT} for large subsystem sizes $r=la_0$ by
means of RG-improved perturbation
theory\cite{collins,affleck98,BarzykinAffleck} in the anisotropic
current-current interactions \fr{HPTLL}. The Euclidean action 
corresponding to \fr{HPTLL} is
$S= \frac{1}{2} \int \mathrm{d}^2z (\partial_\mu\Phi)^2  +S_\text{int}$
with
\begin{align}
  S_\text{int}&=\int \mathrm{d}^2z\left[
  -g_2 \partial \varphi(z)\bar\partial\bar\varphi(\bar
  z)+ \frac{g_1+g_2}{4\pi
  a_0^2} \cos\left(\sqrt{8\pi}(\varphi(z)+\bar\varphi(\bar
  z))\right)+\frac{g_1-g_2}{4\pi
  a_0^2} \cos\left(\sqrt{8\pi}(\varphi(z)-\bar\varphi(\bar
  z)\right)\right].
 \label{interactionaction}
\end{align}
To second order in perturbation theory in $S_{\rm int}$ we have
\begin{align}
\avg{T\mathcal{O}} = \avg{T\mathcal{O}}_0
 - \avg{T\mathcal{O}S_\text{int}}
 +\frac{1}{2}\left(\avg{T\mathcal{O}S_\text{int}^2}_0-\avg{T\mathcal{O}}_0\avg{TS_\text{int}^2}_0\right)+\dots,
\end{align}
where $T$ is the imaginary time ordering operator and $\avg{.}_0$ is the
path integral average with respect to the gaussian action.
The perturbative expansion of the transverse generating function
$G^x_r(\theta)$ thus reads
\begin{align}
G^x_r(\theta)=\avg{Te^{-i\frac{\theta}{\sqrt{2\pi}}(\Phi(r)-\Phi(0))}}=
\left(\frac{a_0}{r}\right)^{\frac{\theta^2}{4\pi^2}}
+{\cal T}_1+\sum_{a=1}^3{\cal T}_{2,a}\ ,
\label{Gxr}
\end{align}
where ${\cal T}_1$ and ${\cal T}_{2,a}$ denote the contributions at
first and second order. The first order contribution can be evaluated
following Ref.~\onlinecite{LudwigCardy}, which gives a logarithmic
divergence in the short-distance cutoff $b$
\bea
{\cal T}_1=-\left(\frac{a_0}{r}\right)^{\frac{\theta^2}{4\pi^2}} \frac{g_2\theta^2}{8\pi^2} \log\Big({\frac{r}{b}}\Big)\ .
\eea
Due to electro-neutrality the various interaction terms
in \eqref{interactionaction} do not mix in second order perturbation
theory. The contribution proportional to $g^2_2$ can again be
evaluated following Ref.~\onlinecite{LudwigCardy} with the result
\begin{align}
{\cal
T}_{2,1}=-\left(\frac{a_0}{r}\right)^{\frac{\theta^2}{4\pi^2}}\frac{g_2^2}{2}
\left[\Big(\frac{\theta^2}{8\pi^2}\Big)^2 \log^2\Big({\frac{r}{b}}\Big)
+\Big(\frac{\theta }{4\pi}\Big)^2\log\Big({\frac{r}{b}}\Big)\right].
\end{align}
The contribution ${\cal T}_{2,2}$ proportional to $(g_1+g_2)^2$ is of
the form
\begin{align}
{\cal T}_{2,2}&=
\left(\frac{a_0}{r}\right)^{\frac{\theta^2}{4\pi^2}}
\left(\frac{g_1+g_2}{8\pi}\right)^2 
\int d^2z\int d^2w\
\frac{1}{\abs{z-w}^4}\left[\left(\frac{\abs{r-z}\abs{w}}{\abs{r-w}\abs{z}}\right)^{\frac{\theta}{\pi}}-1\right].
\label{T22}
\end{align}
The logarithmically divergent parts of \fr{T22} can be determined by
adapting the results of Refs~\onlinecite{LeDoussal,DL} to our real-space
cutoff regularization scheme. This gives
\begin{align}
{\cal T}_{2,2}&=\left(\frac{a_0}{r}\right)^{\frac{\theta^2}{4\pi^2}}
\left(\frac{g_1+g_2}{8\pi}\right)^2
 \left[\theta^2\log^2\Big(\frac{r}{b}\Big)
+\theta^2\left(2-2\gamma_E-\psi\Big(\frac{\theta}{2\pi}\Big)-\psi\Big(-\frac{\theta}{2\pi}\Big)\right)\log\Big(\frac{r}{b}\Big)\right],
\end{align}
where $\gamma_E$ is the Euler-Mascheroni constant and $\psi(x)$ is the
Digamma function. To lighten notations in what follows we define
\be
c_2(\theta)=2-2\gamma_E-\psi\left(\frac{\theta}{2\pi}\right)-\psi\left(-\frac{\theta}{2\pi}\right).
\ee
The third and final contribution in second order of perturbation theory
is proportional to $(g_1-g_2)^2$
\begin{align}
{\cal
T}_{2,3}=\left(\frac{a_0}{r}\right)^{\frac{\theta^2}{4\pi^2}}
\left(\frac{g_1-g_2}{8\pi}\right)^2
\int d^2z\int d^2w\
 \frac{1}{\abs{z-w}^4} \left[\left(\frac{r-z}{r-\bar
z}\frac{r-\bar w}{r-w}\frac{\bar z}{z}\frac{w}{\bar
w}\right)^{\frac{\theta}{2\pi}}-1\right].
\label{T230}
\end{align}
While the leading $\log^2(r/b)$ contribution can be easily extracted
analytically, we resorted to a numerical integration for determining
the subleading $\log(r/b)$ term
\begin{align}
{\cal  T}_{2,3}&=\left(\frac{a_0}{r}\right)^{\frac{\theta^2}{4\pi^2}}
\left(\frac{g_1-g_2}{8\pi}\right)^2
\left(-\theta^2\log^2\Big(\frac{r}{b}\Big)+c_3(\theta)\log\Big(\frac{r}{b}
\Big)\right).
\label{T23}
\end{align}
Our numerical results for ${\cal  T}_{2,3}$ are well-described by the
functional form \fr{T23} as can be see in Fig.~\ref{fig:FitTermT3}.

\begin{figure}[ht]
\begin{center}
\includegraphics[width=0.6\textwidth]{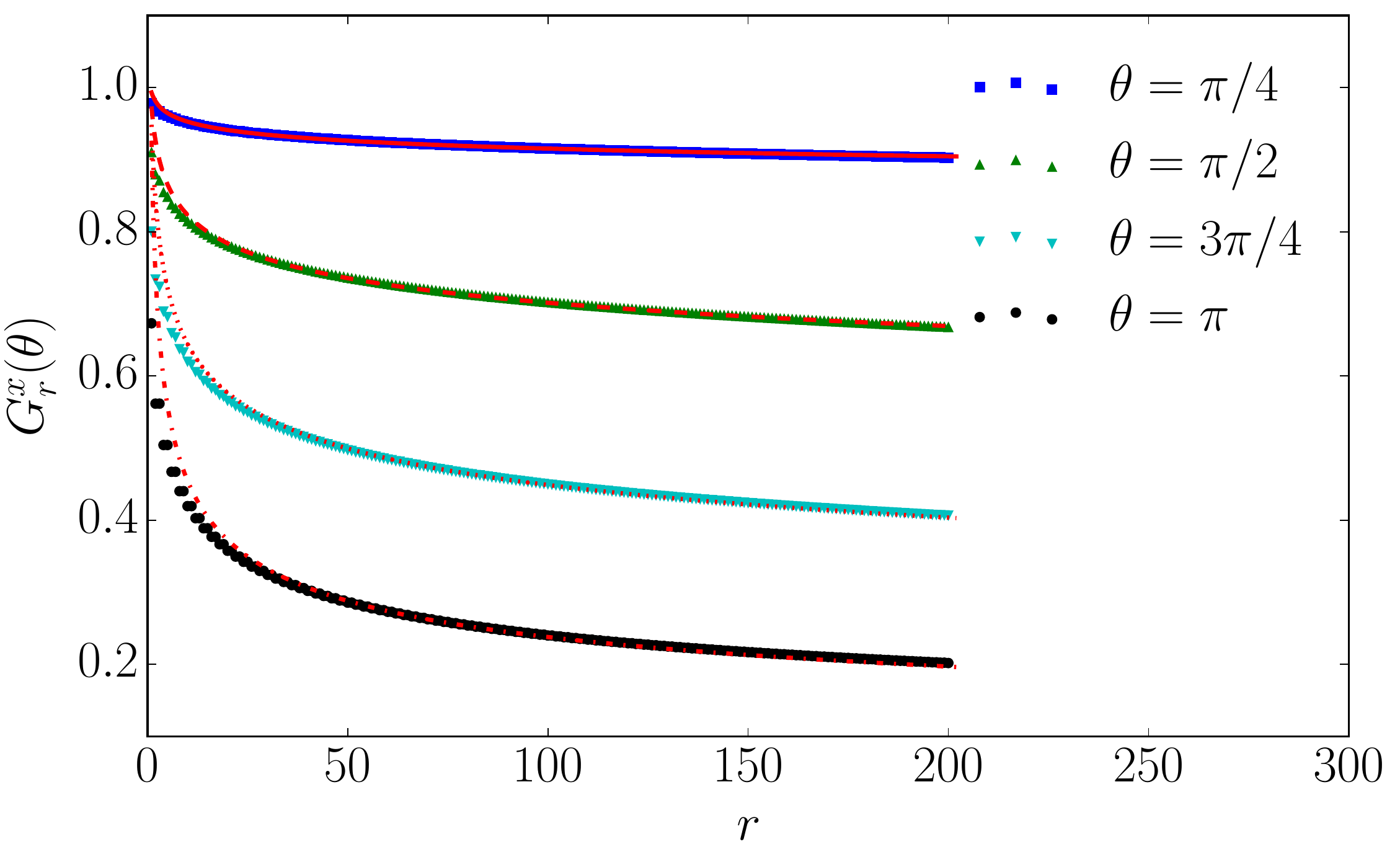}
\caption{\label{fig:Gxcomp}
Comparison of RG-improved perturbation theory \fr{Gxrfinal} 
to iTEBD results for $\Delta=0.95$ and several values of $\theta$.
}
\end{center}
\end{figure}

We now want to RG-improve the perturbative result \fr{Gxr} for
$G^x_r(\theta)$ by solving the Callan-Symanzik equation
\begin{align}
  \left(\frac{\partial}{\partial\log(r/b)}+\sum_i \beta_i(\{g\}_{j}) \frac{\partial}{\partial
  g_i}+2\gamma_\phi \right) G^{x}_r(\theta) = 0 \ .
\label{CS}
\end{align}
The anomalous dimension is extracted from the perturbative expansion
and is given by
\begin{align}
  \gamma_\phi = g_2 \frac{\theta^2}{16\pi^2}-
  g_2^2 \frac{\theta^2}{64\pi^2} -\frac{c_2(\theta)}{128\pi^2} 
  (g_1+g_2)^2 -\frac{c_3(\theta)}{128\pi^2} (g_1-g_2)^2 \ .
\end{align}
Solving the Callan-Symanzik equation by the method of characteristics
we obtain
\begin{align}
  G^{x}_r(\theta)\propto\left(\frac{b}{r}\right)^{\frac{\theta^2}{4\pi^2}}
\exp\Big(-2\int_0^{\log   r/b} \mathrm{d}l\ \gamma_\phi\Big)\; 
F\big(r,\{g(r)\}_j\big)\ .
\label{solutionCS}
\end{align}
The running couplings at scale $\log(r/b)$ are obtained by
integrating \eqref{eq:runningcoupl}  
\bea
g_1(l)&=&\frac{g_1(0)+\mu\tanh\mu l}{1+\frac{g_1(0)}{\mu}\tanh\mu l}\ ,\quad
g_2(l)=\frac{1}{\cosh\mu
l}\frac{g_2(0)}{1+\frac{g_1(0)}{\mu}\tanh\mu l}\ .
\eea
We now expand \fr{solutionCS} in powers of the coupling constant and
match the result to the perturbative expression \fr{Gxr}. This
provides us with an expansion of the function $F$
in \fr{solutionCS}. Putting everything together we arrive at the
following expression for the RG improved correlator
\bea
G^{x}_r(\theta,\{g\}_j)&=& 
\mathcal{A}\left(\frac{b}{r}\right)^{\left(\frac{\theta}{2\pi}\right)^2+
\left(\frac{\mu}{8\pi}\right)^2(c_2(\theta)+c_3(\theta))}\nn
&\times&
\left(\frac{g_2(r)+g_1(r)}{g_2(0)+g_1(0)}\right)^{\frac{\theta^2}{8\pi^2}}
e^{\frac{\theta^2+c_2(\theta)+c_3(\theta)}{32\pi^2}\big(g_1(r)-g_1(0)\big)+
\frac{c_2(\theta)-c_3(\theta)}{32\pi^2}\big(g_2(r)-g_2(0)\big)}.
\label{Gxrfinal}
\eea
The overall amplitude
$\mathcal{A}(\theta)$ is obtained by fitting to iTEBD results for
$G^x_\ell(\theta)$. This leaves us with one free parameter, namely the
initial coupling $g_1(0)$. We fix this by fitting \fr{Gxrfinal} to
iTEBD results for one value of $\theta$. In Fig.~\ref{fig:Gxcomp} we
compare \fr{Gxrfinal} obtained in this way to numerical results
obtained by iTEBD for $\Delta=0.95$.
The agreement is seen to be quite satisfactory. For larger subsystem
sizes we expect $G^x_\ell(\theta)$ to exhibit exponential decay in
the subsystem size $\ell$. This clearly goes beyond RG-improved
perturbation theory. It is an interesting problem how to obtain the
corresponding correlation length in the framework of the perturbed
Luttinger liquid \fr{HPTLL}. 

\section{Full counting statistics for the Heisenberg ferromagnet
(\texo{$\Delta=-1$}) in the zero magnetization sector \texo{$S^z=0$}} 

At $\Delta=-1$ the model \fr{H_XXZ} undergoes a first order quantum
phase transition, where the total magnetisation of the ground state
discontinuously jumps from $S_{z} = 0$, for $\Delta > -1$, to $S_{z}
= \pm L$, for $\Delta < -1$. However, if we restrict ourselves to the
sector of the Hilbert space with zero total magnetisation, the ground
state for $\Delta=-1$ is continuously connected the ground state for
$\Delta+1=0^+$. The ground state in the $S^z=0$ sector at $\Delta=-1$
on a lattice with $L$ sites is
\be
\left.|{\rm  GS}\rangle \right|_{\Delta=-1}
=
\underbrace{\prod_{j\, {\rm odd}} (2S^{z}_{j})}_{U}
\,
\frac{1}{{\cal N}}
[S^{-}]^\frac{L}{2}|\!\uparrow_1\dots\uparrow_L\rangle\ ,
\label{GS-1}
\ee
where $S^{-} = \sum_{j=1}^{L} S^{x}_{j} -i S^{y}_{j}$. Here ${\cal N}$
is a normalization factor and the unitary transformation with $U$ maps
the Hamiltonian \fr{H_XXZ} at $\Delta=-1$ to the isotropic
ferromagnet. We note that the state
$[S^{-}]^\frac{L}{2}|\uparrow_1\dots\uparrow_L\rangle\ $  
admits an exact MPS representation with auxiliary dimension $L+1$
in terms of the vector-valued matrix $\mathbf{\Gamma}_{\alpha \, \beta} =  
\delta_{\alpha \, \beta}|\!\up\rangle  + \delta_{\alpha+1 \, \beta} |\!\down\rangle$,
and boundary vectors $v^{l}_{\alpha} = \delta_{\alpha \, L+1}$,
$v^{r}_{\alpha} = \delta_{\alpha \, 1}$. 

\subsection{Full counting statistics}
The ground state \fr{GS-1} has the following useful representation
\be
\left.|{\rm
GS}\rangle \right|_{\Delta=-1}=\frac{1}{2^L}\sum_{\sigma_1,\dots,\sigma_L}
\left[\prod_{j\ {\rm
odd}}\sigma_j\right]|\sigma_1,\sigma_2,\dots,\sigma_L\rangle\ ,
\ee
which makes it possible to obtain closed-form expressions for the
generating functions on the (staggered) subsystem
magnetization. A straightforward combinatorial analysis gives the
following results for the generating functions for a finite chain of
$L$ sites
\bea
G^z_\ell(\theta)&=&\big(\cos(\theta/2)\big)^\ell\
{}_2F_1\Big(\frac{1-\ell}{2},-\frac{\ell}{2};\frac{1-L}{2};-\tan^2(\theta/2)\Big),\ \nn
F^z_\ell(\theta)&=&\big(\cos(\theta/2)\big)^\ell\
{}_2F_1\Big(\frac{1}{2},-\bigg\lfloor\frac{\ell}{2}\bigg\rfloor;\frac{1-L}{2};-\tan^2(\theta/2)\Big),\ \nn
G^x_\ell(\theta)&=&\big(\cos(\theta/2)\big)^\ell\
{}_3F_2\Big(\frac{1}{2},-\bigg\lfloor\frac{\ell}{2}\bigg\rfloor,-\frac{L}{2};1,\frac{1-L}{2};-\tan^2(\theta/2)\Big),\ \nn
F^x_\ell(\theta)&=&\big(\cos(\theta/2)\big)^\ell\
{}_3F_2\Big(\frac{1-\ell}{2},-\frac{\ell}{2},-\frac{L}{2};1,\frac{1-L}{2};-\tan^2(\theta/2)\Big)\ .
\label{FandG0}
\eea
In the limit $L\to\infty$ \fr{FandG} simplify to
\bea
G^{z}_{\ell}(\theta) & = & F^{z}_{\ell}(\theta) = \big(\cos(\theta/2)\big)^{\ell},\nn
G^{x}_{\ell}(\theta) & = & \big(\cos(\theta/2)\big)^{\ell} \,
 \left._{2}F_{1} \left(1/2,- \lfloor \ell/2 \rfloor;1;-\tan^2(\theta/2)\right) \right. , \nn
F^{x}_{\ell}(\theta) & = & P_{\ell} \left( \cos(\theta/2) \right),
\label{FandG}
\eea
were $P_{\ell}(z)$ are Legendre polynomials. In the vicinity of
$\theta=0$ the generating functions \fr{FandG} exhibit scaling for
large subsystem sizes $\ell$ 
\bea
G^{z}_{\ell}(\theta) & = & F^{z}_{\ell}(\theta) = 
{\rm e}^{ - z^2 / 8}, \quad z= \theta \ell^{1/2},\nn
G^{x}_{\ell}(\theta) & = & {\rm e}^{ - z^2 / 16} I_{0}(z^2/ 16), \nn
F^{x}_{\ell}(\theta) & = & J_{0}(\widetilde{z}/2), \quad \widetilde{z} = \theta \ell ,
\label{FandGscaling}
\eea
where $I_{n}(z)$ and $J_n(z)$ are (modified) Bessel functions. We note that
the universal scaling function in eqn (\ref{fz1}) reduces
to \fr{FandGscaling} in the limit $\Delta\to -1$.

\subsubsection{Probability distribution functions}
\begin{figure}[ht]
\begin{center}
\includegraphics[width=0.3\textwidth]{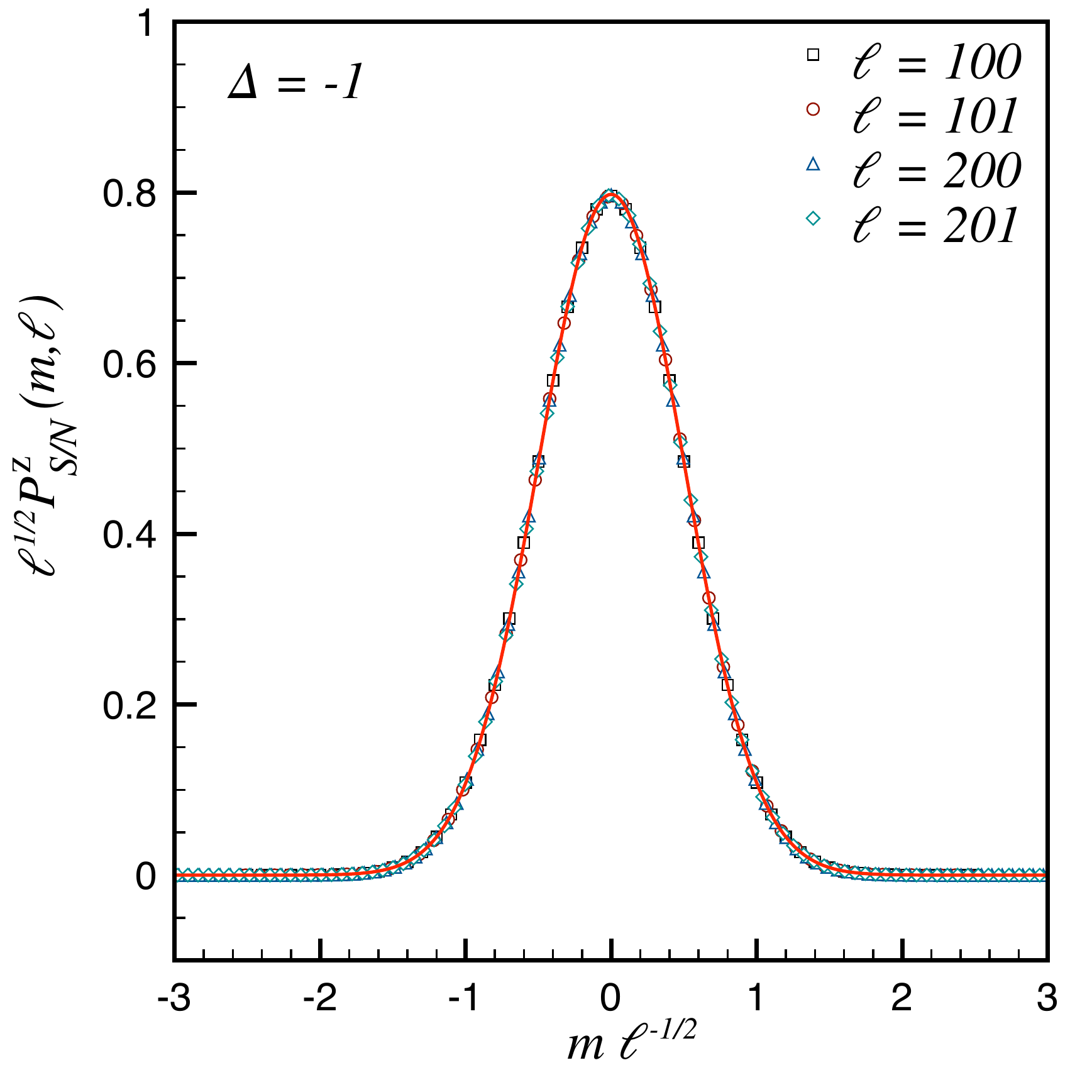}
\qquad
\includegraphics[width=0.3\textwidth]{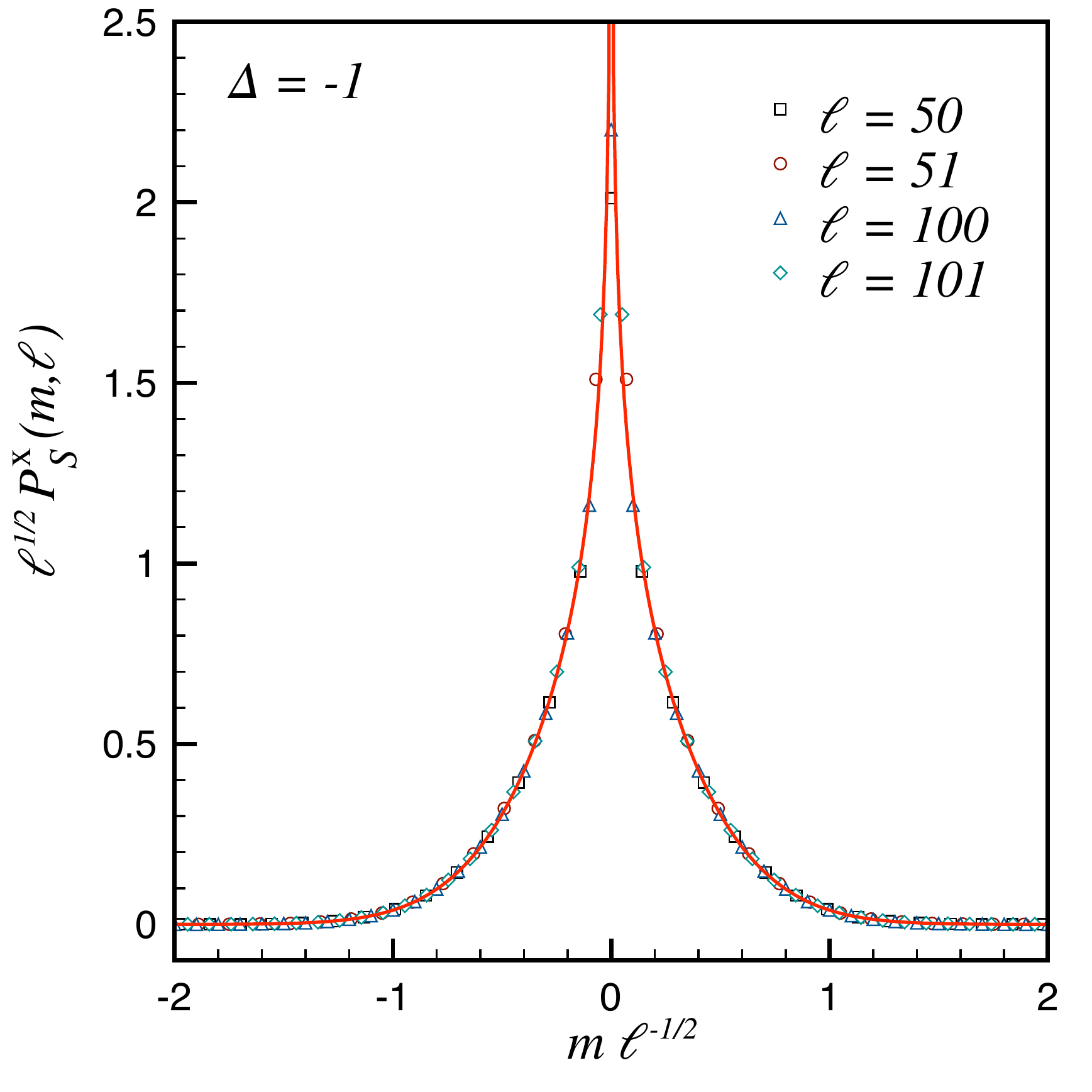}
\qquad
\includegraphics[width=0.3\textwidth]{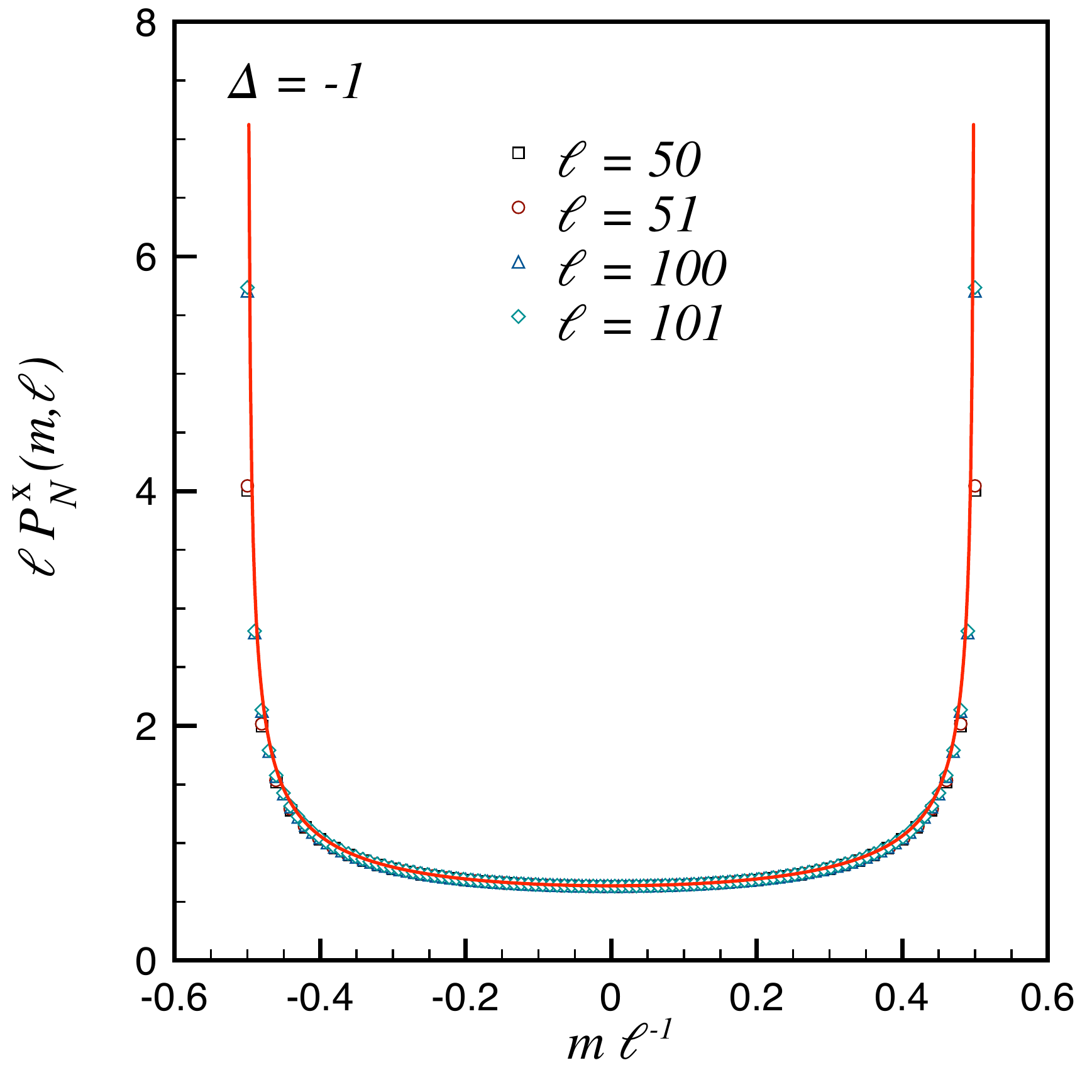}
\caption{\label{fig:PD-1}
Rescaled probability distribution functions  for
$\Delta=-1$. The even/odd effect
in $m$ is more pronounced for even $\ell$. 
The full red lines represent the analytic expressions
(\ref{eq:PDelta-1}) describing the scaling limit.
}
\end{center}
\end{figure}

From \fr{FandGscaling} we can extract the following analytic
expressions  for the probability distribution functions at $\Delta = -1$
\bea
P^{z}_{S}(m,\ell) & = & P^{z}_{N}(m,\ell) = \ell^{-1/2} \sqrt{2/\pi} \; {\rm e}^{ - 2 m^2/\ell },\nn
P^{x}_{S}(m,\ell)  & = &  \ell^{-1/2} \sqrt{2/\pi^3} \; {\rm e}^{ - m^2/\ell } K_{0}(m^2/\ell), \nn
P^{x}_{N}(m,\ell) & = & \ell^{-1} \frac{2}{\pi \sqrt{ 1- 4 m^2 / \ell^2 }
}, \label{eq:PDelta-1} 
\eea
where  $K_{n}(z)$ are modified Bessel functions. In
Fig.~\ref{fig:PD-1} we compare \fr{eq:PDelta-1} to iTEBD results for
finite $\ell$. The agreement is clearly excellent.

\section{Summary and Conclusions}
We have carried out a detailed study of the probability distributions
of the components of the smooth and staggered subsystem magnetizations
in the ground state of the critical spin-1/2 Heisenberg XXZ chain. 
We have shown through a combination of field theory and numerical 
calculations that appropriate ratios of the moments of these
probability distributions are universal. The probability distributions
of the longitudinal staggered subsystem magnetization is essentially
Gaussian. This is in contrast to the transverse component
$N^x(\ell)$, which can be thought of as the order parameter of the
magnetic quasi long-range order in the XXZ chain. The corresponding
probability distribution in the ground state $P^x(m,\ell)$ is very
broad and for attractive and weakly repulsive interactions it exhibits
two slight maxima at fairly large values of $N^x(\ell)$. We have shown
that the dominant features for large subsystem sizes can be understood
in terms of a mapping to the boundary sine-Gordon field theory. The
subleading contributions also exhibit scaling, and their calculation
in a field theory framework is an interesting open problem. The
behaviour of the smooth subsystem magnetization $S^\alpha(\ell)$ is
rather different from the staggered one. The generating function for
the moments of the longitudinal component decays as a power law in
subsystem size and can be accurately determined using Luttinger liquid
theory. The corresponding probability distribution $P_S^z(m,\ell)$ is
extremely narrow and centred around zero. This is perhaps not
surprising as $S^z(L)$ is conserved as a result of the U(1) symmetry
of the XXZ Hamiltonian. The probability distribution for the
transverse component $S^x(\ell)$ is narrow and exhibits a single
maximum at zero as well. The corresponding generating function of the
moments $G^x_\ell(\theta)$ decays exponentially in $\ell$. We have
shown that close to the antiferromagnetic point $\Delta=1$ its
behaviour for intermediate values of $\ell$ can be determined by
renormalization group improved perturbation theory. This calculation
does not account for the exponential decay seen for large values of
$\ell$. The description of the large-$\ell$ regime by field theory
methods remains an interesting open problem.

\acknowledgments
We are grateful to Pasquale Calabrese for collaboration in the early
stages of this project and for numerous important discussions. We thank
Paul Fendley and Austen Lamacraft for enlightening conversations. This
work was supported by the EU Horizon 2020 research and innovation
programme under Marie Sklodowska-Curie Grant Agreement No. 701221
(MC), the EPSRC under grant EP/N01930X/1 (FHLE) and by the Clarendon
Scholarship fund (SG). 
\appendix

\section{Variance of the subsystem magnetization}
\label{app:variances}

The two-point functions in the XXZ chain have been determined in the
framework of perturbed Luttinger liquid theory by Lukyanov and
Terras\cite{lukyanovterras}. After inversion of the spin quantization
axes on all odd sites their result is given by \fr{corrs2}, where the
amplitudes are
\bea
A&=&\frac{1}{2(1-\eta)^2}\, 
      \bigg[\frac{\Gamma(\frac{\eta}{2-2\eta})}
                 {2\sqrt{\pi} \Gamma(\frac{1}{2-2\eta})}
      \bigg]^{\eta}
   \exp\bigg\{-\int_0^{\infty}\frac{dt}{t} 
      \Big(\frac{\sinh(\eta t)}{\sinh(t)\cosh((1-\eta)t)}
           -{\eta}\, e^{-2 t}\, \Big) \bigg\}\ ,\nn
{\tilde A}&=&\frac{2}{\eta (1-\eta)}\, 
      \bigg[\frac{\Gamma(\frac{\eta}{2-2\eta})}
                 {2\sqrt{\pi} \Gamma(\frac{1}{2-2\eta})}
      \bigg]^{\eta+\frac{1}{\eta}}\ \exp\bigg\{-\int_0^{\infty}\frac{dt}{t} 
      \Big(\frac{\cosh(2 \eta t) e^{-2 t}-1}
                {2\sinh(\eta t)\sinh(t)\cosh((1-\eta)t)}
   +\frac{1}{\sinh(\eta t)}-
\frac{\eta^2+1}{\eta}\, e^{-2 t}\, \Big) \bigg\}\ ,\nn
B&=&
\left[\frac{\, \Gamma(1/\eta)}{\Gamma\big(1-1/\eta\big)}\right]^2
             \bigg[ \frac{\Gamma\big(1+\frac{\eta}{2-2\eta}\big)}
                         {2\sqrt{\pi} \Gamma\big(1+\frac{1}{2-2\eta}\big)}
             \bigg]^{4/\eta-4}
   \bigg\{\, \frac{2\pi^2}{\sin^2(2\pi/\eta)}
         -\frac{\eta^2}{(1-\eta)(2-\eta)}-\psi'(1/\eta)
         -\psi'(3/2-1/\eta)\, \bigg\},\nn
{\tilde B}&=&(1-\eta)^2\ 
\frac{4\, \Gamma(1/\eta)}{\Gamma\big(1-1/\eta\big)}
             \bigg[ \frac{\Gamma\big(1+\frac{\eta}{2-2\eta}\big)}
                         {2\sqrt{\pi} \Gamma\big(1+\frac{1}{2-2\eta}\big)}
             \bigg]^{2/\eta-2}\ 2^{\frac{4}{\eta}-5}\ 
          \frac{\Gamma(\frac{1}{\eta}-\frac{1}{2})\,
                \Gamma(1-\frac{1}{\eta})}
               {\Gamma(\frac{3}{2}-\frac{1}{\eta})\, 
                \Gamma(\frac{1}{\eta})}\ .
\eea
The variance $s_x$ \fr{sx} is obtained from the two-point function by
\be
s_x=\frac{1}{4}+\lim_{L\to\infty}\frac{2}{L}\sum_{j> k}\langle{\rm
GS}|S^x_jS^x_k|{\rm GS}\rangle.
\label{sx2}
\ee
The key identities for calculating $s_x$ are
\be
\sum_{n=1}^\infty\frac{1}{n^\gamma}=\zeta(\gamma)\ ,\qquad
\sum_{n=1}^\infty\frac{(-1)^n}{n^\gamma}=\big(2^{1-\gamma}-1\big)\zeta(\gamma)\ ,
\ee
where $\zeta(x)$ is the Riemann zeta function. These identities show
that all terms in the expansion of the two-point functions in fact
contribute to $s_x$, irrespective of how fast their power law decays
are. However, for large $\gamma$ the dominant contribution comes from
the $n=1$ terms in the corresponding sums, i.e. from the
non-universal short-distance behaviour. This shows that it is useful
to take the short-distance behaviour of the correlators into account as
precisely as possible. For example, we can decompose $s_x$ as
\be
s_x=\frac{1}{4}+2\langle
S^x_2S^x_1\rangle+2\sum_{n=2}^\infty 
\langle S^x_{n+1}S^x_1\rangle\ ,
\ee
Using the Lukyanov-Terras result \fr{corrs2} we the obtain the following
approximate expression
\bea
s_x&\approx&\frac{1}{4}+2\langle S^x_2S^x_1\rangle+
\frac{A}{2}\left[(2^{1-\eta}-1)\zeta(\eta)+1\right]
-\frac{AB}{2}\left[\zeta\left(\eta+\frac{4}{\eta}-4\right)
\big(2^{5-\eta-4/\eta}-1\big)+1\right]\nn
&&-\frac{\tilde{A}}{2}\left[\zeta\left(\eta+\frac{1}{\eta}\right)-1\right]
-\frac{\tilde{A}\tilde{B}}{2}\left[
\zeta\left(\eta+\frac{3}{\eta}-2\right)-1\right].
\eea
The nearest neighbour correlator can be simply obtained from the
ground state energy per site and equals
\bea
\langle
0|S^x_{j+1}S^x_j|0\rangle&=&
-\frac{1}{4\pi\sin(\pi\nu)}\int_{-\infty}^\infty\frac{dz}{\sinh z}
\frac{\sinh\big((1-\nu)z\big)}{\cosh(\nu z)}
+\frac{\cos(\pi\nu)}{4\pi^2}\int_{-\infty}^\infty\frac{dz}{\sinh z}
\frac{z\cosh(z)}{\big(\cosh(\nu z)\big)^2}\ ,
\label{corrs1}
\eea
where we have defined $\nu=\frac{1}{\pi}{\rm arccos}(-\Delta)$. We
note that the exact next-nearest-neighbour spin-spin correlators are
also available in the literature and can be taken into account in the
same way. The amplitude $B$ has an unphysical singularity as
$\eta\rightarrow 
2/3$. This merely means that the perturbative calculation of
Ref.~\onlinecite{lukyanov} needs to be redone for $\eta\approx
2/3$. As the contribution of the $B$ term to $s_x$ becomes important
only as we approach the XXX point, we simply drop it in the
following. We can compare the field theory results to direct DMRG
computations of the variance. In Table~\ref{tab:1} we show the results
for several values of $\Delta$. We see that in the attractive regime
we have good agreement, while in the repulsive regime the agreement is 
worse. 

\begin{table}[ht]
\center
\caption{Numerical values for the variance of transverse fluctuations
 extracted from the two-point function.}
\label{tab:1}
\begin{tabular}{|c|c|c|c|c|c|c|c|c|}
\hline\hline
$\Delta$ & $-0.8$ & $-0.6$  & $-0.4$ & $-0.2$& $0$ & $0.2$ & $0.4$ &
$0.6$\\
\hline
$s_x$ &  $0.101$ & $0.081$ & $0.064$ & $0.049$ & $0.036$ & $0.024$ &
$0.013$ & $0.005$
\\
\hline
\text{DMRG} & $0.101$ & $0.079$ & $0.063$ & $0.049$ & $0.037$ & $0.026$ &
$0.017$ & $0.009$\\
\hline\hline
\end{tabular} 
\end{table}


\begin{thebibliography}{99}
\bibitem{remark1}
This calculation is easily extended to the case of a non-zero magnetic
field. Here $\mathbb{B}$ is a block-Toeplitz matrix.
\end{thebibliography}
\end{document}